\documentclass{svjour3}    

\usepackage[a4paper, margin=1in]{geometry}
\usepackage{todonotes}
\usepackage{svg}
\usepackage{array}\newcolumntype{P}[1]{>{\centering\arraybackslash}p{#1}}
\usepackage{ifthen}
\usepackage{xcolor}
\usepackage{natbib}
\usepackage{verbatim}
\usepackage{placeins}
\usepackage{lscape}
\usepackage{multirow}
\usepackage{graphicx}
\usepackage{float}
\usepackage{lipsum}  
\usepackage{todonotes}
\usepackage{hyperref} 
\usepackage{multicol}
\usepackage{booktabs}
\usepackage{amsmath}
\usepackage{rotating}
\usepackage[commandnameprefix=ifneeded, xcolor=blue]{changes}
\usepackage{marginnote}

\definecolor{lavenderblush}{rgb}{0.93, 0.88, 0.90}

\colorlet{shadecolor}{lavenderblush}

\newcommand{\mytitle}{The competent Computational Thinking test (cCTt): a valid, reliable and gender-fair test for longitudinal CT studies in grades 3-6}

\newcommand{\myshorttitle}{The competent Computational Thinking test (cCTt): a valid, reliable and gender-fair test}

\newboolean{ISanonymous}
\setboolean{ISanonymous}{false}

\ifthenelse{\boolean{ISanonymous}}{\newcommand*{\Anonymous}{}}

\makeatletter
\newcommand\notsotiny{\@setfontsize\notsotiny{6.31415}{7.1828}}
\makeatother

\newcommand{\KWA}{Computational Thinking}
\newcommand{\KWB}{Assessment}
\newcommand{\KWC}{Primary School}
\newcommand{\KWD}{Validation}
\newcommand{\KWE}{Developmental appropriateness}
\newcommand{\KWF}{Psychometrics}

\DeclareUnicodeCharacter{0301}{\'{e}}

\newcommand{\MOBOTS}{MOBOTS Group, EPFL}
\newcommand{\LEARN}{Centre LEARN, EPFL}

\newcommand{\CHILI}{CHILI Laboratory, EPFL}
\newcommand{\RJC}{Computer Science Department, Universidad Rey Juan Carlos}

\newcommand{\UNED}{Faculty of Education, Universidad
Nacional de Educación a Distancia (UNED), Madrid, Spain}

\newcommand{\AuthorLH}{Laila El-Hamamsy}
\newcommand{\AuthorBB}{Barbara Bruno}
\newcommand{\AuthorFM}{Francesco Mondada}
\newcommand{\AuthorJDZ}{Jessica Dehler Zufferey}
\newcommand{\AuthorMZC}{María Zapata-Cáceres}
\newcommand{\AuthorEMB}{Estefanía Martín-Barroso}

\newcommand{\AuthorMRG}{Marcos Román-González}

\newcommand{\EmailLH}{laila.elhamamsy@epfl.ch}
\newcommand{\EmailBB}{barbara.bruno@epfl.ch}
\newcommand{\EmailFM}{francesco.mondada@epfl.ch}
\newcommand{\EmailJDZ}{jessica.dehlerzufferey@epfl.ch}
\newcommand{\EmailMZC}{maria.zapata@urjc.es}
\newcommand{\EmailEMB}{estefania.martin@urjc.es}

\newcommand{\EmailMRG}{mroman@edu.uned.es}

\begin{document}

\journalname{Technology, Knowledge and Learning
}

\title{\mytitle}
\titlerunning{\myshorttitle}

\ifthenelse{\boolean{ISanonymous}}
    {\author{Anonymous authors}}
    {\author{\AuthorLH$^{1, 2,}$\footnote[1]{Corresponding author : Laila El-Hamamsy, laila.elhamamsy@epfl.ch} \and
        \AuthorMZC$^{3}$ \and
        \AuthorEMB$^{3}$ \and
        \AuthorFM$^{1,2}$ \and
        \AuthorJDZ$^{2}$ \and
        \AuthorBB$^{4}$ \and
        \AuthorMRG$^{5}$}}

 \ifthenelse{\boolean{ISanonymous}}
    {\authorrunning{First author et al.}}
    {\authorrunning{L. El-Hamamsy et al. }} 

 \ifthenelse{\boolean{ISanonymous}}
    {
        \institute{Anonymous affiliations}
    }
    {
        \institute{$^1$ \MOBOTS \\
                   $^2$ \LEARN \\
                   $^3$ \RJC \\
                   $^4$ \CHILI \\
                   $^5$ \UNED \\
                   \email{\EmailLH, \EmailMZC, \EmailEMB, \EmailFM, \EmailJDZ, \EmailBB, \EmailMRG, }}
    }

\date{Received: date / Accepted: date}

\maketitle

\begin{abstract}

The introduction of computing education into curricula worldwide requires multi-year assessments to evaluate the long-term impact on learning. However, no single Computational Thinking (CT) assessment spans primary school, and no group of CT assessments provides a means of transitioning between instruments. 
This study therefore investigated whether the competent CT test (cCTt) could evaluate learning reliably from grades 3 to 6 (ages 7-11) using data from 2709 students. The psychometric analysis employed Classical Test Theory, Item Response Theory, {Measurement Invariance analyses which} include Differential Item Functioning, normalised z-scoring, and PISA's methodology to establish proficiency levels. 
The findings indicate that the cCTt is valid, reliable and gender-fair for grades 3-6, although more complex items would be beneficial for grades 5-6. Grade-specific proficiency levels are provided to help tailor interventions, with a normalised scoring system to compare students across and between grades, and help establish transitions between instruments. 
To improve the utility of CT assessments among researchers, educators and practitioners, the findings emphasise the importance of i) developing and validating gender-fair, grade-specific, instruments aligned with students' cognitive maturation, and providing ii) proficiency levels, and iii) equivalency scales to transition between assessments. 
To conclude, the study provides insight into the design of longitudinal developmentally appropriate assessments and interventions.


\end{abstract}

\section{Introduction and related work}

\subsection{The relevance of research on Computational Thinking assessments}

Research around Computational Thinking has been increasing significantly over the past two decades with studies touching {``different countries, subjects, research issues, and teaching tools hav[ing] also become more diverse in recent years''} \citep[p.1]{hsu_how_2018}. 
While there is no universally accepted definition of CT, \citet{brennan_new_2012}'s operational definition helps decompose CT into three dimensions: i) the \emph{concepts} that designers engage with as they program, ii) the \textit{practices} that they develop as they engage with these concepts, and finally iii) \textit{the perspectives} that they form regarding the world and themselves. Many researchers have advocated that CT is a competence that is not specific to CS, that all should acquire \citep{wing_computational_2006}, and that has potential for learning and meta-cognition \citep{yadav_computational_2022}, with recent studies having demonstrated the link between CT and other abilities \citep{xu_neural_2021, xu_relations_2022, li_development_2021, tsarava_cognitive_2022}. Therefore, it is not surprising to see an increasing number of countries looking to or presently introducing Computational Thinking (or the closely related Computer Science, or even more broadly Digital Education) in their curricula throughout K-12 \citep{weintrop_assessing_2021, european_reviewing_2022}. 
However, to be able to teach CT, guide students and provide feedback from the teachers' perspective \citep{hsu_how_2018}, or design and validate CT-interventions from the researchers' perspective, it is essential to have reliable and validated CT assessments spanning K-12 \citep{european_reviewing_2022}. 
Unfortunately, the use of validated CT assessments is something that \citet{tang_assessing_2020} noted was lacking in approximately 50\% of CT-related studies. From the practitioners' perspective, assessment issues need to be resolved for successful integration of CT in K-12 curricula \citep{cutumisu_scoping_2019}. 
This is because the {``purpose of an assessment is to facilitate student learning''} \citep[p.543]{guggemos_computational_2022} and {``validated assessments [...] measure students’ progress in meeting the learning outcomes prescribed by the programs of study''} \citep[p.652]{cutumisu_scoping_2019}. 
It thus becomes paramount to develop and guide  {``researchers and practitioners in choosing developmentally appropriate CT assessments''} \citep[p.651]{cutumisu_scoping_2019}. 
It is therefore not surprising to see that CT assessment {``is at the forefront of CT research [and] gathering the greatest interest of researchers''} \citep[p.16]{tikva_mapping_2021}.\\

\subsection{The lack of validated and reliable assessments at all levels of schooling, namely primary school}

According to \citet{tang_assessing_2020}'s meta review, CT assessments are provided in four formats. The first are \emph{portfolios}, which are the most common assessment format, but are likely to conflate with programming abilities, cannot be used in pre-post assessments and are difficult to scale up, in addition to being difficult to standardise and thus provide evidence of validity and reliability. The second are \emph{interviews}, which suffer from the same limitations as Portfolio assessments. The third are \emph{surveys}, which assess dispositions and attitudes towards CT (e.g. the Computational Thinking Scale, \citealp{korkmaz_validity_2017}), but do not provide insight into competencies. Finally, we find \emph{traditional tests} that should be used in combination with other assessment methods \citep{grover_systems_2015, roman-gonzalez_combining_2019} as they lack insight into the students' thought processes and, when too closely tied to a specific environment, may conflate with programming abilities. Tests however have the advantage of being psychometrically validated, and being usable in pre-post test designs and in large scale studies which is why we focus on this assessment format. 
Unfortunately, few CT tests have undergone extensive validation procedures \citep{tang_assessing_2020, cutumisu_scoping_2019} (e.g. through psychometric analyses). 
For instance, while Bebras tasks are often employed in CT-related research as they provide a large pool of items spanning K-12 with varying difficulty, they have undergone limited psychometric validation \citep{hubwieser_playing_2014, bellettini_how_2015}. 
Some researchers have even created their own ad-hoc Bebras-based assessments \citep{rojas-lopez_learning_2018, del_olmo-munoz_computational_2020} without providing evidence of reliability and validity. 
Even more preoccupying is that \citet{araujo_exploring_2017} {found that Bebras scores only ``moderately correlated'' with students' grades in an introductory programming course and concluded that it is ``not very likely that CT measures can be derived from the Bebras test as it is currently designed''} \citep[p.1]{araujo_exploring_2017}. 
In the past few years, several CT test-based assessments have been developed to be agnostic from specific programming environments and evaluated for validity and reliability.
Considering the increase of CT-studies and CT-curricula throughout K-12 worldwide \citep{weintrop_assessing_2021}, it is important that validated assessments span the full range of formal education. 
As i) a single validated assessment, the CTt \citep{roman-gonzalez_which_2017, roman-gonzalez_combining_2019} covers most of secondary school (grades 5-10, ages 10-16), and ii) as most efforts to develop and validate assessments for CT have focused on secondary and tertiary education \citep{zapata-caceres_computational_2020, roman-gonzalez_combining_2019, tsarava_cognitive_2022}, we choose to focus here on CT-assessments for primary school. \\

\subsection{An increasing number of primary school Computational Thinking assessments but without the means to do longitudinal assessments}
\label{sec:gap_longitudinal}

Considering instruments for primary school that provide evidence of reliability and validity, and excluding those that are i) dependent on specific programming environments \citep{marinus_unravelling_2018, kong_validating_2022}, ii) were administered to small samples \citep{marinus_unravelling_2018, parker_development_2021, chen_assessing_2017}, or iii) require manual annotations \citep{chen_assessing_2017, gane_design_2021}, we have identified the following psychometrically validated CT assessments. 
Firstly, the TechCheck \citep{relkin_techcheck_2020} and its variants \citep{relkin_techcheck-k_2021,relkin_cross-grade_2022} are validated instruments with good pyschometric properties and are developmentally appropriate for K-2 students (ages 5-7). 
Secondly, the Computational Thinking Assessment for Chinese Elementary Students (CTA-CES, \citealp{li_development_2021}) was designed and validated for Chinese students in grades 3-6 (ages 9-12). 
Unfortunately, the authors did not do a grade-specific analysis to see how the instrument performed for each grade, despite observing significant differences between students in grades 3-4 and 5-6. 
Provided cultural differences which may also exist between Chinese students and students in other regions of the world, it would be interesting to have other instruments covering such a wide range of grades in primary school. 
Finally, the Beginners' CT test (BCTt, \citealp{zapata-caceres_computational_2020}) was developed for students in grades 1-6 on the basis of the CT test (CTt) for secondary school (grades 5-10, \citealp{roman-gonzalez_which_2017}). 
The BCTt uses a similar approach as the CTt to assess CT, with a focus on CT-concepts \citep{brennan_new_2012}, but employing {``simplified and friendlier''} tasks \citep[p.5]{tsarava_cognitive_2022}. 
During the validation of the BCTt \citep{zapata-caceres_computational_2020} a ceiling effect was observed for upper grades. 
The competent CT test (cCTt) was thus developed and demonstrated good reliability and validity for students in grades 3-4 through Classical Test Theory and Item Response Theory \citep{elhamamsy_competent_2022}, and was shown to be better suited for grades 3-4 than its counterpart \citep{el-hamamsy_comparing_2022}. 
One important element to note is that while the existing instruments increasingly cover the full range of primary school education, there is a lack of continuity or links between them which would permit having multi-year longitudinal assessments. 
This is despite the interest that researchers and practitioners involved in the evaluation of CT-related curricular reforms may have for such CT assessments~(\citealp{tsarava_cognitive_2022},e.g. in the context of analysing the impact and sustainability of CT-related curricular reforms). 
Indeed, to the best of our knowledge:

\begin{enumerate}
    \item \emph{No single validated CT assessment currently spans primary school} like the CT test (CTt, \citealp{roman-gonzalez_which_2017, roman-gonzalez_combining_2019}) does in secondary school for grades 5-10 (ages 10-16). 
    This is not surprising given the significant differences often found even between 2 consecutive grades which require adapting the instruments to improve their validity. 
    This was in particular the case of the TechCheck \citep{relkin_techcheck_2020} (for which the researchers created two new versions \citep{relkin_techcheck-k_2021,relkin_cross-grade_2022} to improve the validity for students throughout K-2), and the competent CT test (cCTt, \citealp{elhamamsy_competent_2022}) which adapted the Beginners' CT test (BCTt, \citealp{zapata-caceres_computational_2020}) to improve validity and reliability of the instrument for students in grades 3-4. 
    
    \item \emph{No group of validated CT assessments provide a means of easily passing from one assessment to another when following students over multiple years}, e.g. by providing equivalency scales allowing to switch between one and the next. 
    This is neither the case of the TechCheck and its variants in K-2, nor the CT test (CTt, \citealp{roman-gonzalez_which_2017, roman-gonzalez_combining_2019}) and its variants the Beginners' CT test (BCTt, \citealp{zapata-caceres_computational_2020}) and the competent CT test (cCTt, \citealp{elhamamsy_competent_2022}). 
\end{enumerate}

\subsection{{Aim of the present study}}

{To address the issues in the CT literature when it comes to conducting longitudinal studies in primary school (see section \ref{sec:gap_longitudinal}), the aim of this study is to expand the psychometric validation of the cCTt (originally validated for grades 3-4) to a large cohort of grade 3-6 students and to introduce additional pyschometric analyses to continue to contribute to the cCTt's validation}. As such, in the present article, we are interested in the following research question:

\begin{itemize}
    \item \textbf{RQ}: Is the cCTt valid, reliable and fair with respect to gender for students in grades 3-6 (ages 7-11)? And how do the psychometric properties compare across these grades?
\end{itemize}

{The investigation builds on the methodology of the original cCTt validation for 2 additional grades (5-6) with additional psychometric analyses (see Table \ref{tab:novelty}) that serve three main objectives and contribute to the literature on CT assessments by:} \\

\begin{table}[h]
\caption{{Contributions of the present study compared to the original cCTt validation and evaluation methodology with metrics}}
\label{tab:novelty}
\begin{tabular}{p{2.75cm}P{1.75cm}P{1.75cm}p{8cm}}
\toprule
Evaluation Methodology & Initial cCTt study & Present study & Metrics \\ \midrule
Reliability through Classical Test Theory & Grades 3-4 & Grades 3-6 & Item difficulty, point-biserial correlation, Cronbach's $\alpha$ and the drop $\alpha$ \\
Reliability through Item Response Theory & Grades 3-4 & Grades 3-6 & 2-parameter logistic IRT models with item difficulty and discrimination, Item Characteristic Curves, Item Information Curves, Test Information Functions \\
Construct validity through Confirmatory Factor Analysis & Grades 3-4 & Grades 3-6 & $\chi^2$ test, Comparative Fit Index (CFI), Tucker Lewis Index (TLI),  Root Mean Square Error or Approximation (RMSEA) and Standardized  Root Mean Square Residual (SRMR) and factor loadings to establish the model fit's adequacy \\
Test-level Measurement invariance analyses through Confirmatory Factor Analysis & x & Grades 3-6 & Confirmatory Factor Analysis model comparisons with increasing constraints (loadings, intercepts and residuals per group; namely gender) and $\chi^2$ difference tests to establish whether the construct-related properties of the test vary by group \\
Item-level Measurement invariance analyses through Differential Item Functioning & x & Grades 3-6 & Differential Item Functioning using the Mantel-Haenszel, Logistic Regression Likelihood Ratio Test (LRT), and Generalized Lord's $\chi^2$ to determine whether response patterns differ according to groups \\
Normalized scoring & x & Grades 3-6 & Z-scoring and percentiles to facilitate cross-study and cross-assessment comparisons \\
Student proficiency profiles & x & Grades 3-6 & IRT models and Wright Maps to determine which items are indicative of students belonging to a given proficiency profile\\ \bottomrule
\end{tabular}
\end{table}

(1) \emph{Determining whether the cCTt can be used to cover 4 years of primary school with a single instrument} for longitudinal assessments, and including student profiles to help researchers, practitioners and educators understand the impact of their interventions and adapt accordingly. Indeed, for researchers, it would be possible to determine how an intervention affects individuals or groups of individuals over extended periods of time, therefore providing more reliable insight into the relevance of an intervention beyond short term interventions which are presently the most common in the field. For educators on the other hand, such assessments may help establish student profiles and help target their classroom interventions and offer tailored support that is adapted to students' needs \citep{guggemos_computational_2022}. Finally, for practitioners, to evaluate the longitudinal impact of widespread computing-related curricular reforms, it is essential to have validated assessments throughout K-12 to follow students' progress over time. This not only helps establish the impact of such reforms, but also helps determine how to adjust the learning objectives per grade and the pedagogical content developed by curriculum designers. In all cases, we argue that i) these three types of stakeholders and their needs should be accounted for when developing and validating CT assessments, and that ii) it is essential to have families of assessments that cover K-12, with the possibility of carrying over information from past years and from other instruments to have access to baseline performance assessments.  \\
    
(2) \emph{Providing a first step towards establishing means of intra- or inter-assessments comparisons} through normalised z-scoring to establish percentiles (see section \ref{sec:CTT}) and IRT analyses to establish {student proficiency profiles (see section \ref{sec:IRT})}. Intra-assessment equivalencies help compare performance across grades. Inter-assessments equivalencies on the other hand may serve two purposes. One is to compare performance between different families of assessments which may be relevant when comparing the outcomes of studies having used different types of assessments. The other, is to be able to link performance between consecutive assessments that are part of a same family. This is particularly relevant for example to link the performance of the cCTt and CTt in longitudinal studies, notably considering that certain percentiles have already been published for students in grades 5-6 and 7-8 (see Table 4 in \citealp{roman-gonzalez_which_2017} for the aggregate grade 5-6 and 7-8 percentiles and Table 6.22 in \citealp{roman_gonzalez_code_2016} for the grade specific percentiles). The present study therefore provides a first step towards conducting a comparative study between the cCTt and the CTt and establishing an equivalency scale between them. The latter is indeed only possible once we have identified whether a comparison would be beneficial in grades 5-6, and at which point an equivalency scale is necessary to switch between the cCTt and CTt. \\

(3) \emph{Establishing the fairness of the instrument} with respect to gender through {measurement invariance analyses (see section \ref{sec:MeasurementInvariance_theory}) at the test-level with Confirmatory Factor Analysis (see section \ref{sec:TestlevelMeasurementInvariance_theory}) and at the item-level with }Differential Item Functioning (see section \ref{sec:DIF_theory}). This is particularly important when considering that significant differences have been found between boys' and girls' scores when validating CT assessments \citep{elhamamsy_competent_2022, roman-gonzalez_which_2017, kong_validating_2022} and during interventions \citep{mouza_multiyear_2020}. However, without conducting gender-related {measurement invariance analyses},  it is not possible to establish whether the differences found are due to the instrument being biased, or true differences between boys' and girls' abilities. Given that gender gaps are often related to stereotypes and stereotype threat, these may start as early as 2-3 years old \citep{bers_state_2022}, with several studies having found evidence of computer science related gender gaps starting in kindergarten \citep{sullivan_girls_2016, master_gender_2021}, it is critical to have validated assessments that have proven their gender-fairness in order to be sure that targeted interventions help address the gender divide in computing.

\section{Methodology}
\label{sec:methodology}
\subsection{The competent CT test (cCTt)}

\begin{figure}[!htbp]
    \centering
    \includegraphics[width=0.47\textwidth]{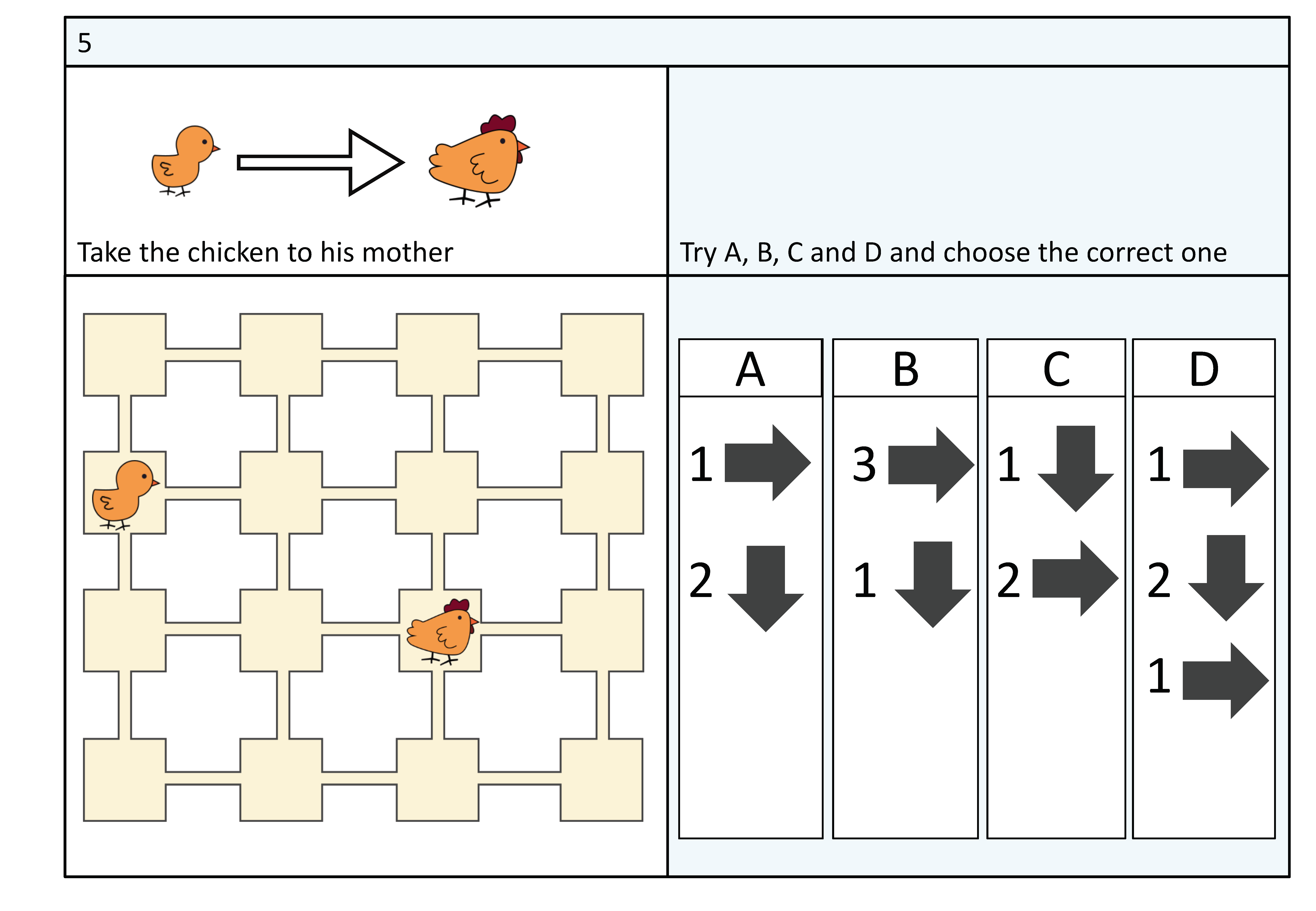}
    \includegraphics[width=0.47\textwidth]{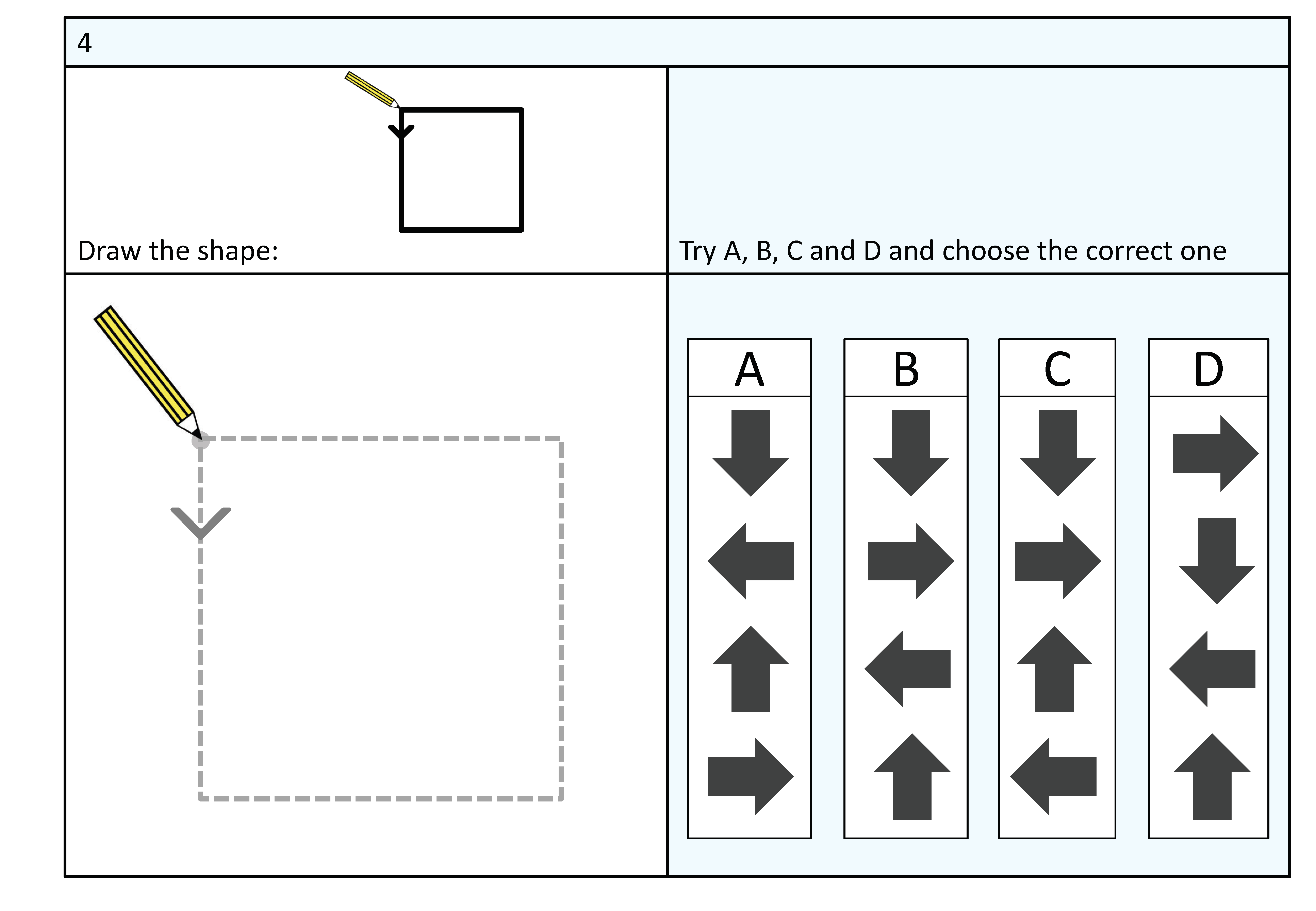}
    \caption{Two main question formats of cCTt: grid (left) and canvas (right) (Figure taken from \citealt{elhamamsy_competent_2022}).}
    
    \label{fig:cCT_format}
\end{figure}

The cCTt\footnote{Please note that the cCTt items are presented in \citet{elhamamsy_competent_2022} and an editable version is available upon request to the co-authors of the article.} is a psychometrically validated 25-item multiple choice CT assessment for upper primary school (originally validated for grades 3-4, \citealp{elhamamsy_competent_2022}). 
The cCTt is derived from the BCTt \citep{zapata-caceres_computational_2020}, itself an adaptation of the CT test for primary school \citep{roman-gonzalez_which_2017, roman-gonzalez_combining_2019}, which is considered to be agnostic from existing programming languages and adapted to students without prior experience in CS or CT. 
The cCTt proposes items of progressive difficulty targeting the CT-concepts defined by \citet{brennan_new_2012} by employing grid-type and canvas-type questions (see Fig.~\ref{fig:cCT_format}) to evaluate notions of sequences, simple loops (only one instruction is repeated), complex loops (two or more instructions are repeated), conditional statements, while statements and combinations of these concepts (see Table~\ref{tab:cCTt_description}). 
The instrument was validated in two stages \citep{elhamamsy_competent_2022}.

\begin{table}[!htbp]
    \centering
    \caption{cCTt number of questions per block and question types (Table adapted from \citealt{elhamamsy_competent_2022})}

    \label{tab:cCTt_description}
    \begin{tabular}{l|cccc}
    \toprule
     & & \multicolumn{2}{c}{cCTt} & \\ \midrule
    Blocks & Grid (3x3) & Grid (4x4) & Canvas & Total \\
    Sequences &  1 & 1 & 2 & 4 \\
    Simple loops &  0 & 4 & 0 & 4 \\
    Complex loops &  0 & 5 & 2 & 7 \\
    Conditional statements &  1 & 3 & 0 & 4 \\
    While statements &  1 & 3 & 0 & 4 \\
    Combinations &  0 & 2 & 0 & 2 \\ \midrule
    Total & 3 & 18 & 4 & 25 \\ \bottomrule
    \end{tabular}
\end{table}

In the first stage, experts evaluated the face, construct and content validity of the instrument through a survey and focus group. In the second stage, the test was administered to students and analysed through Classical Test Theory and Item Response Theory. The psychometric analysis of the students' data showed that the test has adequate reliability (Cronbach's \begin{math}\alpha=.85\end{math}), a wide range of item difficulties, and adequate discriminability for students in grades 3-4 \citep{elhamamsy_competent_2022}. The objective of the present study is to :
\begin{itemize}
    \item to extend this validation procedure to students in grades 5-6
    \item {to expand the initial validation procedure to include (i) gender-related measurement invariance analyses to establish the instruments' gender fairness, (ii) normalized z-scoring as a first step towards establishing equivalency scales, and (iii) student proficiency profiles to indicate what students should be able to master or not at a given level and help teachers tailor support provided to students \citep{guggemos_computational_2022}.}
\end{itemize}  
 
\subsection{Participants and data collection}

To validate the cCTt in grades 5-6 we used data collected from a Computer Science curricular reform project in \ifdefined\Anonymous Anonymous region in anonymous country (details removed for peer review)\else the Canton Vaud in Switzerland\fi. Within this project, all in-service grade 1-6 teachers were trained to introduce CS into their practices prior to the data collection, but do so with varying degrees. There is no imposed amount of activities to teach. Therefore, some teachers choose to teach no activities, while others teach the activities of their choosing, with most just teaching one or two activities per year \ifdefined\Anonymous (Anon. a 202X, Anon. b 202X). \else \citep{el-hamamsy_computer_2021, el-hamamsy_tacs_2022}. \fi 
The teachers in grades 5-6 are from 7 schools in urban and rural areas and they are therefore considered to be representative of the region and were therefore asked to participate in the assessment of their students' CT competencies. The objective was to conduct a pre- post-test experimental design to evaluate the impact of CS activities taught in between in the context of a novel CS curricular reform. 
While the study itself is not the focus of the present article, the data from the pre-test acquired between November 2021 and January 2022 is of interest as the cCTt was administered to 1209 grade 5-6 students (585 in grade 5, 624 in grade 6, see Table \ref{tab:participants})\footnote{The data is publicly accessible on Zenodo \citep{elhamamsy_cCTtdataset36_2023}
}. The administration of the instrument followed the protocol established by the parent-BCTt \citep{zapata-caceres_computational_2020} and its adaptation for the cCTt. The cCTt administration was done by accompaniers who were hired and trained to go into the schools and administer the test to all the students. {The test was administered electronically, and the students' responses were graded as either 0 (incorrect) or 1 (correct) according to the cCTt's grading scheme provided in \citet{el-hamamsy_datasetcCTt_2022}.}

\begin{table}[htbp!]
\centering
\caption{Number of Participants According to Age and Gender. {Grade 3-4 students would have had teachers trained to teach CS education for 3 years, while the grade 5-6 students would have teachers trained to teach CS education for 4 years. Please note however that CS education in the region is taught on a voluntary basis with most teachers teaching only 1-2 two-hour CS activities per year, and a lower proportion of teachers being willing to teach CS in grades 5-6 \ifdefined\Anonymous (Anon. a 202X, Anon. b 202X). \else \citep{el-hamamsy_computer_2021, el-hamamsy_tacs_2022}. \fi}}
\footnotesize
\label{tab:participants}
\begin{tabular}{l|llll|l}
\toprule
\multirow{2}{*}{\textbf{Gender}} &  \multicolumn{4}{c|}{\textbf{Grade}}&  \multirow{2}{*}{\textbf{Total}} \\ \cline{3-4} 
{} &   3 &   4 &   5 &   6 &  \\ 
\midrule
\textbf{Boys}     &  376 &  379 &  289 &  317 &  1361 \\
\textbf{Girls}     &  333 &  369 &  296 &  307 &  1305 \\ \midrule
\textbf{Total} &  709 &  748 &  585 &  624 &  2666 \\
\bottomrule
\end{tabular}
\end{table}

In order to provide a full picture of the psychometric properties of the cCTt for grades 5-6, a detailed comparison is made with data collected from grade 3-4 students in the same region which was used for the initial cCTt validation in \citet{elhamamsy_competent_2022} and is publicly available on Zenodo \citep{el-hamamsy_datasetcCTt_2022}. Please note that i) no student took the test twice, they are all unique and ii) the students are considered to be comparable as they are from the same administrative region and therefore follow the same curriculum. This implies that the cohorts are equivalent an their performances in the cCTt can be compared. \\

\begin{figure}[h]
    \centering
    \includegraphics[width=0.8\textwidth]{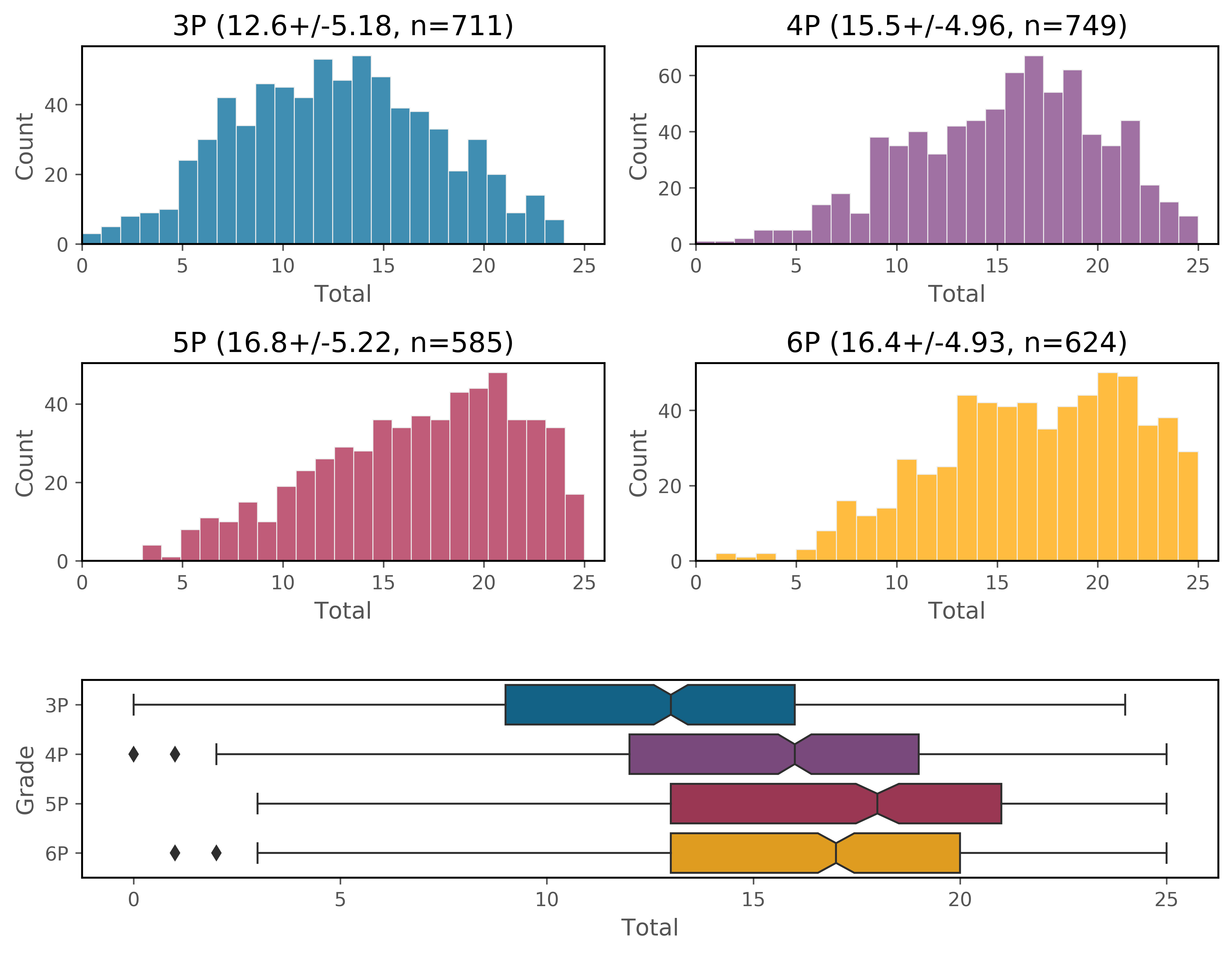}
    \caption{Distribution of scores across grades}
    \label{fig:score_distrib}
\end{figure}

The distribution of scores obtained per student and grade can be seen in Fig.~\ref{fig:score_distrib} with the descriptive statistics being provided in Table~\ref{tab:cCTt_stats}. The {skewness} and kurtosis values are within the acceptable range for normal univariate distribution according to \citep{gravetter2020essentials} although the {skewness} to the left increases between grades 3, 4 and 5-6, {indicating the start of a ceiling effect in grades 5-6 which will have implications on the cCTt's properties in these grades as we will see in the section \ref{sec:results}. Despite the start of a ceiling effect in grades 5-6, a one-way ANOVA reveals significant differences according to students' grades ($F(3)=95$, $p<0.0001$). Dunn's test for multiple comparisons in Table~\ref{tab:Dunn_grade_comp} indicates that there are significant differences between all grades, except between grades 5 and 6 where a plateau appears to have been reached. The fact that grade 6 students had similar (albeit slightly lower scores) than grade 5-6 students is likely due to the context in which the data collection took place. Indeed, students in grade 6 experienced two years with teachers who were still new and being trained to teach CS education (when they were in grades 3 and 5), and received less CS education in grades 5-6 compared to grades 3-4. }

\begin{table}[h]
    \centering
    \caption{Descriptive statistics of the cCTt per grade. Please note that acceptable limits to prove normal univariate distribution are $[-2;+2]$ for {Skewness} and $[-7;+7]$ for Kurtosis \citep{gravetter2020essentials}, with values close to 0 being desirable.}    
    \label{tab:cCTt_stats}
    \footnotesize
    \begin{tabular}{lcccccccc}
    \toprule
    Grade &      N &  Mean &  Std. error of the mean &  Std. deviation &  {Skewness} &  Kurtosis &  Min &   Max \\
    \midrule
    3 &  711 &  12.6 &                   0.19 &            5.18 &    0.021 &    -0.60 &  0 &  24 \\
    4 &  749 &  15.5 &                   0.18 &            4.96 &   -0.35 &    -0.41 &  0 &  25 \\
    5 &  585 &  16.8 &                   0.22 &            5.22 &   -0.49 &    -0.55 &  3 &  25 \\
    6 &  624 &  16.4 &                   0.20 &            4.93 &   -0.42 &    -0.46 &  1 &  25 \\
    \bottomrule
    \end{tabular}
\end{table}

\begin{table}[h]
    \centering
    \caption{Dunn's test for multiple comparisons between grades with Benjamini-Hochberg p-value correction (minimum Cohen's $D=0.13$ to achieve a statistical power of .8).}
    \label{tab:Dunn_grade_comp}
    \footnotesize
    \begin{tabular}{rccc}
    \toprule
    {} &  3P &  4P & 5P  \\
    \midrule
    4P & $\Delta=2.87$pts, $p<0.0001$, $D=0.57 $ &
    \\
    5P & $\Delta=4.22$pts, $p<0.0001$, $D=0.81 $ & $\Delta=1.35$pts, $p<0.0001$, $D=0.27 $ 
    \\
    6P & $\Delta=3.80$pts, $p<0.0001$, $D=0.75 $ & $ \Delta=0.93$pts, $p=0.0013$, $D=0.19 $ & $\Delta=0.42$pts, $p=0.1345$, $D=0.08$ (n.s.) \\
    \bottomrule
    \end{tabular}
\end{table}

\subsection{Psychometric analysis}
\label{sec:psychometric analysis}

The objective of the study is to establish the psychometric validity (i.e. does the instrument measure exactly what it aims to measure? \citealp{souza_psychometric_2017}) and reliability (i.e. does the instrument reproduce a result consistently in time and space? \citealp{souza_psychometric_2017}) of the cCTt for students in grades 5-6, and to compare these results to those obtained with data from students in grades 3-4 for whom the instrument has already been validated. \\

Two complementary approaches \citep{o_a_comparative_2016, de_champlain_primer_2010} to analyse the validity and reliability are leveraged with the rationale and methodologies being detailed in the following sections: 

\begin{enumerate}
    \item Classical Test Theory (see section~\ref{sec:CTT}), to provide the instruments' difficulty, reliability and discrimination ability. However, Classical Test Theory often suffers from test-dependency and sample dependency \citep{hambleton_comparison_1993, devellis_classical_2006}, in addition to not being able to separate the test and person characteristics. 
    \item Item Response Theory (IRT, see section~\ref{sec:IRT}), to provide item difficulty and discriminability in a more test- and sample- independent way through the latent ability scale \citep{hambleton_comparison_1993, dai_comparison_2020, jabrayilov_comparison_2016, xie_item_2019}. More specifically, IRT looks to estimate the probability of a student getting a given item correct and intends to be generalisable beyond the sample of students being measured. This thus makes it possible to conduct the inter-grade comparisons from the perspective of the latent ability scale (see section \ref{sec:IRT}).
\end{enumerate}

{To these we add test-level and item-level invariance analyses (see section \ref{sec:MeasurementInvariance_theory}) to determine whether the cCTt is gender-fair, and whether the properties of the cCTt change according to grade.} \\

The Classical Test Theory, IRT and invariance analyses are conducted in R (version 4.2.1, \citealp{r_core_team_r_2019}) using the following packages: lavaan (version 0.6-11, \citealp{rosseel_lavaan_2012}), CTT (version 2.3.3, \citealp{CTT_R}), psych (version 2.1.3, \citealp{revelle_psych_2021}), ltm (version 1.1.1, \citealp{rizopoulos_ltm_2006}), subscore (version 3.3, \citealp{subscore_Q3}), difR (version 5.1, \citealp{difR_package}), WrightMap (version 1.3, \citealp{wrightMap}) and TAM (version 4.1-4, \citealp{TAM_4.1-4}). Statistical analyses are conducted with one-way and two-way ANOVA, with Benjamini-Hochberg p-value correction to reduce the Type I error rate. When reporting the statistics, the minimum effect size required to achieve a power of .8 (considering the significance level - 0.05, sample size  - dependent on the test, number of groups - dependent on the test) is taken into account. 

\subsubsection{Classical Test Theory}
\label{sec:CTT}

Classical Test Theory focuses on test scores to evaluate the reliability of the considered instrument \citep{hambleton_comparison_1993} through 3 main metrics.

The first metric is the \emph{item difficulty index} which is defined as the proportion of \emph{correct} responses obtained per item. Please beware that according to this definition, which is commonly employed in the literature, items with low difficulty indices are hard questions while items with high difficulty indices are easy questions. Numerous thresholds have been employed in the literature to the purpose of identifying which items are too easy and which are too hard, but these are often arbitrary. As items with difficulties between $.4$ and $.6$ are considered to have maximum discrimination indices \citep{vincent_role_2020}, the thresholds often vary around these values. To be consistent with the thresholds employed in the validation of the cCTt for grades 3-4 \citep{elhamamsy_competent_2022},  we consider an item with a difficulty index that exceeds \begin{math}.85\end{math} as too easy, while items with a difficulty index below \begin{math}.25\end{math} are too hard and could be considered for revision.

The second metric is the \emph{point biserial correlation} which measures the discrimination between high ability examinees and low ability examinees. A point biserial correlation above \begin{math}.15\end{math} is recommended, with good items generally having point biserial correlations above \begin{math}.25\end{math} \citep{varma2006preliminary}. In this article, we consider a threshold of \begin{math}.2\end{math}, which is commonly employed in the field \citep{elhamamsy_competent_2022, chae_relationship_2019}. 

The third and final metric is the \emph{reliability of the scale} which is often computed using Cronbach's \begin{math}\alpha\end{math}, a measure of internal consistency of scales \citep{bland1997statistics}. Scales which are consistent will have high Cronbach's \begin{math}\alpha\end{math} while scales which are inconsistent, and thus less reliable, have low Cronbach's \begin{math}\alpha\end{math}. 
In the context of assessments when Cronbach's \begin{math}\alpha\end{math} is between \begin{math}.7<\alpha<.9\end{math} reliability is considered high, and between \begin{math}.5<\alpha<.7\end{math} it is considered moderate \citep{hinton_spss_2014, taherdoost_validity_2016}. The drop alpha which provides an estimate of the reliability of the scale should a given item be removed may also be computed. As such, we gain insight into whether removing a specific question would help improve the internal consistency of the test. 

To these, we further introduce percentiles computed through \emph{z-scoring} as done by \citet{relkin_cross-grade_2022} for the TechCheck and its variants. This approach allows us to compare the CT skills if students within and across grades and may therefore serve as a first step towards establishing equivalency scales between instruments. 

Unfortunately, as mentioned previously, Classical Test Theory tends to be sample-dependent \citep{hambleton_comparison_1993, elhamamsy_competent_2022}, meaning that comparing the results from two different populations may lead to inconsistent results. That is why Classical Test Theory should be complemented by other validation procedures which are considered sample-independent, such as IRT which is described below \citep{bean_item_2021}.   \\

\subsubsection{Item Response Theory (IRT)}
\label{sec:IRT}

Item Response Theory is a sample-independent validation procedure which considers that students have a given ability which is supposed to lead to consistent performance, independently of the test \citep{hambleton_comparison_1993}. By computing the probability of a person with a given ability to answer each question correctly (measured in standard deviations from the mean), IRT is more likely to generalise beyond a specific sample of learners \citep{xie_item_2019} and provide consistency between two different populations \citep{dai_comparison_2020, jabrayilov_comparison_2016}.

\paragraph{IRT pre-requisits.}

Prior to applying IRT, one must verify whether we meet the unidimensionality criteria, and if not, to what degree this is misspecified, as the larger the misspecification, the bigger the impact on the estimated parameters. One approach that can be employed to verify unidimensionality is Confirmatory Factor Analysis \citep{kong_validating_2022}. As the data is binary, we employ an estimator which is adapted to this data type (Diagonally Weighted Least Squares). The goodness of fit of CFA models can be estimated using multiple metrics which can be either global (i.e. the distance between the CFA model and a perfect model, \citealp{xia_rmsea_2019}) or local (i.e. distance between the CFA model and the baseline model with the worst fit, \citealp{xia_rmsea_2019}). {Fit indices to assess CFA models should include:}
\begin{itemize}
    \item {The \begin{math}\chi^2 / df \end{math} ratio \citep{alavi_chi-square_2020, prudon_confirmatory_2015, elhamamsy_competent_2022} which should be $<3$ for good fit, $<5$ for acceptable fit \citep{kyriazos_applied_2018} }
    \item {The root mean square error of approximation or RMSEA which should be \begin{math}<.06\end{math} for good fit, \begin{math}<.08\end{math} for acceptable fit \citep{xia_rmsea_2019, hu_cutoff_1999, chen_assessing_2017}}
    \item {The standardised root mean square residual or SRMR which should be \begin{math}<.08\end{math} \citep{xia_rmsea_2019, hu_cutoff_1999}}
    \item {The comparative fit index (CFI) and Tucker Lewis index (TLI) which should be \begin{math}>.95\end{math} for good fit,  \begin{math}>.90\end{math} for acceptable fit \citep{kong_validating_2022}}
    \item {Cronbach's $\alpha$ for reliability of the scale for each factor which should be $>.7$ }
    \item {The factor loadings for each item which should be $>.3$}
\end{itemize}

Then, when analysing the results of IRT (as provided by the lavaan package), ANOVA may be employed to determine which model fits best (1PL, 2PL, 3PL or 4PL) by comparing the difference in log likelihood and degrees of freedom, and determining whether the difference is significant. Once the best model type has been selected, one should verify the local independence between pairs of item residuals for the selected model type with Yen's Q3 statistic \citep{yen_effects_1984} and make adjustments to ensure the independence between them. The resulting IRT model should then be evaluated using multiple fit indices \citep{alavi_chi-square_2020}{, including:}

\begin{itemize}
    \item {Yen's Q3 statistic \citep{yen_effects_1984} which should be $<.2$ for good fit, and $<.3$ for acceptable fit \citep{christensen_critical_2017}.}
    \item {Item discrimination values which are considered very low if in [$0.01$, $0.34$], low if in [$0.35$;$0.64$], moderate if in [$0.65$;$1.34$], high if in [$1.35$;$1.69$], and very high if $> 1.70$ \citep{baker_basics_2001}.}
    \item {Item difficulty values which are considered very easy if $<-2$, easy if in [$-2$;$-0.5$], medium if in [$-0.5$;$0.5$], hard if in [$0.5$,$2$], very hard if $>2$ \citep{hambleton_fundamentals_1991}.}
\end{itemize}

\paragraph{IRT Models.}

Several IRT models exist for binary response data: 1-Parameter Logistic (1-PL) where only difficulty varies across items, 2-Parameter Logistic (2-PL) where both difficulty and discrimination vary across items, 3-Parameter Logistic (3-PL) which considers that students may be able to guess the right answer, and the 4-Parameter Logistic (4-PL) which considers that even students with high ability may not respond correctly to a question. While we tested all four models, only the 1-PL and 2-PL models converged to stable solutions (see section \ref{sec:IRT_best_model}). As such, we only detail the characteristics of 1-PL and 2-PL models for the reader. Instruments are expected to have questions of varying difficulty, to be able to provide information over the spectrum of latent abilities. Instruments evaluated using 2-PL, 3-PL and 4-PL models should have good discriminability so that the items are better able to detect differences between the abilities of the respondents.  \\

The results of IRT are typically presented using three characteristic plots {which we illustrate below using questions Q7, Q8, Q16, and Q19 from the grade 3-4 cCTt data taken from \citet{el-hamamsy_comparing_2022}}\footnote{{Please note that these items were arbitrarily chosen for illustrative purposes with the objective of having difficulty and discrimination parameters that differed sufficiently between them.}}. 

\begin{figure}[!htpb]
    \centering
    \includegraphics[width=0.9\textwidth]{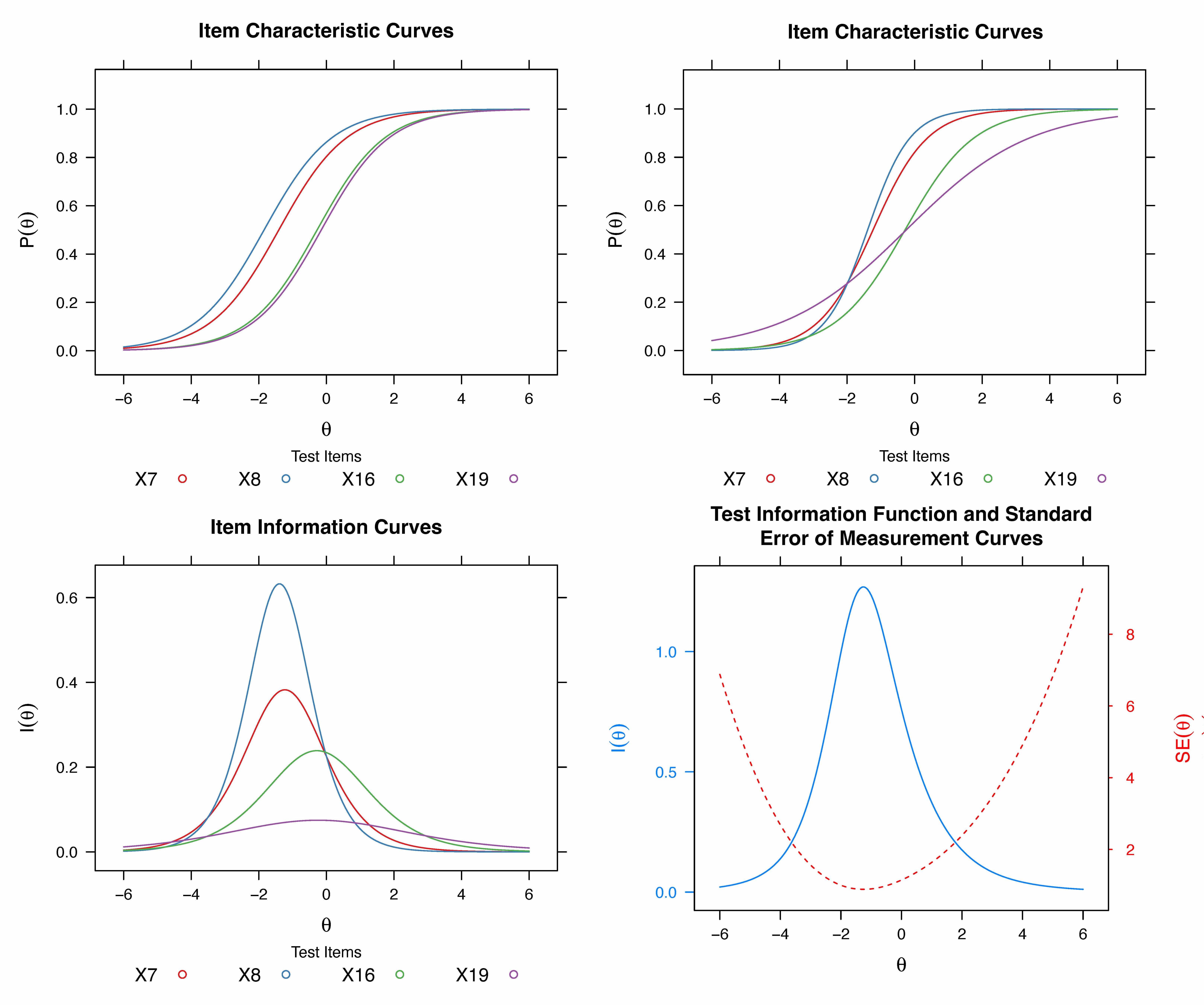}
    \caption{IRT Theory plots, taken from \citet{el-hamamsy_comparing_2022}\\
    \textbf{(A - top left)} Item Characteristic Curves for four items of equal discrimination (slope) and varying difficulty (using a 1-PL model on the cCTt test data). The item's difficulty (\begin{math}b_i\end{math}) is the x-value (\begin{math}\theta\end{math}) where the ICC reaches a \begin{math}y=.5\end{math} probability of answering correctly, and represents the number of standard deviations from the mean the question difficulty is. Items to the left of the graph are considered easier while items on the right are considered harder. \\
    \textbf{(B - top right)} Item Characteristic Curves (ICC) for four items (blue, red, green, purple) of varying difficulty and discrimination (using a 2-PL model on cCTt test data). In this example, blue and red items are of equal difficulty \begin{math}b_i\end{math} (\begin{math}y=0.5\end{math} crossing) and relatively similar discrimination \begin{math}a_i\end{math}, while items green and purple are of equal difficulty and varying discrimination. As the blue item is steeper, it has a higher discrimination than the red, green and purple items.\\
    \textbf{(C - bottom left)} Item Information Curves (IICs) for the items in \textbf{(B)}. The bell shaped curves represent the amount of information \begin{math}I_i\end{math} provided for each of the test's items according to the student's ability \begin{math}\theta\end{math}. These IICs vary in both maximum value (dependent on the item's discriminability, i.e. the ICC slope), and the x-value at which they reach it (the item's difficulty). Here, the blue and red curves, as well as the green and purple curves, have the same difficulty (they both reach their maximum around x=-2 and x=0 respectively), but are of different discriminability: the blue item discriminates more than the red, the red more than the green and the green more than the purple (steeper ICC slope, and higher maximum IIC value).\\
    \textbf{(D - bottom right)} Test Information Function (TIF, in blue) for the four items from Fig. \ref{fig:IRT_theory} \textbf{(B)} and \textbf{(C)}, and the standard error of measurement (SEM, in red). The TIF (blue) is the sum of the instrument's IICs from Fig. \ref{fig:IRT_theory} \textbf{(B)} and \textbf{(C)}, and the SEM is the square root of the variance. The TIF shows that the instrument displays maximum information around -2 and provides more information in the low-medium ability range than in the high ability range. The SEM (red) is at its lowest where the test provides the most information (maximum of the TIF) and at its highest where the test provides the least information (minimum of the TIF).
    }
    \label{fig:IRT_theory}
\end{figure}

The first {type of plots} are \textbf{Item Characteristic Curves} (ICCs, see Fig.~\ref{fig:IRT_theory} A and B) which are logistic (i.e. S-shaped) curves that indicate the probability of a student to answer an item correctly (y-axis, \begin{math}P(\theta)\end{math}) according to their latent ability (x-axis, \begin{math}\theta\end{math}). For all types of models (1-PL, 2-PL, 3-PL and 4-PL), each item is considered to have a given difficulty (varying latent ability, i.e. the x-value, starting which students of higher ability have a 50\% probability, y-value, of answering correctly). {According to \citet{de_ayala_theory_2022}, typical items have difficulties between -3 and +3. Easy items have scores below -2, average items having scores between -2 and +2 and hard items have scores above +2.}

When considering 2 parameter logistic (2-PL) models, the items also have varying discriminability as can be seen through varying ICC slopes (see Fig.~\ref{fig:IRT_theory} B): items with high discriminability will have a steep ICC slope, while items with low discriminability will have a gentle slope. 

The second plot consists of bell shaped \textbf{Item Information Curves}, or IICs (see Fig.~\ref{fig:IRT_theory} C) which indicate the amount of information provided by each item for a given latent ability. The maximum of each IIC is reached for the item's difficulty, i.e. the ability starting which students have at least a 50\% probability of answering correctly. Generally, items with high discriminability (steep ICC slopes) provide a lot of information at the item's difficulty. {According to \citet{de_ayala_theory_2022}, reasonably good discrimination values range from approximately 0.8 to 2.5.}

The last plot is the \textbf{Test Information Function} (TIF) with Standard Error of the Measurement (SEM). The TIF is the sum of the Item Information Curves of the items in the test (see Fig.~\ref{fig:IRT_theory} D). That is to say the TIF is the sum of the information provided over the latent ability scale by all the items of the test. The maximum of the TIF indicates where the instrument is better able to discriminate. The range of abilities where the test provides the least information also have the highest standard error of the measurement, meaning that the test is also less reliable for these ability estimates. Instruments may thus easily be compared according to the TIF scale to determine where they are able to provide more information about the students' ability. The reliability of the test at a given ability level may also be computed using the following formula $r=1-SEM(\theta)^2$ where $\theta$ is the ability.

\paragraph{Establishing student proficiency profiles. }

As an {``assessment is not an end in itself'' and given that ``it should contribute to promoting student learning}'' \citep[p.541]{guggemos_computational_2022}, proficiency profiles are established through IRT in order to improve the utility of the cCTt for researchers, educators and practitioners. The objective is to provide profiles {``ranging from very low levels of proficiency to very high levels of proficiency''} \citep[p.276]{oecd_pisa_2017}, which each {``describ[ing] what students typically know and can do at [said] level of proficiency''} \citep[p.276]{oecd_pisa_2017}. These profiles are established as was done by \citet{guggemos_computational_2022} for the CTt, by drawing inspiration from the approach employed by the OECD for the PISA assessments \citep{oecd_pisa_2014, oecd_pisa_2017} and the fact that a student of a given ability ``are increasingly more likely to complete tasks located at progressively lower points on the scale, and increasingly less likely to complete tasks located at progressively higher points on the scale'' \citep[p.276]{oecd_pisa_2017}. 
Therefore, based on the outputs of IRT models, multiple proficiency levels are constructed with respect to the test's item difficulties on the logit scale. To define the levels, the OECD proposed to consider proficiency levels of a width of 0.8 with the following criteria: 
\begin{itemize}
    \item Students of a given ability have a 62\% chance of answering an item of said difficulty correctly 
    \item Students at the bottom of a proficiency level (which is bounded by two difficulties on the logit scale) should have a 62\% chance of answering the questions at the bottom of the level correctly, a 42\% chance of answering the questions at the top of the level correctly, and an average 52\% correct response rate for all the items in that level
    \item Students at the top of a proficiency level (which is bounded by two difficulties on the logit scale) should have a 62\% chance of answering the questions at the top of the level correctly, a 78\% chance of answering the questions at the bottom of the level correctly, and an average 70\% correct response rate for all the items in that level
\end{itemize}

To achieve this therefore requires computing an adjusted difficulty per item, which for a 2PL model can be done using equation \ref{equ:adjusted_diff}, where $P_i$ represents the probability of answering an item correctly (and here should be equal to .62), $\theta$ represents the ability of the student to reach a probability $P_i$ of answering the item correctly, $b$ represents an item's difficulty, $a$ represents the item's discrimination. 

\begin{equation}\label{equ:adjusted_diff}
\begin{split}
    P(\theta, a, b) = \frac{e^{a(\theta-b)}}{1+e^{a(\theta-b)}}  \\
    \Leftrightarrow 0.62 (1+e^{a(\theta-b)}) = e^{a(\theta-b)} \\
    \Leftrightarrow \log{\frac{.62}{.38}} = a(\theta-b)) \\
    \Leftrightarrow \theta = \frac{1}{a}\log{\frac{.62}{.38}}+b\\
\end{split}
\end{equation}

\subsubsection{Measurement invariance}
\label{sec:MeasurementInvariance_theory}
{
Measurement invariance, also referred to as measurement equivalence, refers to the fact that an instruments' properties remains the same across groups \citep{teresi2006overview}. There are two complementary means of evaluating of invariance: test-level analyses of invariance to establish the equivalence of constructs across groups \citep{putnick2016measurement} and item-level analyses of invariance to establish whether response patterns for invidiual test items differ between groups. 
}

\paragraph{Test-level measurement invariance} 
\label{sec:TestlevelMeasurementInvariance_theory}

{is verified using Confirmatory Factor Analysis (CFA) where the factor structure here corresponds to each block of questions. Measurement invariance through CFA is done is multiple steps which we will explain for the case of gender. }

{First configural invariance is verified by running the CFA and grouping according to gender to estimate loadings, intercepts and residuals for boys and girls. If the model fits well then configural invariance is achieved. The model obtained here is referred to as the configural model.} 

{Second we test for metric invariance by constraining the loadings to be equal for boys and girls. This model, which we refer to as the metric model, is then compared with the configural model from the first step using a $\chi^2$ difference test. If the configuration model and the metric model do not differ significantly then metric invariance is achieved as the loadings do not differ significantly between boys and girls. }

{Third we test for scalar invariance by implementing the scalar model which constrains both loadings and intercepts of the model to be equal for boys and girls. Running the $\chi^2$ difference test between the scalar model and the metric model from step two should be non significant to indicate that scalar invariance is achieved. }

{Fourth we test strict invariance where we have a strict model which constrains loadings, interecepts and residuals between both groups. We then compare the strict model to scalar model and if the different is non significant then strict invariance is achieved.}


{Please note that we used Satorra Bentler's method of calculating the $\chi^2$ difference test between these nested models \citep{satorra2010ensuring}.}

\paragraph{Item-level measurement}
\label{sec:DIF_theory}
is verified using Differential Item Functioning (DIF). DIF is a statistical approach that is usually employed to determine whether there are biases in response patterns between groups for certain items (e.g. according to gender as done by \citealp{rachmatullah_toward_2022, sovey_gender_2022} or countries as done by \citealp{rachmatullah_toward_2022}). Similarly to IRT, DIF attempts to determine whether members of different groups who have the same underlying ability have a different probability of answering a question correctly. DIF therefore indicates whether an instruments' items are {``consistent and fair for all participants''} \citep[p.2]{sovey_gender_2022} and is an indicator of the validity of the instrument. More specifically, provided responses of students from different groups (e.g. gender or demographics), an item that is identified as being DIF should be reformulated in order for students of a given ability in both groups have the same probability of answering the questions correctly.

In the present context we employ DIF to determine whether i) there are differences in the response patterns between students in grades 3-4 and grades 5-6, and whether grade-specific IRT models should be employed instead of a single model, and ii) the instrument is fair with respect to gender as instruments such as these are often employed to determine whether gender gaps exist and whether interventions help address them given the lack of diversity in computing-related fields \citep{rachmatullah_toward_2022}. While it would have been interesting to establish fairness in terms of socio-economic status, this type of information is considered sensitive in the region. 

The DIF analysis was conducted with the following parameters:
\begin{itemize}
    \item the IRT model (1-PL, 2-PL, 3-PL or 4-PL) that is most appropriate for the four grades
    \item purification, an iterative approach that removes items flagged as DIF before repeating the search to ensure that all DIF items are identified \citep{magis_generalized_2011}
    \item Benjamini-Hochberg p-value correction to reduce the false discovery rate due to multiple comparisons
    \item multiple DIF detection methods {as the estimates vary across tests and our objective was to ensure that there were no items that were incorrectly flagged as non-DIF \citep{magis2020package}: \\
    (1) The Generalised Mantel-Haenszel $\chi^2$ statistic which is a non parametric test based on a chi-squared test on a 2x2 contingency table which looks at the number of correct and incorrect responses between the reference and focal group. This test is one of the most applied DIF detection methods and has the advantage of providing a value for magnitude of the difference between both groups, but does not account for respondents' ability and additional factors. \\
    (2) the generalised logistic regression Likelihood-ratio test (LRT) which computes the likelihood test statistic between two models, the first which constrains all parameters to be equal between the groups, and the second which allows the parameters to vary between groups. The advantage of such methods is that they can be used to control for other contextual or demographic-related factors.  \\
    (3) The generalised Lord's $\chi^2$ statistic, also referred to as the Wald test, which is computed based on the estimates of the difficulty and discrimination parameters for each group. Such an approach therefore accounts for respondents' ability but does not provide an estimate of the magnitude of the DIF between groups. The advantage of IRT methods is that they account for test-taker parameters (e.g. ability, item difficulty, discrimination, and guessing).} 
\end{itemize}

\section{Results}
\label{sec:results}

\subsection{Classical Test Theory for cCTt sample dependent reliability}
\label{sec:CTT_results}

\begin{figure}[htbp!]
    \centering
    \includegraphics[width=0.47\textwidth]{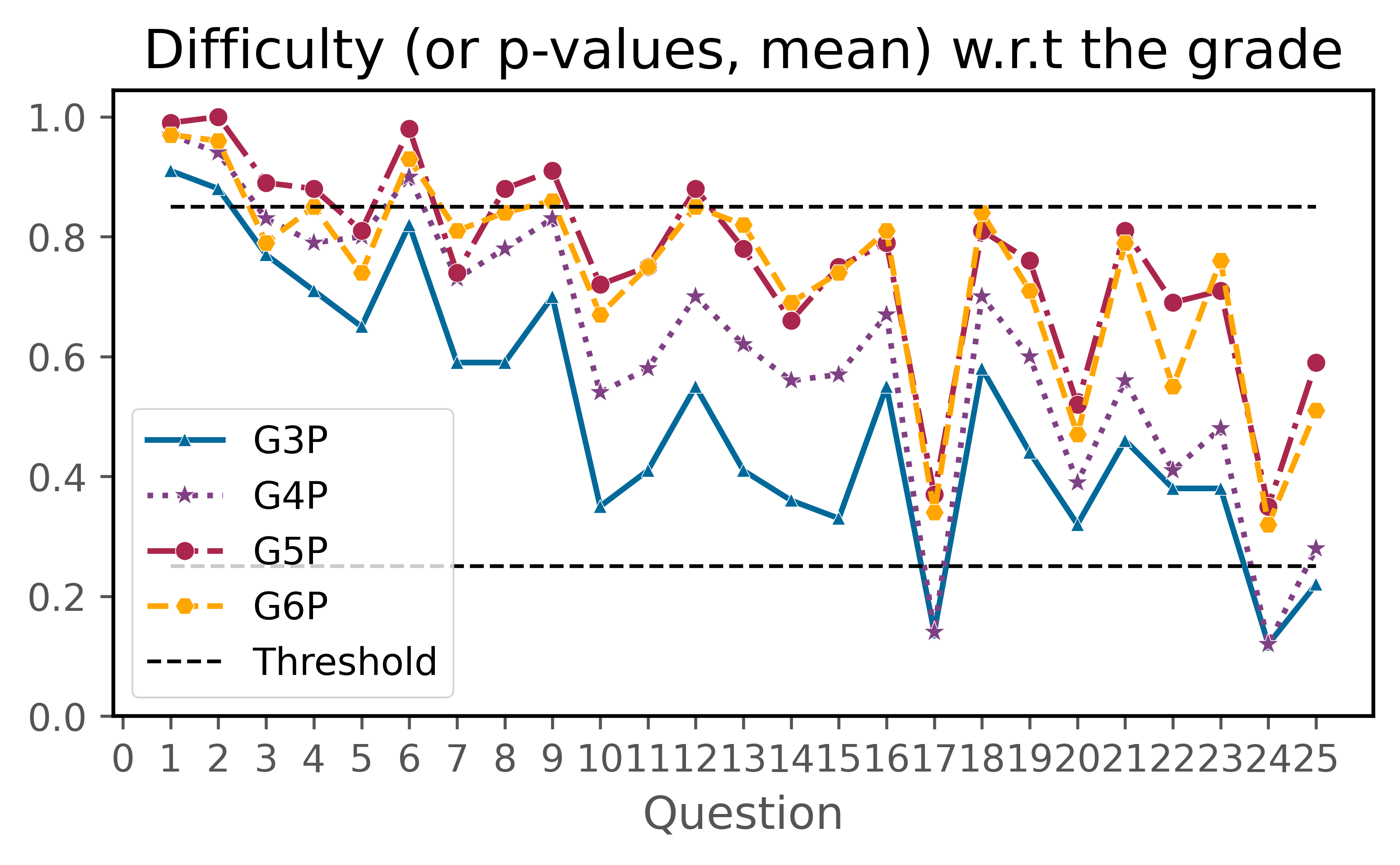}
    \includegraphics[width=0.47\textwidth]{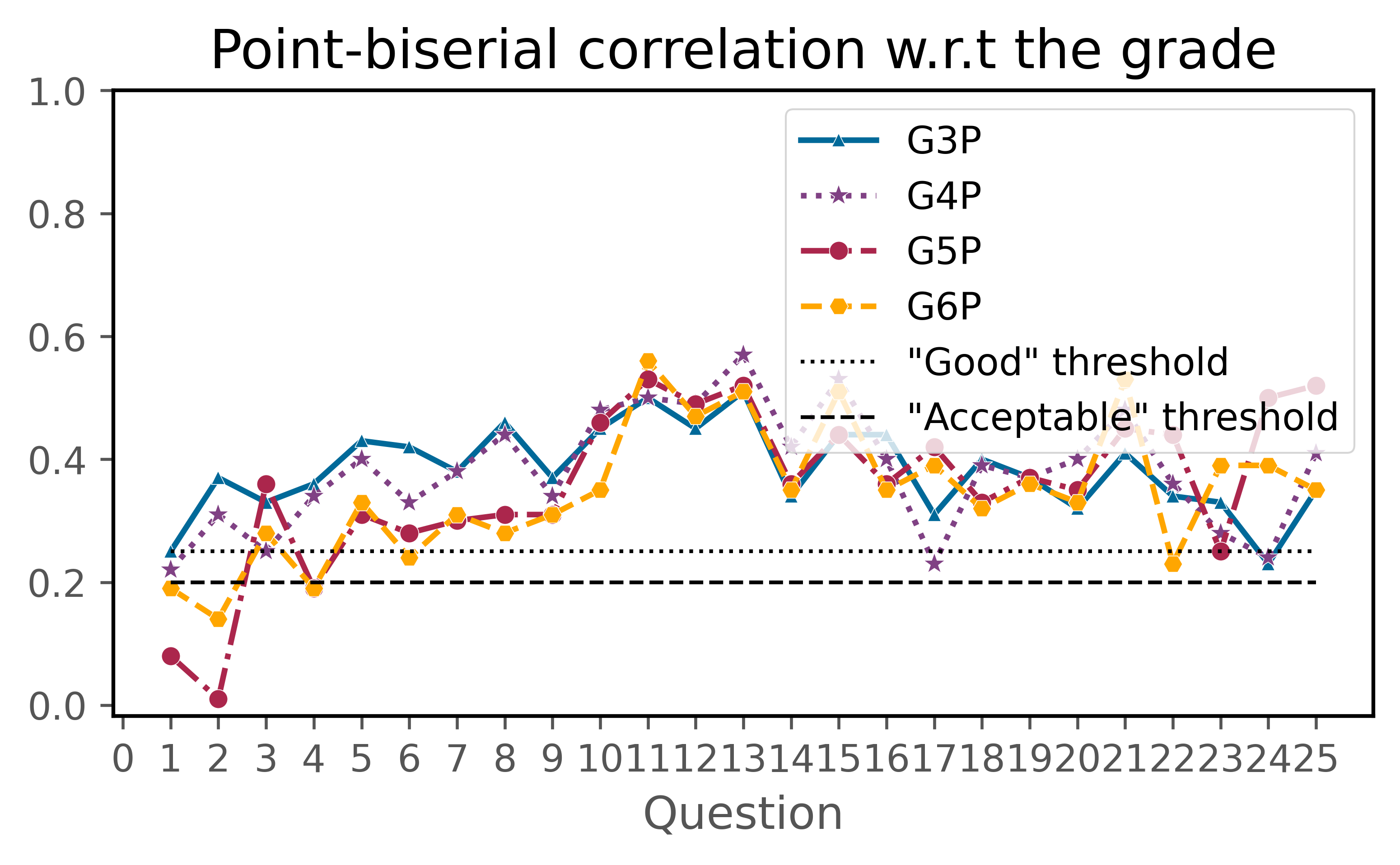}
    \caption{Classical Test Theory - Item Difficulty Index (left) and Point-biserial correlation (right). Please note that items with a difficulty index above the $.85$ threshold are considered too easy while items below the $.25$ threshold are considered too difficult. Similarly, items with a point-biserial correlation above the $.2$ threshold are considered acceptable while those above $.25$ are considered good.}
    \label{fig:CTT_item_difficulty_PBC}
\end{figure}

Fig.~\ref{fig:CTT_item_difficulty_PBC} reports the Classical Test Theory analysis results (difficulty indices and point biserial correlations) for all questions according to the students' grade (see Table \ref{tab:CTT_full_cCTt} in appendix \ref{app:CTT} for the detailed values). {As a reminder the difficulty index provides the average number of students having correctly responded to a question (ranges from $0$ to $1$ and should neither be too low or too high) while the point-biserial correlation indicates to what extent a question is able to discriminate well between students (ranges between $-1$ and $1$ and should be above $.2$).}\\

Starting with item difficulty indices, the trends observed for students in grades 3-4 appear consistent with those observed in grades 5-6 {(see Fig.~\ref{fig:CTT_item_difficulty_PBC}). Concretely, the individual test items appear to maintain their relative difficulties but are easier for older students. The result is that there are no more items that are classified as too difficult in grades 5-6}, but a larger number which are too easy compared to grades 3-4 (see Fig.~\ref{fig:CTT_item_difficulty_PBC}). 

When considering the point biserial correlation, items where students {have near perfect scores (difficulty scores close to 1) also have low point biserial correlations. This means that answering these items incorrectly is not representative of the students' overall performance on the test, and is more likely due to oversights}. 

Finally, when considering the reliability provided by Cronbach's $\alpha$: 
\begin{itemize}
    \item The cCTt exhibits good overall reliability for each grade ($\alpha_{3}=.84$, $\alpha_{4}=.84$, $\alpha_{5}=.83$, $\alpha_{6}=.82$). 
    \item {The individual item's drop $\alpha$, i.e. the reliability of the instrument should an item be removed, are always lower than the overall $\alpha$ for that grade (see Table \ref{tab:CTT_full_cCTt} in appendix \ref{app:CTT}). This means that removing an item will not improve the reliability of the test and that they should not be removed.}
\end{itemize}

Taking all of these elements into account, it would appear that the following number of questions could be revised to improve the validity of the instrument: 

\begin{itemize}
    \item 5 in grade 3: Q1 and Q2 which are too easy, Q17, Q24 and Q25 which are too hard
    \item 5 in grade 4: Q1, Q2, and Q6 which are too easy, Q17 and Q24 which are too hard
    \item 7 in grade 5: Q1-Q4, Q6, Q8, Q9 which are too easy, notably considering that Q1, Q2 and Q4 have low point-biserial correlations
    \item 6 in grade 6: Q1-Q2, Q4, and Q6, Q8, Q9 which are too easy, notably considering that Q1, Q2 and Q4 have low point-biserial correlations
\end{itemize}

While the two first items of the instrument would be the most important to revise, these could be considered as a means for the students to familiarise with the test and could simply be removed from the final score. This is particularly relevant for students in grades 5-6 as the point-biserial correlation is below the acceptable limit for these grades. Furthermore, given the mastery that students appear to have on sequences in grades 5-6, and the scores obtained on more advanced CT-concepts, it may be relevant to introduce more questions on advanced CT-concepts in their stead.

\subsection{Item Response Theory (IRT) for cCTt sample-agnostic reliability}

\subsubsection{Applicability of IRT}

{One key criteria to be able to apply IRT on a given dataset is that the data be unidimensional. As explained previously, one means of doing so is to run a Confirmatory Factor Analysis (CFA) with a single dimension encompassing all the test items. The advantage of doing so is that we know to what extent we meet unidimensionality criteria for IRT analysis, and what items may need to be removed in order to reach acceptable unidimensionality. Please note that the results of the unidimensional CFA here should not be interpreted as being the most representative of the underlying structure of our data. Indeed, the 6 factor CFA model which groups the questions according to the blocks defined in the test design (sequences, loops, complex loops, if-else statements, while statements, combinations of instructions) offers a significantly better fit (see multi-dimensional CFA results in appendix \ref{app:CFA_multidimensional}). The results of the CFA here should only be interpreted as ``do we have sufficient undimensionality to be allowed to run unidimensional IRT?''}. {The unidimensional CFA fit indices in Table \ref{tab:CFA_fit_unidim} indicate that we are closer to unidimensionality when removing Q2 which is why we apply the IRT analyses in the following sections without Q2.} \\

\begin{table}[h]
\centering
\caption{cCTt Unidimensional CFA Robust Fit Indices}
\footnotesize
\label{tab:CFA_fit_unidim}
\begin{tabular}{p{2.2cm}P{1.3cm}P{1.3cm}P{1.3cm}P{1.3cm}|P{1.3cm}P{1.3cm}P{1.5cm}P{1.3cm}}
\toprule
cCTt Questions & \multicolumn{4}{c|}{Q1-Q25} & \multicolumn{4}{c}{Q1 + Q3-Q25} \\
Grade & 3 & 4 & 5 & 6 & 3 & 4 & 5 & 6 \\ \midrule
df & 275 & 275 & 275 & 275 & 735.53 & 939.86 & 368.79 & 399.77 \\
$\chi^2$ & 805 & 992 & 667 & 420 & 252 & 252 & 252 & 252 \\
p-$\chi^2$ & 0 & 0 & 0 & 0 & 0 & 0 & 0 & 0 \\
$\chi^2$/df & 2.93 & 3.61 &	2.43 & 1.53 & 2.92 & 3.73 & 1.46 & 1.59 \\
CFI & .90 & .87 & .82 & .92 & .90 & .88 & .94 & .93 \\
TLI & .89 & .86 & .80 & .92 & .89 & .87 & .93 & .92 \\
RMSEA & .052 & .059 & .064 & .038 & .052 & .06 & .036 & .04 \\
RMSEA upper 90\% ci & .048 & .055 & .064 & .031 & .048 & .056 & .028 & 0.033 \\
RMSEA lower 90\% ci  & .056 & .063 & .070 & .046 & .056 & .065 & .044 & .048 \\
SRMR & .09 & .11 & .13 & .11 & .09 & .11 & .11 & .11 \\
Factor Loading & $\beta_{Qi}>.40$ & $\beta_{Qi}>.39$ & $\beta_{Qi}>.31$ & $\beta_{Qi}>.30$  & $\beta_{Qi}>.40$ & $\beta_{Qi}>.39$ & $\beta_{Qi}>.34$ & $\beta_{Qi}>.42$ \\
estimates &  & &  & except $\beta_{Q4}=.29$ &  & & except $\beta_{Q1}=.25$ & except $\beta_{Q4}=.29$ \\

Factor Loading & $p<.001$ & $p<.001$ & $p<.001$ & $p<.001$ & $p<.001$ & $p<.001$ & $p<.001$ & $p<.001$ \\
p-values &  &  & & except $p_{Q2}=.015$ & & & & \\ \bottomrule
\end{tabular}
\end{table}

\subsubsection{Identifying the most appropriate model}
\label{sec:IRT_best_model}
The 3-PL model did not converge to a stable solution for students in grades 3, 4 and 6, and the 4-PL model did not converge at all for any of the grades. As the objective is to compare the instruments, and use a single model type for the analysis, we fit the 1PL and 2PL models for each grade. Using ANOVA to compare the 1PL and 2PL models indicates that the 2PL model significantly improves the fit in all cases\footnote{The 2PL model \added(without Q2) was significantly better than the 1PL for all grades. Indeed : 
\begin{itemize}
    \item {In grade 3, the 1PL model has a higher AIC and BIC than the 2PL model ($AIC_{1PL}=18636$, $BIC_{1PL}=18750$; $AIC_{2PL}=18592$, $BIC_{2PL}=18812$). The Likelihood Ratio Test is significant ($LRT=89.61$, $df=23$, $p<.001$) and therefore indicates that the 2PL model should be preferred over the 1PL model.}
    \item {In grade 4, the 1PL model has a higher AIC and BIC than the 2PL model ($AIC_{1PL}=18081$, $BIC_{1PL}=18197$; $AIC_{2PL}=17970$, $BIC_{2PL}=18191$). The Likelihood Ratio Test is significant ($LRT=157.46$, df= 23, $p<.001$) and therefore indicates that the 2PL model should be preferred over the 1PL model.}
    
    \item {In grade 5, the 1PL model has a higher AIC and BIC than the 2PL model ($AIC_{1PL}=11708$, $BIC_{1PL}=11818$; $AIC_{2PL}=11624$, $BIC_{2PL}=11834$). The Likelihood Ratio Test is significant ($LRT=130.26$, df= 23, $p<.001$) and therefore indicates that the 2PL model should be preferred over the 1PL model.}

    \item {In grade 6, the 1PL model has a higher AIC and BIC than the 2PL model ($AIC_{1PL}=12967$, $BIC_{1PL}=13078$; $AIC_{2PL}=12865$, $BIC_{2PL}=13078$). The Likelihood Ratio Test is significant ($LRT=148.11$, df= 23, $p<.001$) and therefore indicates that the 2PL model should be preferred over the 1PL model.}

\end{itemize}
}.

Yen's Q3 statistic \citep{yen_effects_1984} to measure local independence indicates that none of the pairs of item residuals have a high correlation (all values $<.2$) for grades 3, 5 and 6, and that just 2 items have a statistic between $.2$ and $.3$ (acceptable) for students in grade 4. We can thus consider that local independence is not violated. 

\subsubsection{Comparing the instruments' grade-specific IRT models' properties}

\begin{figure}[htbp!]
    \centering
    \includegraphics[width=0.45\linewidth]{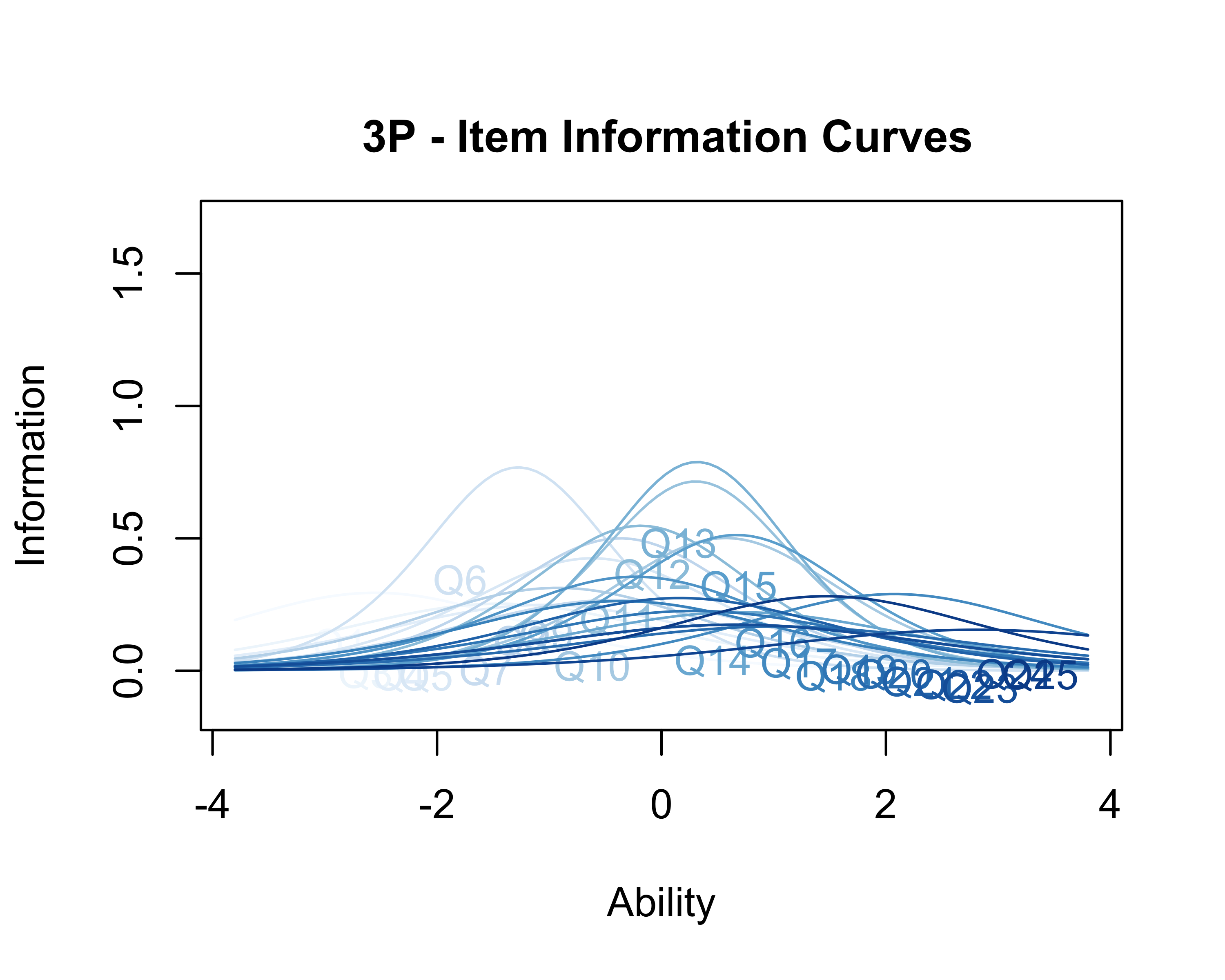}
    \includegraphics[width=0.45\linewidth]{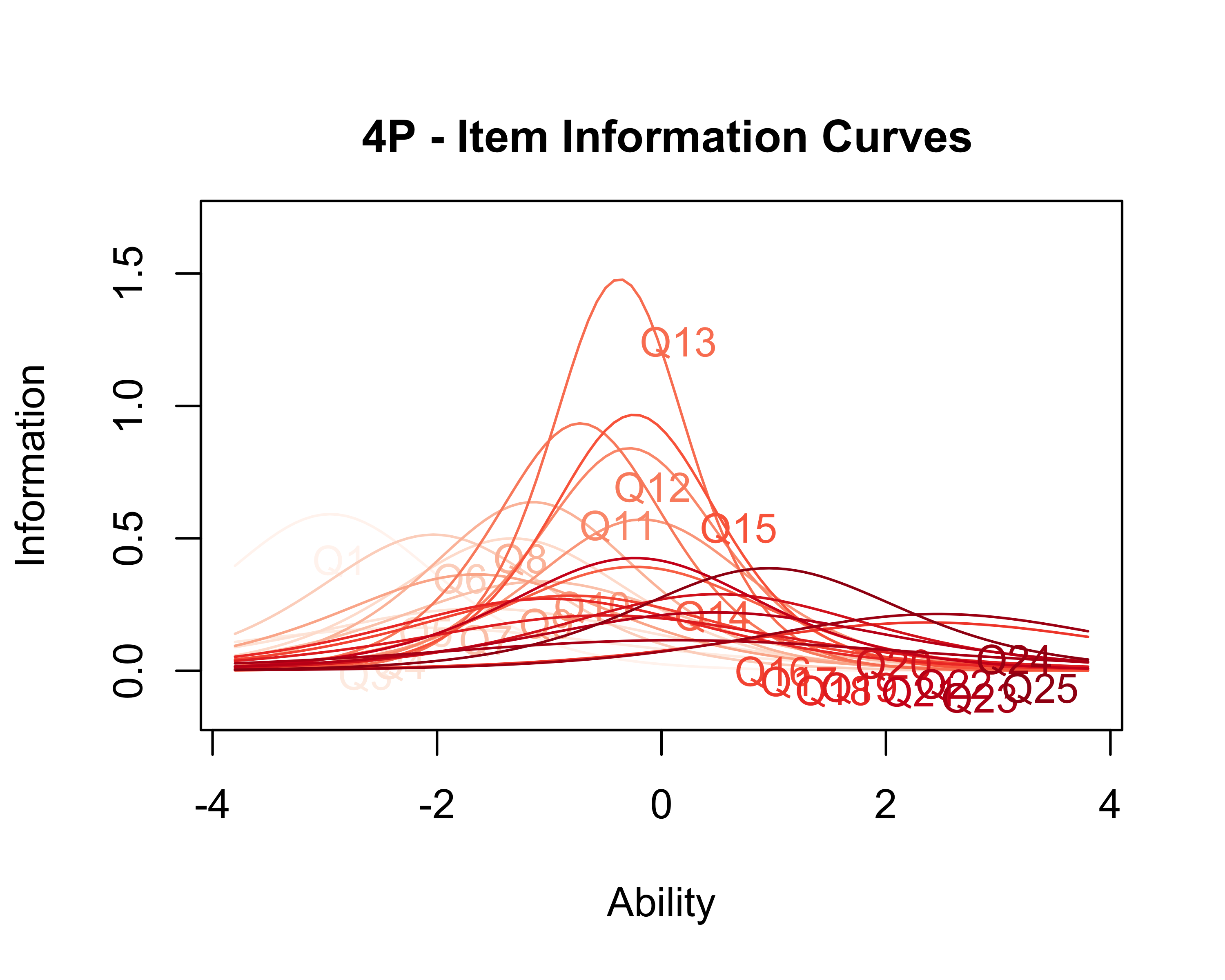}
    
    \vspace{-10pt}
    \includegraphics[width=0.45\linewidth]{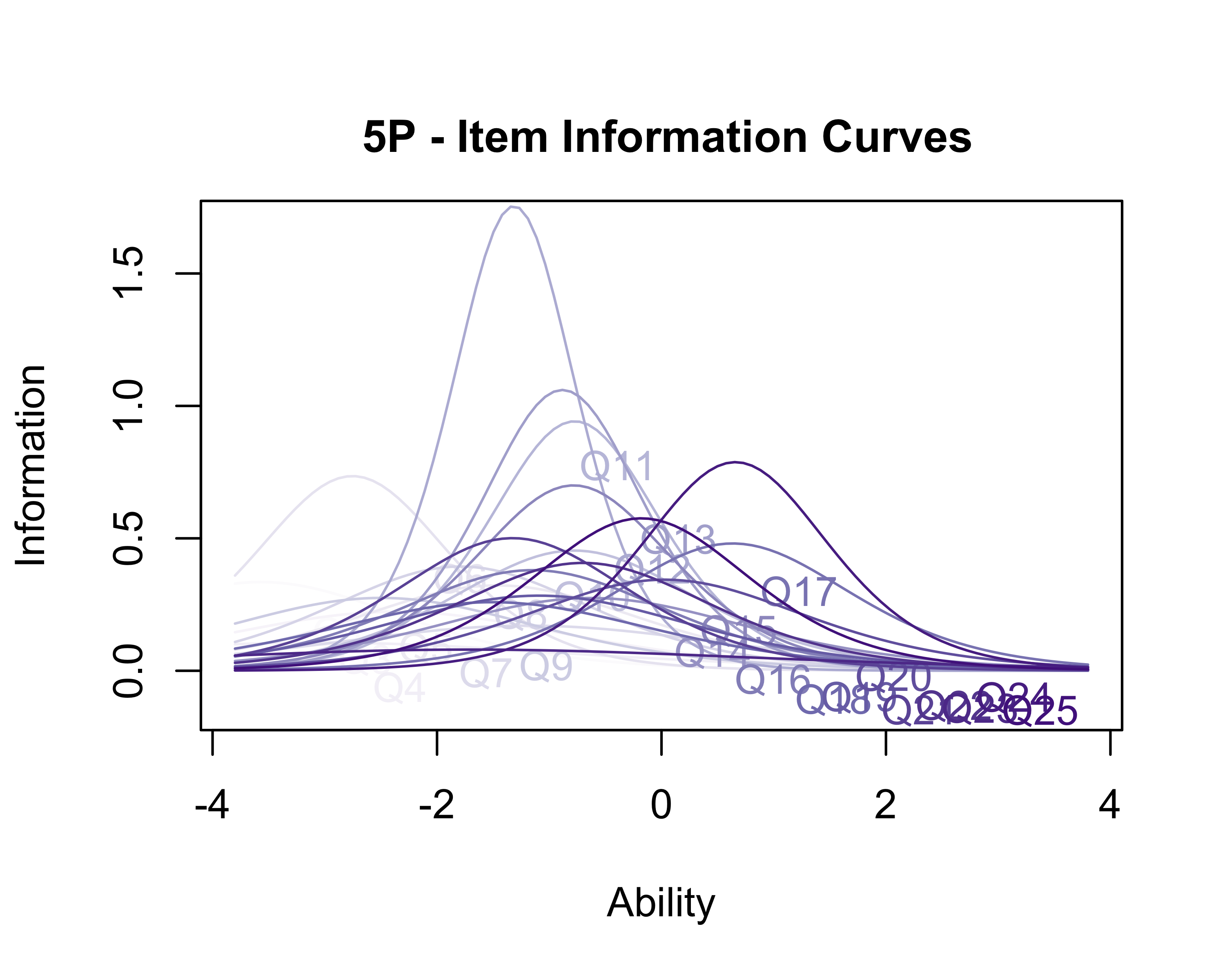}
    \includegraphics[width=0.45\linewidth]{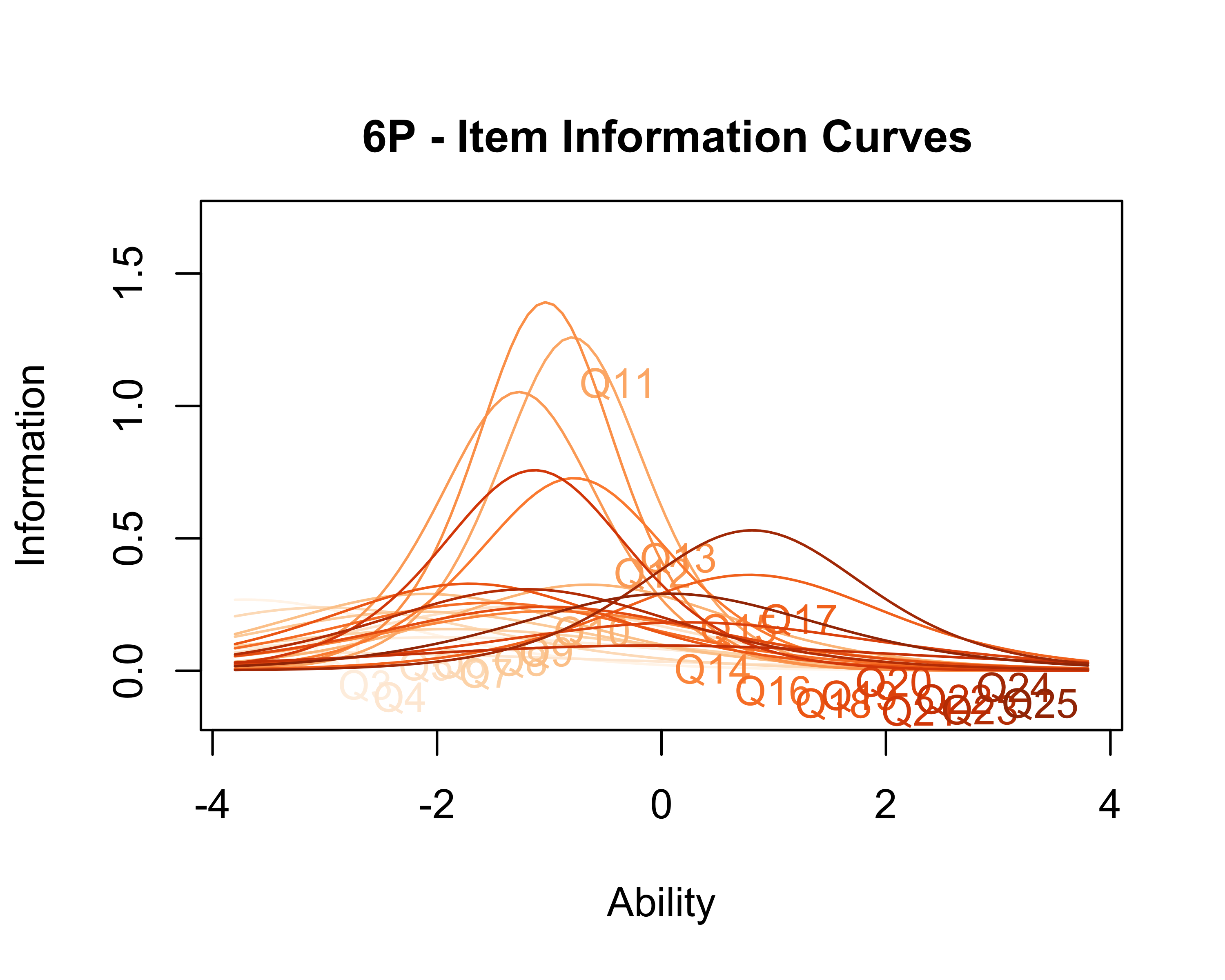}

    \vspace{-5pt}
    \caption{2-PL IRT Item Information Curves per grade.}
    \label{fig:IRT_IIC}
\end{figure}

\begin{figure}[htbp!]
    \centering
    \includegraphics[width=0.45\linewidth]{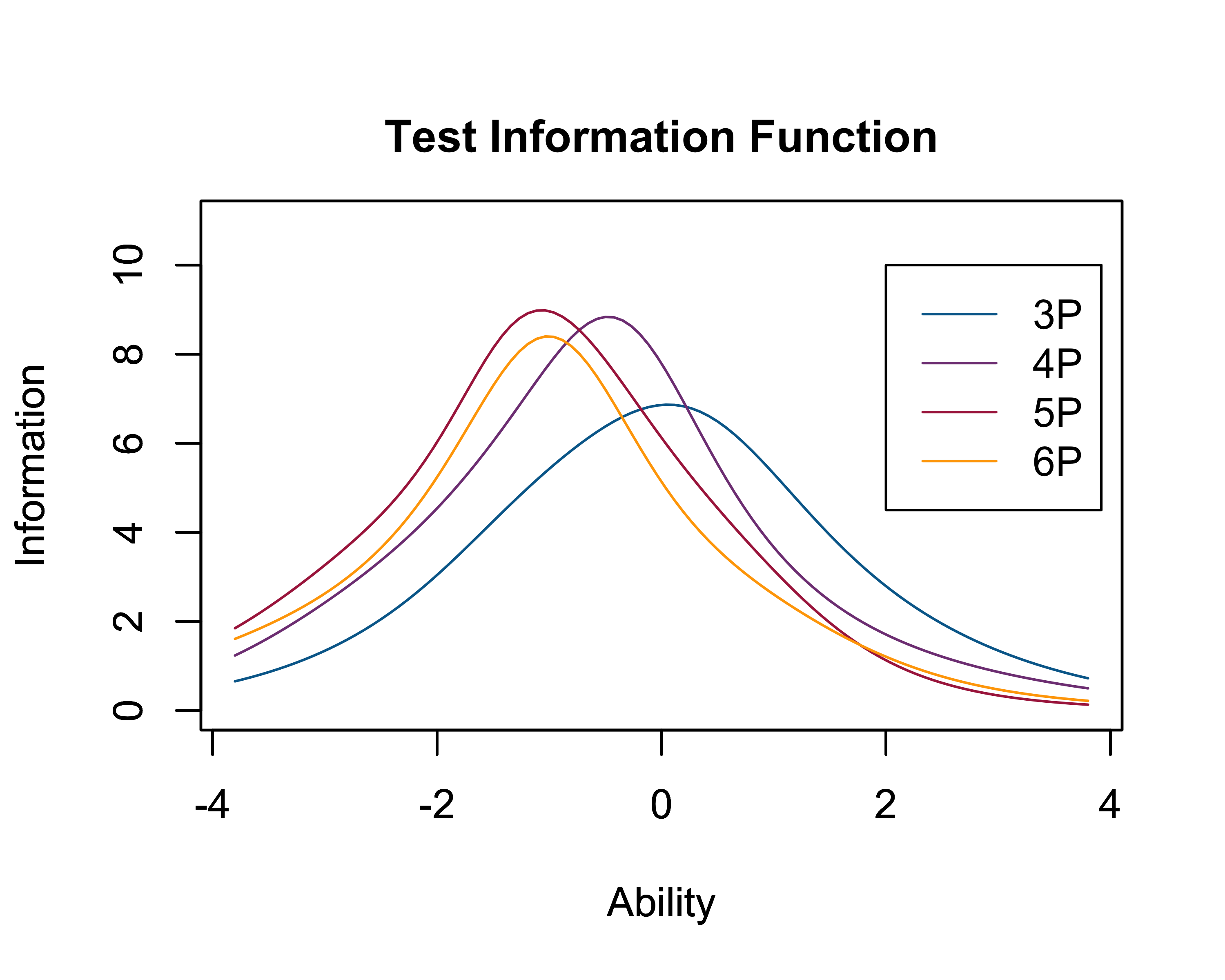}
    \includegraphics[width=0.45\linewidth]{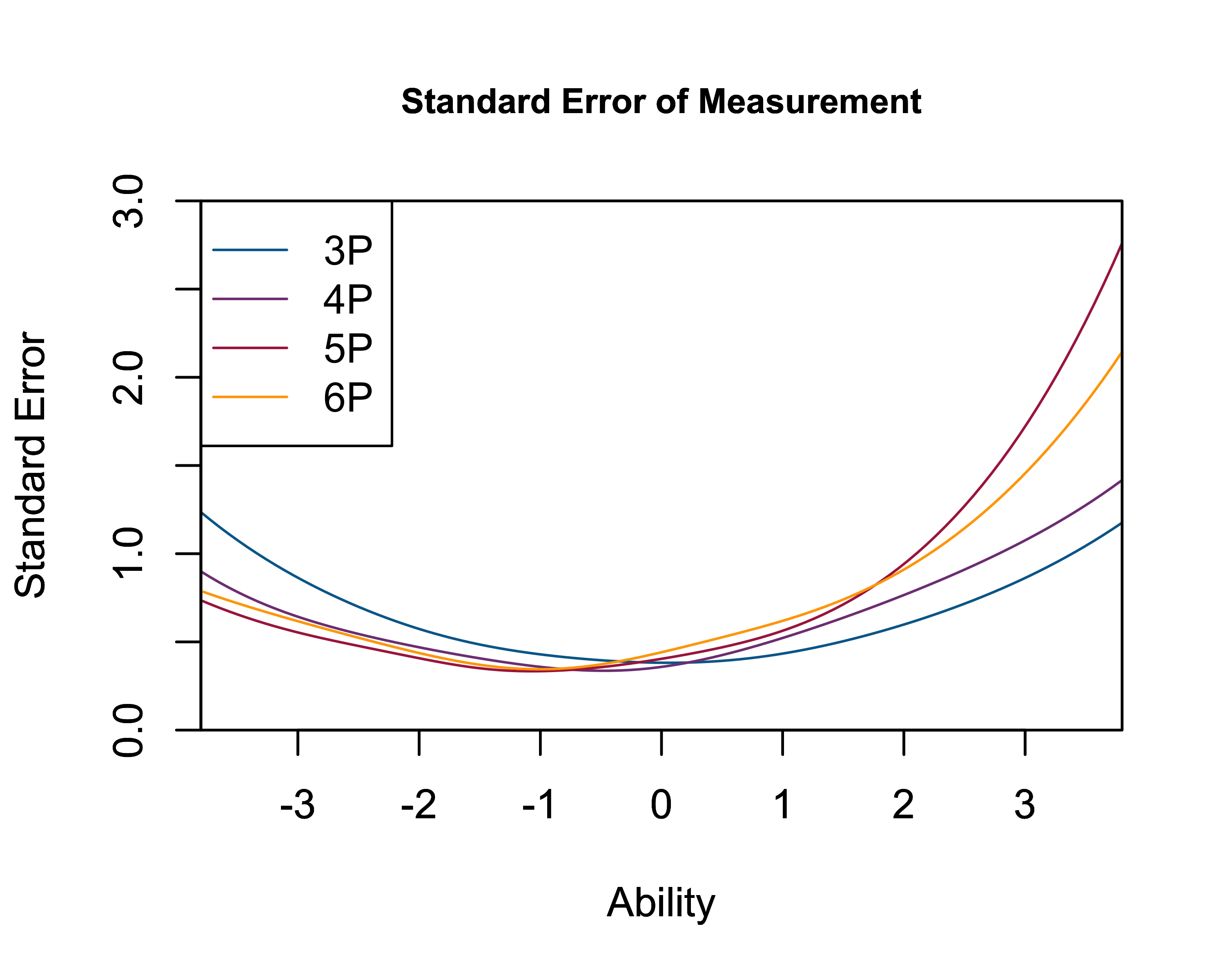}

    \vspace{-5pt}
    \caption{2-PL IRT Test Information Function (left) and Standard Error of Measurement (right) according to the students' grade.}
    \label{fig:IRT_TIF}
\end{figure}

To gain better insight into how the properties of the test differ according to grade, we compare the IRT models for each of the grades in terms of difficulty and discrimination indices. 
The grade specific Item Response Theory models parameters are provided in Table \ref{tab:IRT_parameters} in appendix \ref{app:IRT_parameters}. Fig.~\ref{fig:IRT_ICC} shows the Item Characteristic Curves (ICCs), Fig.~\ref{fig:IRT_IIC} the Item Information Curves (IICs) and Fig.~\ref{fig:IRT_TIF} the Test Information Functions and Standard Error of Measurements (TIFs).

\begin{figure}[htbp!]
    \centering
    \includegraphics[width=0.45\linewidth]{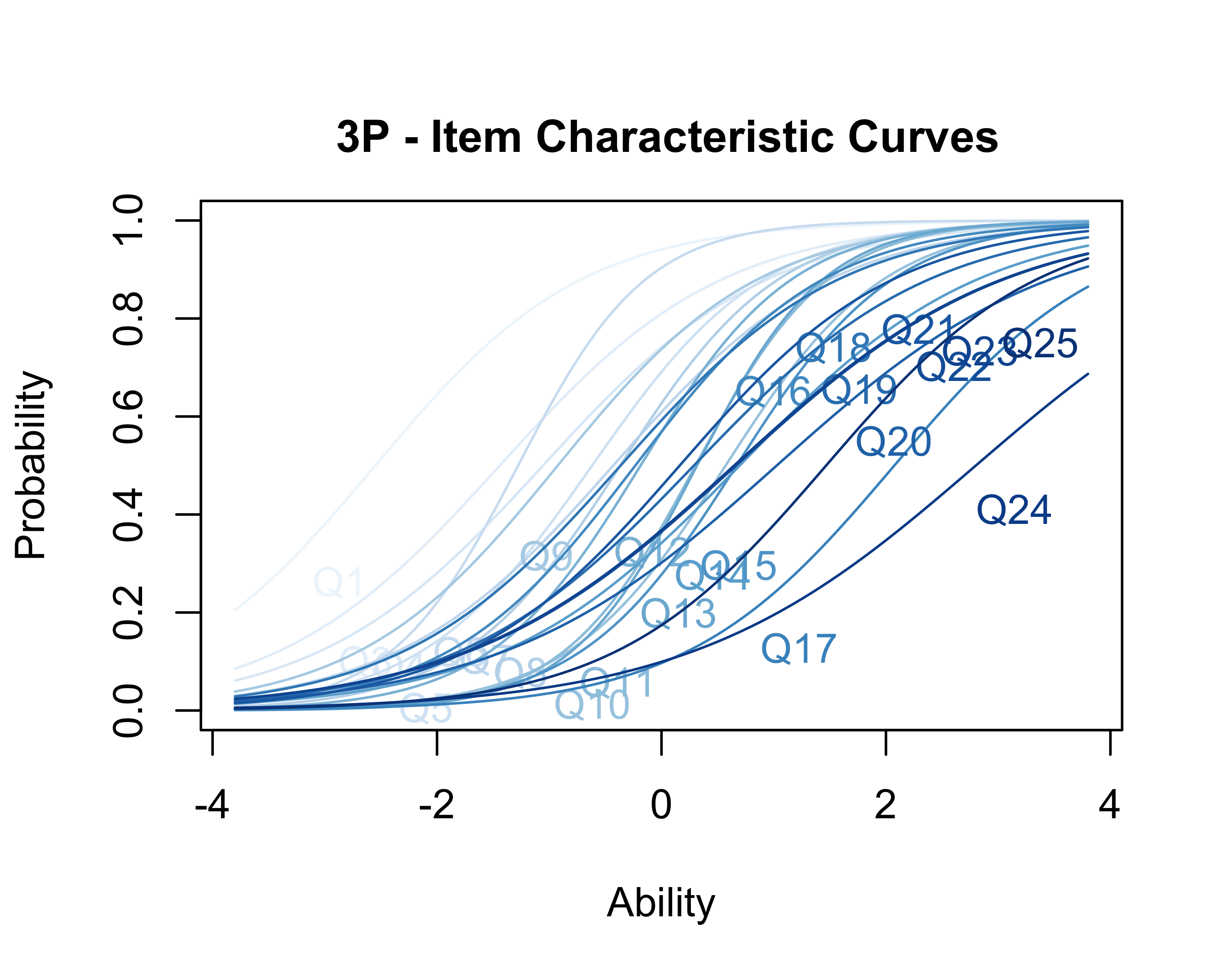}
    \includegraphics[width=0.45\linewidth]{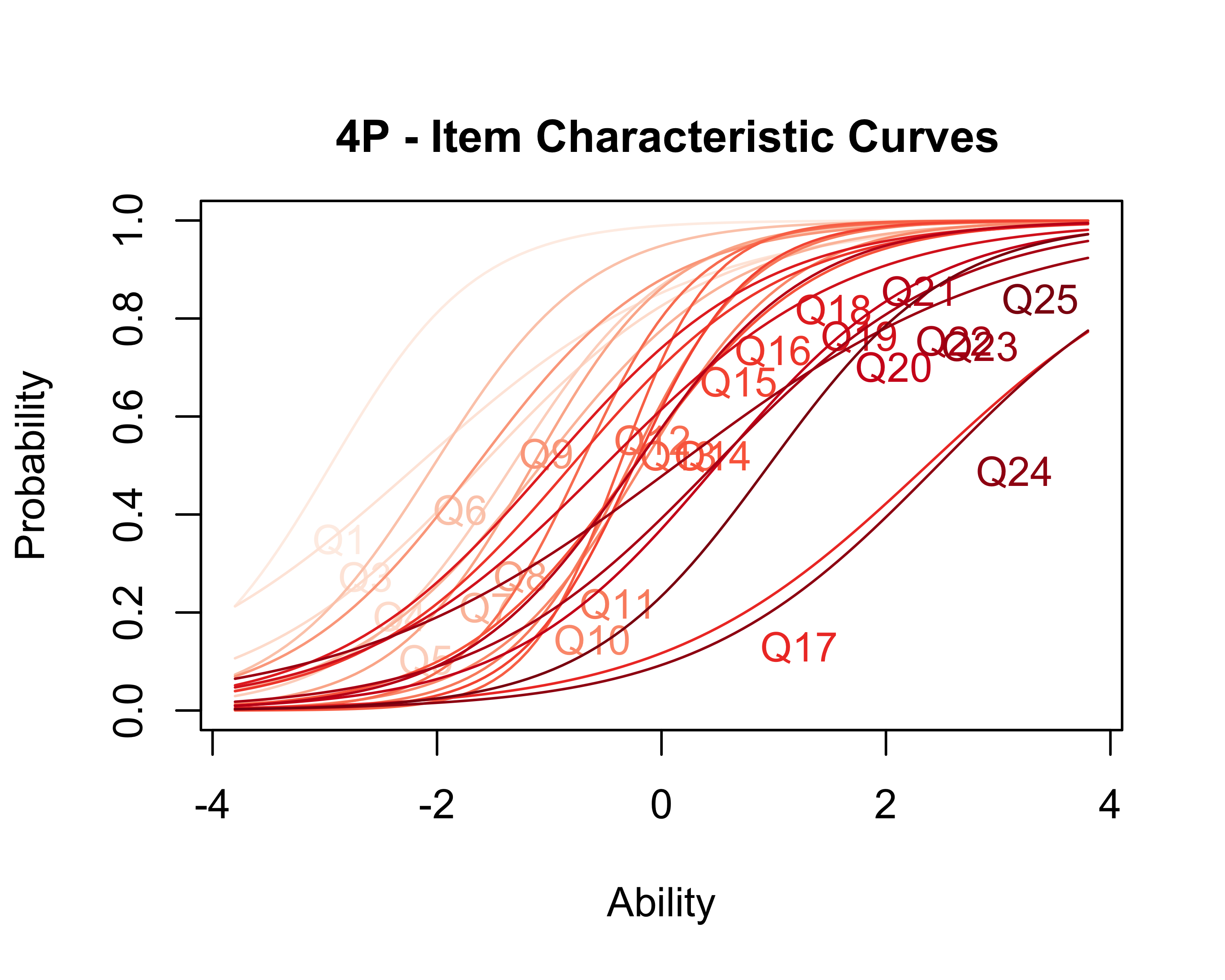}
    
    \vspace{-10pt}
    \includegraphics[width=0.45\linewidth]{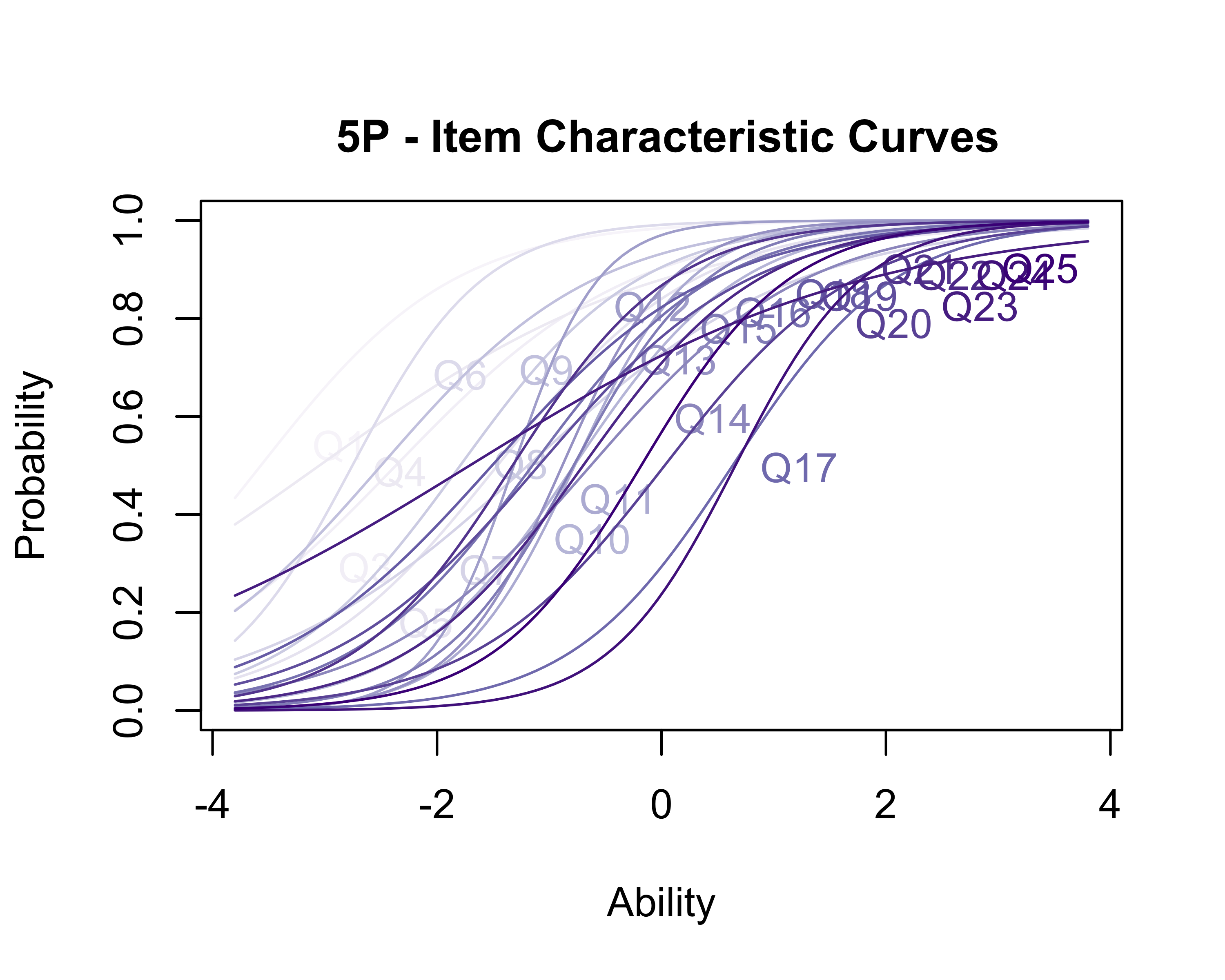}
    \includegraphics[width=0.45\linewidth]{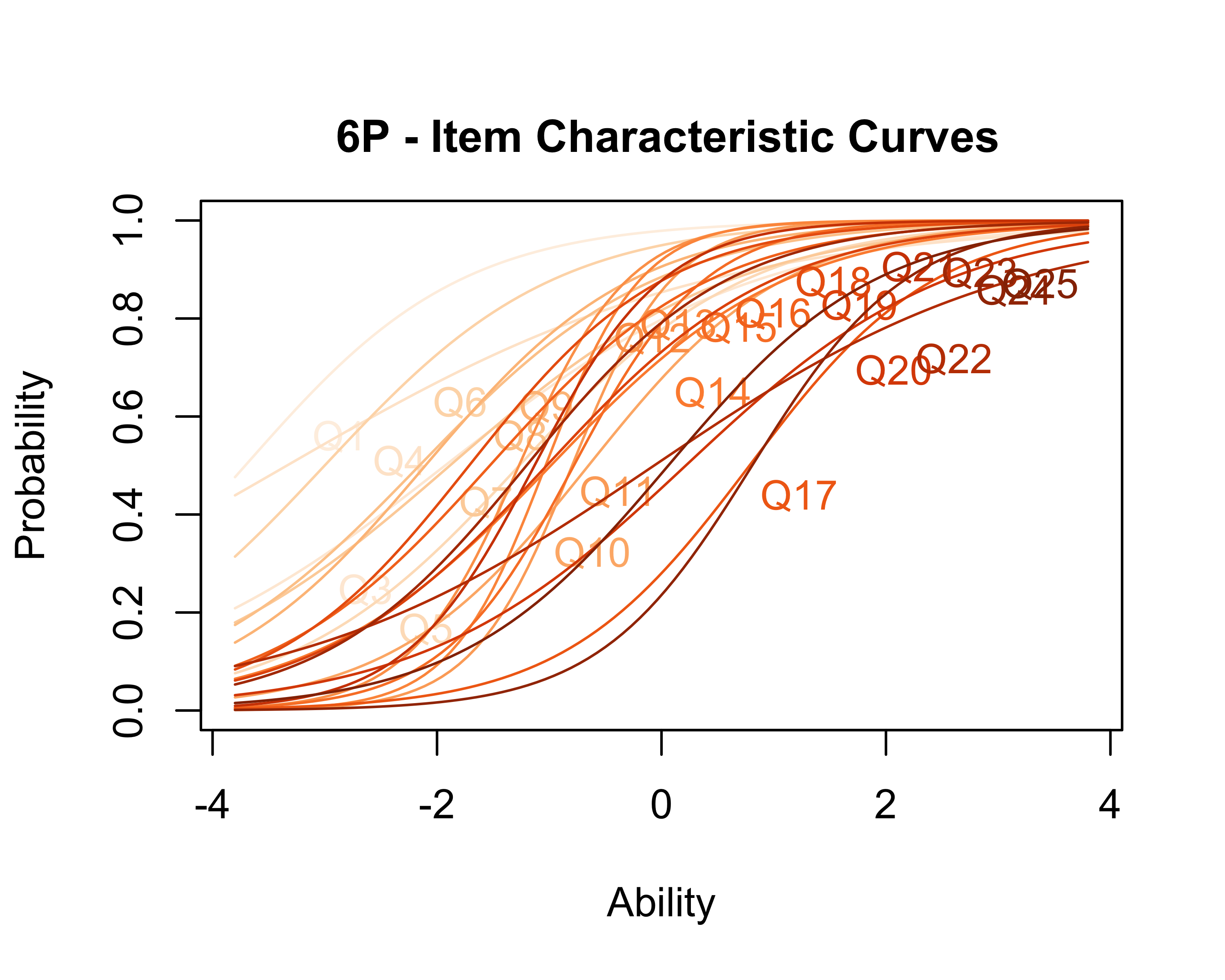}
    \vspace{-5pt}
    \caption{2-PL IRT Item Characteristic Curves per grade.}
    \label{fig:IRT_ICC}
\end{figure}

\FloatBarrier

Item discrimination does not differ significantly according to grade (one-way ANOVA $F(3)=0.77$, $p=.51$) and can be described as follows: 
\begin{itemize}
    \item The average item discrimination for all grades is in the upper-moderate range with. 
    \item The minimum item discrimination value is in the upper-low range for all grades.
    \item The maximum item discrimination value in the very high range. 
\end{itemize}    

On the other hand, the distribution of item difficulties does differ significantly according to grade ($F(3)=7.52$, $p=.00015$).
{On average, the results in terms of item difficulty appear to indicate that the cCTt is easier for older students. More specifically the cCTt can be considered of medium-high difficulty for grade 3 students, medium-low difficulty for grade 4 students, and easy for students in grades 5-6.} 
We therefore use Dunn's test for multiple comparisons with Benjamini-Hochberg p-value corrections to determine which of these differences are significant and account for the minimum effect size required to meet a statistical power of $.8$. The test indicates that the differences are significant between grades 3 and 5 ($\Delta=1.35$, $p=.0007$, $D=1.22$), and 3 and 6 ($\Delta=1.33$, $p=.0007$, $D=1.15$). This is confirmed by the Test Information Function, which indicates that the cCTt provides the most information for medium ability students in grades 3, while it provides more information for medium-low ability students in grades 4-6. \\

\begin{figure}[htbp!]
    \centering
    \includegraphics[height=0.24\textheight]{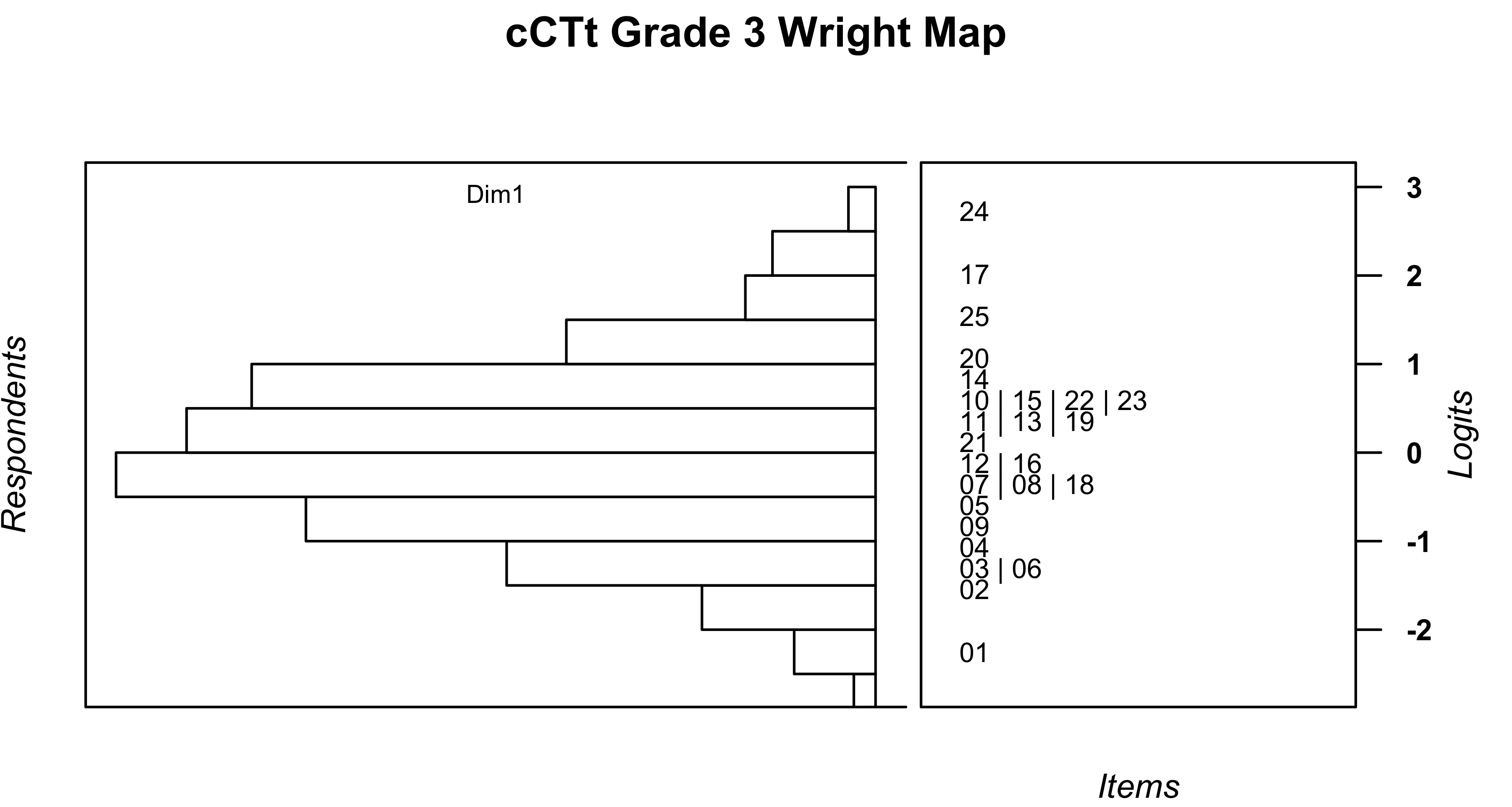}
    \includegraphics[height=0.24\textheight]{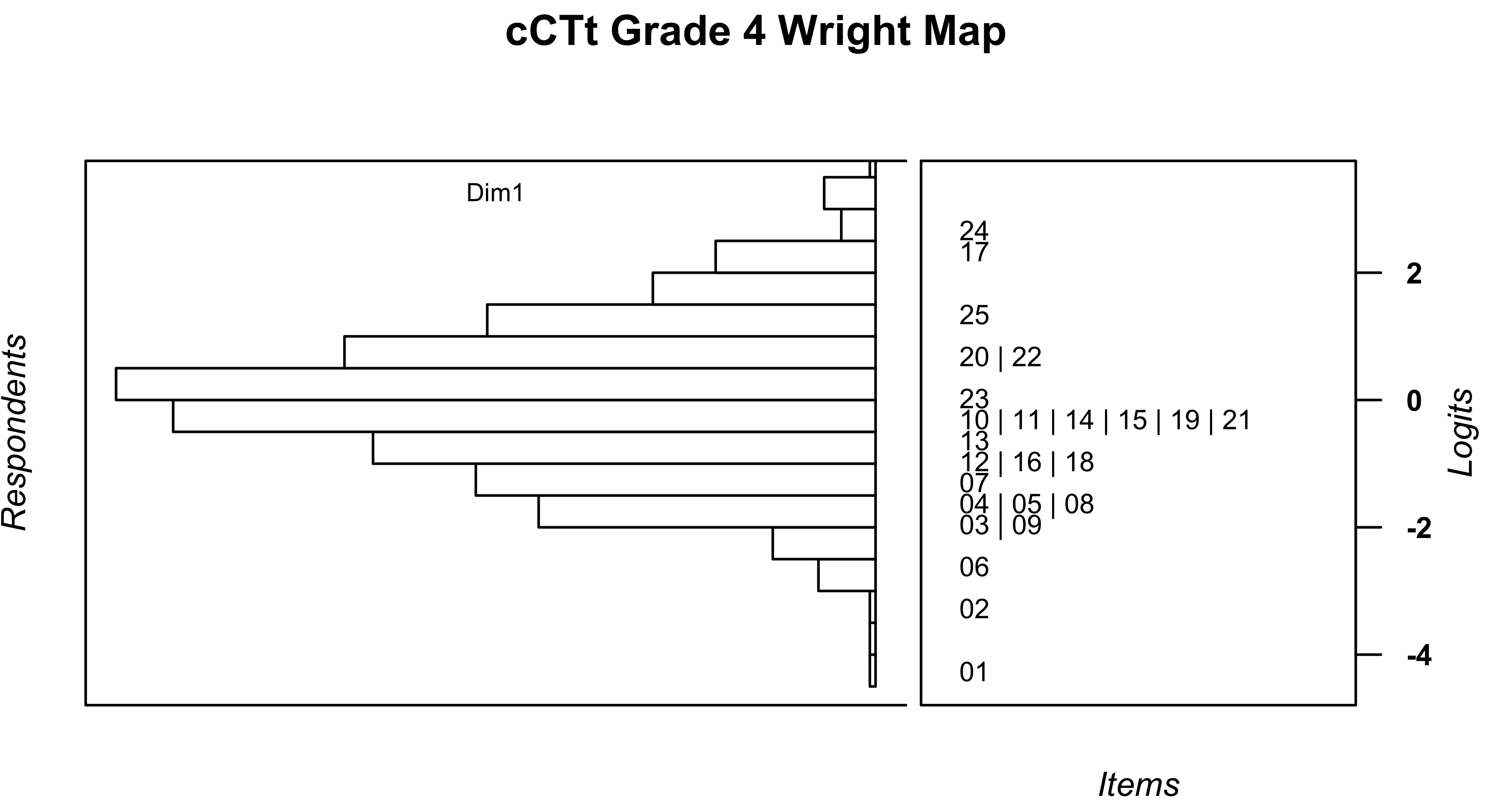}
    \includegraphics[height=0.24\textheight]{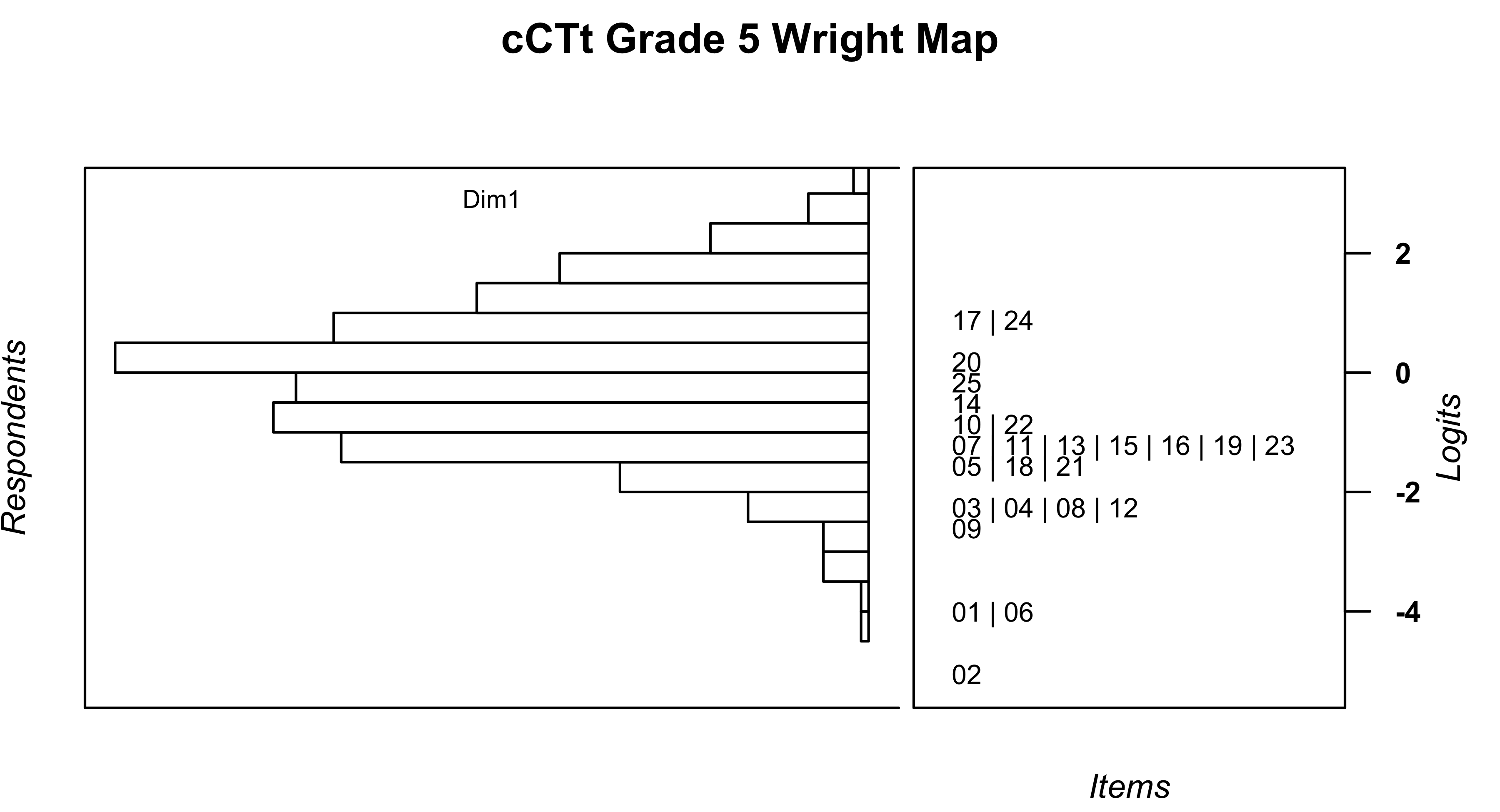}
    \includegraphics[height=0.24\textheight]{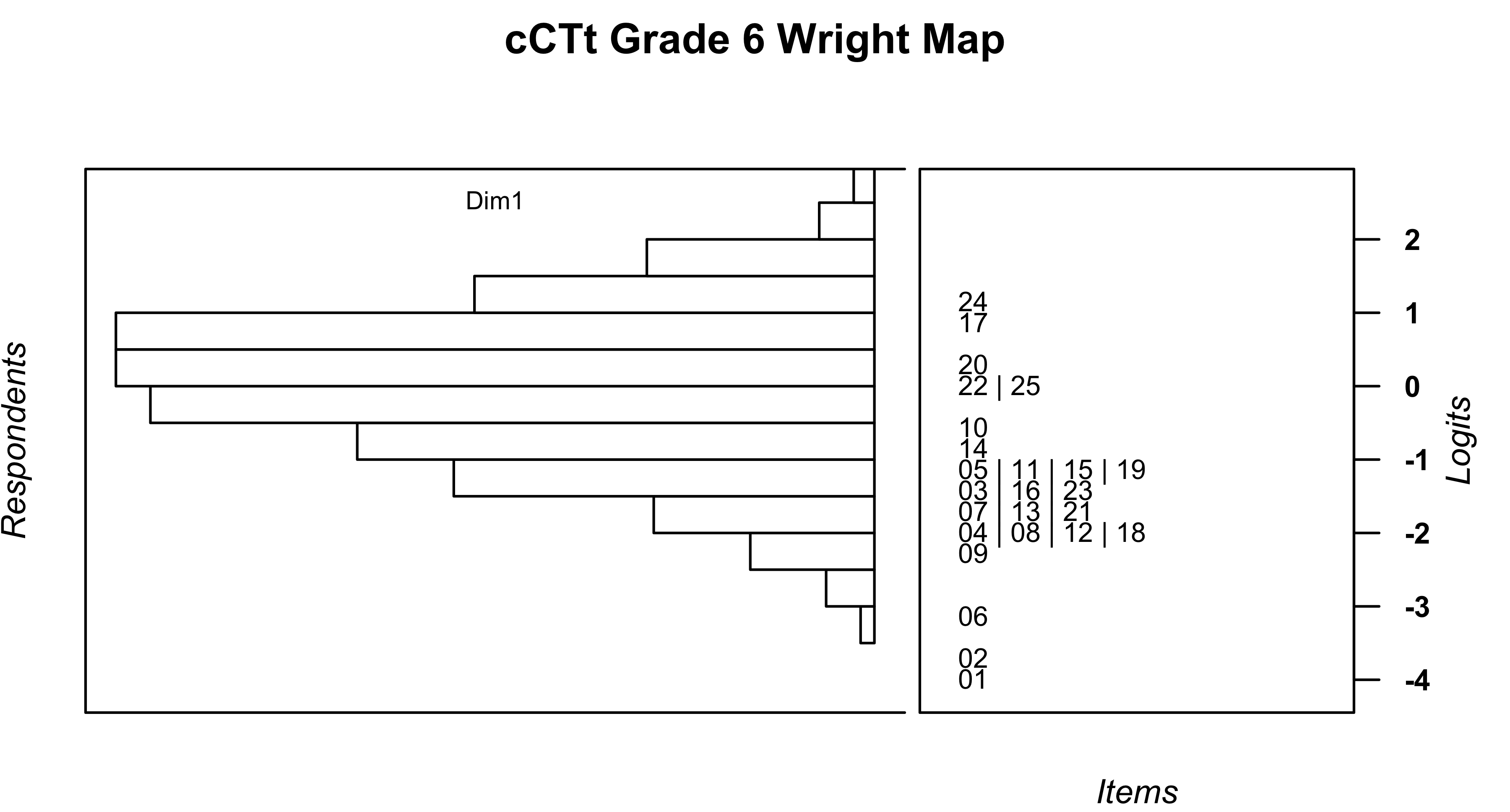}
    \caption{Grade-specific Wright Maps with EAP reliabilities of $.85$ for grade 3,  $.842$ for grade 4, $.80$ for grade 5, and $.78$ for grade 6. {The EAP reliability is therefore }sufficiently high for research purposes.}
    \label{fig:WrightMap}
\end{figure}

A more in depth look into the grade-specific Wright Maps on the 2PL models (see Fig. \ref{fig:WrightMap}) indicates that for grades 3-4 the items are aligned with the ability of the majority of the candidates, while in grades 5-6 the items are aligned with the ability of a smaller proportion of students, an in particular those who are at the lower end of the logit scale. 

\subsection{Measurement Invariance}

\subsubsection{Grade-related measurement invariance}

{We implemented the grade-related invariance analyses to compare the cCTt's properties between grades 3-4 and grades 5-6.}

\paragraph{Test-level grade-invariance.}

{The configural model (which estimates the loadings intercepts and residuals for grades 3-4 and 5-6) has good fit indicating that the cCTt has grade-related configural invariance (scaled $\chi^2/df=821/474<2$, $chi^2_{3-4}=484$, $chi^2_{5-6}=338$, $CFI=.97$, $TLI=.96$, $RMSEA=.026$, RMSEA 90\% $ci=[.023;.029]$ $SRMR=.032$).}
{However, when using Satorra Bentler's scaled chi-squared difference test there are significant difference between:}
\begin{itemize}
    \item {the grade-based metric and configural model ($\Delta \chi^2=65.71$, $\Delta df=18$, $p<.0001$), meaning that there is no metric invariance between grades 3-4 and 5-6.}
    \item {the grade-based configural model and the scalar model ($\Delta \chi^2=62.01$, $\Delta df=18$, $p<.0001$), meaning that there is no scalar invariance between grades 3-4 and 5-6.}
    \item {the grade-based scalar model and the strict model ($\Delta \chi^2=62.01$, $\Delta df=18$, $p<.0001$), meaning that there is no scalar invariance between grades 3-4 and 5-6.}
\end{itemize}

{As we have identified that there is no test-level grade-related measurement invariance, this means that the test-level properties differ between grades 3-4 and 5-6, confirming the importance of having done the grade-specific Classical Test Theory and Item Response Theory analyses. Please note that we do not provide all the models for the test-level invariance analysis as we already provide the grade-specific CFAs in appendix \ref{app:CFA_multidimensional}}. 

\paragraph{Item-level grade-invariance.}
To determine at the item-level whether there are differences in response patterns between grades 3-4 and 5-6, we employed Differential Item Functioning (DIF) for a 2-PL model. 
{We employed three metrics (Mantel-Haenszel test, Logistic Regression Likelihood Ratio Test (LRT), Generalized Lord's $\chi^2$) to determine whether there was grade-related DIF and the magnitude of this DIF.}
The results in Table \ref{tab:GradeDIF} indicate that all the items were flagged by two out of three metrics as DIF, with 16/25 being flagged by all three detection methods as DIF {(see Table \ref{tab:GradeDIF_FULL} and Fig. \ref{fig:GradeDIF} in appendix \ref{app:GradeDIF_FULL} for the detailed statistics). The DIF was also flagged as being ``Large'' for all items by the Mantel-Haenszel test. The grade-related DIF therefore} indicates that there are differences in difficulty or discriminability among the questions depending on the grades the students are in, supporting the importance of {computing grade-specific IRT-parameters and comparing across grades (see section \ref{sec:IRT})}.

\begin{table}[!htbp]
\centering
\caption{Differential Item Functioning between grades 3-4 (reference) and grades 5-6 (focal). Please note that the DifR package does not provide the effect size for Lord's $\chi^2$ statistic. Abbreviations: Stat. = statistic, Adj. P. = adjusted p-value using Benjamini-Hochberg p-value correction).}
\label{tab:GradeDIF}
\footnotesize
\begin{tabular}{lcccc}
\toprule
  & {Mantel-Haenszel (M.-H.) $\chi^2$} & {Logistic Regression } & {Generalized Lord's $\chi^2$} & \#DIF \\ 
  & & Likelihood Ratio Test (LRT) & \\\midrule
Q1 & Large DIF & Negligeable DIF & DIF & 3/3 \\
Q2 & Large DIF & Negligible DIF & DIF & 3/3 \\
Q4 & Large DIF & No DIF & DIF & 2/3 \\
Q5 & Large DIF & Negligible DIF & DIF & 3/3 \\
Q6 & Large DIF & Negligible DIF & DIF & 3/3 \\
Q7 & Large DIF & Negligible DIF & DIF & 3/3 \\
Q8 & Large DIF & Negligible DIF & DIF & 3/3 \\
Q9 & Large DIF & Negligible DIF & DIF & 3/3 \\
Q10 & Large DIF & Negligible DIF & DIF & 3/3 \\
Q11 & Large DIF & Negligible DIF & DIF & 3/3 \\
Q12 & Large DIF & Moderate DIF & No DIF & 2/3 \\
Q13 & Large DIF & Negligible DIF & DIF & 3/3 \\
Q14 & Large DIF & Negligible DIF & DIF & 3/3 \\
Q15 & Large DIF & Negligible DIF & DIF & 3/3 \\
Q16 & Large DIF & Negligible DIF & DIF & 3/3 \\
Q17 & Large DIF & Negligible DIF & No DIF & 2/3 \\
Q18 & Large DIF & Negligible DIF & No DIF & 2/3 \\
Q19 & Large DIF & Large DIF & No DIF & 2/3 \\
Q20 & Large DIF & Negligible DIF & DIF & 3/3 \\
Q21 & Large DIF & Large DIF & No DIF & 2/3 \\
Q22 & Large DIF & Negligible DIF & No DIF & 2/3 \\
Q23 & Large DIF & Negligible DIF & DIF & 3/3 \\
Q24 & Large DIF & DIF & Large DIF & 3/3 \\
Q25 & Large DIF & Negligible DIF & No DIF & 2/3 \\ \bottomrule
\end{tabular}
\end{table}

\FloatBarrier
\subsubsection{Gender-related measurement invariance}

\paragraph{Test-level gender-invariance.}

{Contrary to the case of grade, the test-level gender-invariance analysis indicated that there was measurement invariance at all levels:}
\begin{itemize}
    \item {Configural invariance was achieved as the gender-based configural model met the fit requirements (scaled $\chi^2/df=1024/520<3$, $CFI=.96$, $TLI=.99$, $RMSEA=.030$, RMSEA 90\% $ci=[.027, .033]$, $SRMR=.033$). }
    \item {Metric invariance was achieved as there was no statistical difference between the gender-based metric and configural models ($\Delta \chi^2=7.88$, $\Delta df=19$, $p=.99$).}
    \item {Scalar invariance was achieved as there was no statistical difference between the gender-based scalar and metric models ($\Delta \chi^2=14.73$, $\Delta df=19$, $p=.74$).}
    \item {Strict invariance was achieved as there was no statistical difference between the gender-based strict and scalar models ($\Delta \chi^2=21.06$, $\Delta df=25$, $p=.69$).}
\end{itemize}

{Please note that we do not provide all the models for the test-level invariance analysis as gender-invariance was established and we already provide the grade-specific CFAs in appendix \ref{app:CFA_multidimensional}.} 

\paragraph{Item-level gender-invariance.}
\label{sec:GenderDIF}

{However, despite having test-level invariance,} given the importance of having generalisable instruments that are fair towards all groups of participants, {it was important to employ Differential Item Functioning (DIF) to investigate whether each of the cCTt's items were fair with respect to gender. The synthesised results in Table \ref{tab:GenderDIF} (see Table \ref{tab:GenderDIF_FULL} in appendix \ref{app:GenderDIF_FULL} for the detailed statistics) indicate 22/25 items were never flagged as DIF by any of the three metrics. Only three items were flagged by just one metric, {the Mantel Haenszel metric}, as having a negligeable DIF (Q8, Q14, Q19). {Furthermore, as the Mantel Haenszel metric does not account for respondents' ability and item characteristics, and that more advanced DIF detection methods fo not detect any DIF, we can consider that there is no gender-DIF present at the item-level.} 

\begin{table}[h]
\centering
\caption{Synthesis of the Differential Item Functioning results between genders (girls = focal). Abbreviations: Stat. = statistic, Adj. P. = adjusted p-value using Benjamini-Hochberg p-value correction). {For the full statistics please refer to Table \ref{tab:GenderDIF_FULL} in appendix \ref{app:GenderDIF_FULL}.}}
\label{tab:GenderDIF}
\footnotesize
 \begin{tabular}{lcccc}
 \toprule
  & Mantel-Haenszel (M.-H.) & Logistic Regression & Generalized Lord's $\chi^2$ & \#DIF \\ 
  &  & Likelihood Ratio Test (LRT) & \\ \midrule
Q1 & NoDIF & NoDIF & NoDIF & 0/3 \\
Q2 & NoDIF & NoDIF & NoDIF & 0/3 \\
Q3 & NoDIF & NoDIF & NoDIF & 0/3 \\
Q4 & NoDIF & NoDIF & NoDIF & 0/3 \\
Q5 & NoDIF & NoDIF & NoDIF & 0/3 \\
Q6 & NoDIF & NoDIF & NoDIF & 0/3 \\
Q7 & NoDIF & NoDIF & NoDIF & 0/3 \\
Q8 & Negligible DIF & NoDIF & NoDIF & 1/3 \\
Q9 & NoDIF & NoDIF & NoDIF & 0/3 \\
Q10 & NoDIF & NoDIF & NoDIF & 0/3 \\
Q11 & NoDIF & NoDIF & NoDIF & 0/3 \\
Q12 & NoDIF & NoDIF & NoDIF & 0/3 \\
Q13 & NoDIF & NoDIF & NoDIF & 0/3 \\
Q14 & Negligible DIF & NoDIF & NoDIF & 1/3 \\
Q15 & NoDIF & NoDIF & NoDIF & 0/3 \\
Q16 & NoDIF & NoDIF & NoDIF & 0/3 \\
Q17 & NoDIF & NoDIF & NoDIF & 0/3 \\
Q18 & NoDIF & NoDIF & NoDIF & 0/3 \\
Q19 & Negligible DIF & NoDIF & NoDIF & 1/3 \\
Q20 & NoDIF & NoDIF & NoDIF & 0/3 \\
Q21 & NoDIF & NoDIF & NoDIF & 0/3 \\
Q22 & NoDIF & NoDIF & NoDIF & 0/3 \\
Q23 & NoDIF & NoDIF & NoDIF & 0/3 \\
Q24 & NoDIF & NoDIF & NoDIF & 0/3 \\
Q25 & NoDIF & NoDIF & NoDIF & 0/3\\ \bottomrule
\end{tabular}
\end{table}

Based on the findings at the test- and item-levels, we can conclude that the cCTt provides gender-related measurement invariance and has no significant gender-DIF items. The cCTt can therefore be considered fair with respect to gender, can be used for gender-analyses, and we do not need to establish gender-specific properties for the test. 

\FloatBarrier
\subsection{Proficiency profiles and z-scoring for inter- and intra-grade, and inter-test comparisons}

\subsubsection{Proposing Grade-Specific Student Proficiency Profiles}

Based on the procedure described by PISA \citep{oecd_pisa_2014,oecd_pisa_2017}, and the corrected item difficulties (see Table \ref{tab:IRT_parameters} in appendix \ref{app:IRT_parameters}), we establish grade-specific student proficiency profiles with anchor items that are located at the middle of each proficiency level as done by \citet{guggemos_computational_2022}. These proficiency profiles with their representative items are described in Table \ref{tab:proficiency_levels}.

\begin{landscape}
\begin{table}[h]
\centering
\caption{Grade specific proficiency profiles with anchor items (i.e. an item located approximately at the middle of the proficiency level on the logit scale). The logit bounds are provided for the 62\%  difficulty values.}
\label{tab:proficiency_levels}
\footnotesize
\begin{tabular}{cccP{2cm}cccccccP{1.2cm}}
\toprule
 & \multirow{2}{*}{\begin{tabular}[c]{@{}c@{}}Proficiency \\ level\end{tabular}} & \multirow{2}{*}{\begin{tabular}[c]{@{}c@{}}Logit bounds\end{tabular}} & Items & \multicolumn{7}{c}{Types of tasks the students are able to solve for that proficiency level} & \multirow{3}{*}{\begin{tabular}[c]{@{}c@{}}Percentage \\ of \\ Students\end{tabular}} \\
 &  &  &  & Sequences & \begin{tabular}[c]{@{}c@{}}Simple \\ loops\end{tabular} & \begin{tabular}[c]{@{}c@{}}Complex \\ loops\end{tabular} & \begin{tabular}[c]{@{}c@{}}If-else \\ statements\end{tabular} & \begin{tabular}[c]{@{}c@{}}While \\ statements\end{tabular} & \begin{tabular}[c]{@{}c@{}}Combinations \\ of concepts\end{tabular} & \begin{tabular}[c]{@{}c@{}}Insufficient items \\ for reliable estimates\end{tabular} \\ \midrule
 
\multirow{5}{*}{Grade 3} & Level 0 & \textless{}-1.0 & Q1 &  &  &  &  &  &  & x & 13.1\% \\
 & Level 1 & {[}-1.0, -0.2{]} & Q6, Q3, \textbf{Q4}, Q9, Q5 & x & (x) &  &  &  &  &  & 28.4\% \\
 & Level 2 & {[}-0.2, 0.6{]} & Q8, Q7, Q18, Q12, \textbf{Q16}, Q13, Q11 & x & x & x & (x) &  &  &  & 33.3\% \\
 & Level 3 & {[}0.6, 1.4{]} & Q21, Q19, Q10, \textbf{Q15}, Q14, Q23, Q22 & x & x & x & x & x &  &  & 18.0\% \\
 & Level 4 & \textgreater 1.4 & Q20, Q25, Q17, Q24 & x & x & x & x & x & x & x  & 7.2\% \\ \midrule

\multirow{6}{*}{Grade 4} & Level 0 & \textless{}-1.7 & Q1 &  &  &  &  &  &  & x & 3.1\% \\
 & Level 1 & {[}-1.7, -0.9{]} & Q6, Q3, \textbf{Q9}, Q4, Q5 & x & (x) &  &  &  &  &  & 14.0\% \\
 & Level 2 & {[}-0.9, -0.1{]} & Q8, Q7, \textbf{Q18}, Q12, Q16, Q13 & x & x & x & (x) &  &  &  & 26.7\% \\
 & Level 3 & {[}-0.1, 0.7{]} & Q11, Q15, Q19, Q21, Q10, Q14 & x & x & x & x &  &  &  & 34.8\% \\
 & Level 4 & {[}0.7, 1.5{]} & Q23, Q20, \textbf{Q22}, Q25 & x & x & x & x & x & (x) &  & 15.6\% \\
 & Level 5 & \textgreater{}1.5 & Q17, Q24 & x & x & x & x & x & x & x & 5.7\% \\ \midrule

\multirow{5}{*}{Grade 5} & Level 0 & \textless{}-2.5 & Q1 &  &  &  &  &  &  & x & 0.3\% \\
 & Level 1 & {[}-2.5, -1.7{]} & Q6, Q4, \textbf{Q9}, Q3 & x &  &  &  &  &  &  & 2.6\% \\
 & Level 2 & {[}-1.7, -0.9{]} & \textbf{Q8}, Q12, Q5, Q18, Q21 & x & (x) &  &  &  &  &  & 14.4\% \\
 & Level 3 & {[}-0.9, -0.1{]} & Q23, Q16, Q13, Q19, Q7, Q11, \textbf{Q15}, Q10, Q22, Q14 & x & x & x & x & x &  &  & 26.0\% \\
 & Level 4 & \textgreater -0.1 & Q25, Q20, Q24, Q17 & x & x & x & x & x & x & x & 56.8\% \\ \midrule

\multirow{4}{*}{Grade 6} & Level 0 & \textless -1.7 & Q1, Q6, Q4 &  &  &  &  &  &  & x & 3.7\% \\
 & Level 1 & {[}-1.7, -0.9{]} & Q9, Q8, \textbf{Q18}, Q7, Q3, Q16, Q12 & x & x &  &  &  &  &  & 13.3\% \\
 & Level 2 & {[}-0.9, -0.1{]} & Q21, Q13, Q5, Q23, Q11, Q19, \textbf{Q15}, Q14, Q10 & x & x & x & x & (x) &  &  & 24.7\% \\
 & Level 3 & \textgreater -0.1 & Q25, Q22, Q20, Q24, Q17 & x & x & x & x & x & x & x & 58.3\% \\ \bottomrule
\end{tabular}
\end{table}
\end{landscape}

\subsubsection{Providing Grade-Agnostic Wright Map and student profiles for longitudinal studies and to establish students' cognitive maturation according to age}

While the grade-specific Wright Maps and student profiles provided are interesting to establish students' proficiency at a given level, grade-agnostic profiles provide more direct insight into the cognitive maturation of students as they age. To that effect we construct a grade-agnostic 2PL IRT model (see model fit in Table \ref{tab:IRT_model_fit_grade_agnostic}, and parameters in Table \ref{tab:IRT_parameters_grade_agnostic} in appendix \ref{app:IRT_grade_agnostic}), compute the Wright Map (see Fig. \ref{fig:WrightMapGradeAgnostic} in appendix \ref{app:IRT_grade_agnostic}), and establish grade-agnostic student proficiency profiles for all students in grades 3-6 (see Table \ref{tab:proficiency_levels_grade_agnostic}). These grade-agnostic profiles can therefore be of use for those interested in evaluating the longitudinal development of students' CT-concepts. We also indicate in Table \ref{tab:proficiency_levels_grade_agnostic} the percentage of students per grade at each proficiency level which can provide a baseline for future studies interested in conducting international comparisons (as done by PISA with OECD countries).

\begin{table}[h]
\centering
\caption{Grade-agnostic proficiency profiles with anchor items (i.e. an item located approximately at the middle of the proficiency level on the logit scale). Please note that the logit bounds are based on the 62\% Difficulty values.}
\label{tab:proficiency_levels_grade_agnostic}
\footnotesize
\begin{tabular}{llccccc}
\toprule
 &  & Level 0 & Level 1 & Level 2 & Level 3 & Level 4 \\ \midrule
Logit bounds & \textless -1.6 & {[}-1.6, -0.8{]} & {[}-0.8, 0.0{]} & {[}0.0, 0.8{]} & \textgreater{}0.8 \\ \\
\multicolumn{2}{l}{Items} & Q1 & \begin{tabular}[c]{@{}c@{}}Q6, Q3, \\ Q4, Q9\end{tabular} & \begin{tabular}[c]{@{}c@{}}Q8, Q5, Q18, \\ Q12, Q7, Q16, \\ Q13, Q21, Q11, \\ Q19\end{tabular} & \begin{tabular}[c]{@{}c@{}}Q15, Q10, Q14, \\ Q23, Q22\end{tabular} & \begin{tabular}[c]{@{}c@{}}Q25, Q20, \\ Q17, Q24\end{tabular} \\ \midrule
\multirow{7}{*}{\begin{tabular}[c]{@{}l@{}}Tasks the \\ students are \\ able to solve \\ for  that \\ proficiency \\ level\end{tabular}} & Sequences &  & x & x & x & x \\
 & Simple loops &  & (x) & x & x & x \\
 & Complex loops &  &  & x & x & x \\
 & If-else statements &  &  & x & x & x \\
 & While statements &  &  &  & x & x \\
 & Combinations &  &  &  &  & x \\
 & Not enough items & x &  &  &  & x \\ \midrule
\multirow{4}{*}{\begin{tabular}[c]{@{}l@{}}Percentage of \\ students per \\ proficiency \\ level \& grade\end{tabular}} & Grade 3 & 8.7\% & 30.2\% & 37.4\% & 17.7\% & 5.9\% \\
 & Grade 4 & 4.2\% & 16.7\% & 32.6\% & 31.2\% & 15.5\% \\
 & Grade 5 & 1.0\% & 6.7\% & 22.4\% & 34.0\% & 35.6\% \\
 & Grade 6 & 1.0\% & 7.2\% & 23.4\% & 36.2\% & 32.1\% \\ \bottomrule
\end{tabular}
\end{table}

\subsubsection{Score normalisation for equivalency scales}

Using a normalised scoring approach \citep{relkin_cross-grade_2022}, we compute the percentile to which each student belongs according to the score obtained (after z-scoring) and their grade. As explained by \citet{relkin_normative_2023}, using normalised scores is a more reliable means of comparing student outcomes across grade-specific instruments. {The objective of the z-scoring is not to determine which items to revise}\footnote{The better starting point for decisions regarding adjustments to the items would be the results of the Classical Test Theory and Item Response Theory analyses, particularly since we do not have any gender-related DIF in the present case.}, {but to provide a means of determining in which percentile students are according to the number of questions they have correctly responded to, thus making it possible to compare across grade-specific tests, different types of tests, and as such across studies.} In our case this would mean making it possible to compare the results between the BCTt, cCTt and CTt, and potentially with other similar CT instruments.

The normalised scoring is provided in Table~\ref{tab:ZScoresPercentiles} and is computed in two ways: (i) for the full cCTt-25, (ii) without the first two questions as these questions tended to bee too easy and had low point-biserial correlations. {We provide this second variant of the normalised scoring as the first two items are the most important to revise, particularly in the case of grades 5-6 students (see section \ref{sec:CTT_results}), and therefore recommend that these be considered as practice items when students in these grades are part of the study.}

\begin{table}[h]
    \centering
    \caption{Normalised z-scoring and percentiles according to the score obtained on the cCTt and grade. {We have computed this in two ways: the total over the 25 cCTt items and the adjusted total over Q3-Q25. Abbreviations: z-score (Z), percentile (Pctl)}}
    
    \label{tab:ZScoresPercentiles}
    \footnotesize
    \begin{tabular}{l|
    P{0.5cm}P{0.5cm}P{0.5cm}P{0.5cm}P{0.5cm}P{0.5cm}P{0.5cm}P{0.5cm}|
    P{0.5cm}P{0.5cm}P{0.5cm}P{0.5cm}P{0.5cm}P{0.5cm}P{0.5cm}P{0.5cm}}
    \toprule
      & \multicolumn{8}{c|}{Total (Q1-Q25)} & \multicolumn{8}{c}{Adjusted Total (Q3-Q25)} \\
     Grade & \multicolumn{2}{c}{3} & \multicolumn{2}{c}{4} & \multicolumn{2}{c}{5} & \multicolumn{2}{c|}{6} & \multicolumn{2}{c}{3} & \multicolumn{2}{c}{4} & \multicolumn{2}{c}{5} & \multicolumn{2}{c}{6} \\
     Metric & Z & Pctl & Z & Pctl & Z & Pctl & Z & Pctl & Z & Pctl & Z & Pctl & Z & Pctl & Z & Pctl \\
    \midrule
    0 & -2.4 & 0.3 & -3.1 & 0.1 & -3.2 & 0.0 & -3.3 & 0.0 & -2.2 & 0.6 & -2.8 & 0.2 & -3.0 & 0.0 & -3.1 & 0.0 \\
    1 & -2.2 & 0.8 & -2.9 & 0.3 & -3.0 & 0.0 & -3.1 & 0.2 & -2.0 & 1.8 & -2.6 & 0.5 & -2.8 & 0.2 & -2.9 & 0.3 \\
    2 & -2.0 & 1.8 & -2.7 & 0.5 & -2.8 & 0.0 & -2.9 & 0.5 & -1.8 & 3.2 & -2.4 & 0.8 & -2.6 & 0.3 & -2.7 & 0.6 \\
    3 & -1.9 & 3.0 & -2.5 & 0.9 & -2.7 & 0.4 & -2.7 & 0.7 & -1.6 & 4.9 & -2.2 & 1.5 & -2.4 & 0.9 & -2.5 & 1.0 \\
    4 & -1.7 & 4.3 & -2.3 & 1.6 & -2.5 & 0.9 & -2.5 & 0.8 & -1.4 & 8.0 & -2.0 & 2.9 & -2.2 & 1.7 & -2.3 & 1.4 \\
    5 & -1.5 & 6.7 & -2.1 & 2.3 & -2.3 & 1.6 & -2.3 & 1.1 & -1.2 & 12.6 & -1.8 & 5.1 & -2.0 & 3.1 & -2.0 & 2.2 \\
    6 & -1.3 & 10.5 & -1.9 & 3.5 & -2.1 & 3.2 & -2.1 & 2.0 & -1.0 & 18.7 & -1.6 & 7.1 & -1.8 & 5.7 & -1.8 & 4.0 \\
    7 & -1.1 & 15.6 & -1.7 & 5.7 & -1.9 & 5.0 & -1.9 & 3.9 & -0.8 & 25.5 & -1.4 & 10.2 & -1.6 & 8.1 & -1.6 & 6.8 \\
    8 & -0.9 & 20.9 & -1.5 & 7.6 & -1.7 & 7.2 & -1.7 & 6.2 & -0.6 & 32.1 & -1.2 & 15.0 & -1.4 & 10.8 & -1.4 & 10.3 \\
    9 & -0.7 & 26.6 & -1.3 & 10.9 & -1.5 & 9.3 & -1.5 & 8.3 & -0.4 & 38.3 & -0.9 & 20.3 & -1.2 & 14.1 & -1.2 & 13.8 \\
    10 & -0.5 & 33.0 & -1.1 & 15.8 & -1.3 & 11.8 & -1.3 & 11.5 & -0.2 & 44.9 & -0.7 & 25.0 & -1.0 & 17.9 & -1.0 & 17.5 \\
    11 & -0.3 & 39.1 & -0.9 & 20.7 & -1.1 & 15.4 & -1.1 & 15.5 & 0.0 & 51.8 & -0.5 & 29.9 & -0.8 & 22.6 & -0.8 & 23.3 \\
    12 & -0.1 & 45.8 & -0.7 & 25.5 & -0.9 & 19.6 & -0.9 & 19.4 & 0.2 & 59.0 & -0.3 & 36.0 & -0.6 & 27.7 & -0.6 & 29.8 \\
    13 & 0.1 & 52.8 & -0.5 & 30.4 & -0.7 & 24.3 & -0.7 & 24.9 & 0.4 & 66.6 & -0.1 & 41.9 & -0.4 & 33.3 & -0.3 & 36.3 \\
    14 & 0.3 & 59.9 & -0.3 & 36.2 & -0.5 & 29.1 & -0.5 & 31.8 & 0.6 & 73.1 & 0.1 & 49.0 & -0.2 & 39.6 & -0.1 & 43.4 \\
    15 & 0.5 & 67.1 & -0.1 & 42.3 & -0.4 & 34.6 & -0.3 & 38.5 & 0.8 & 78.5 & 0.3 & 57.8 & -0.0 & 45.8 & 0.1 & 50.0 \\
    16 & 0.7 & 73.2 & 0.1 & 49.6 & -0.2 & 40.6 & -0.1 & 45.1 & 1.0 & 83.4 & 0.5 & 66.1 & 0.2 & 52.2 & 0.3 & 56.7 \\
    17 & 0.8 & 78.6 & 0.3 & 58.2 & 0.0 & 46.7 & 0.1 & 51.3 & 1.2 & 87.2 & 0.7 & 73.9 & 0.4 & 59.5 & 0.5 & 63.9 \\
    18 & 1.0 & 83.6 & 0.5 & 66.2 & 0.2 & 52.9 & 0.3 & 57.4 & 1.4 & 90.8 & 0.9 & 80.7 & 0.6 & 67.0 & 0.7 & 71.5 \\
    19 & 1.2 & 87.4 & 0.7 & 74.0 & 0.4 & 59.7 & 0.5 & 64.2 & 1.6 & 94.4 & 1.1 & 85.6 & 0.8 & 74.7 & 0.9 & 79.5 \\
    20 & 1.4 & 91.0 & 0.9 & 80.7 & 0.6 & 67.1 & 0.7 & 71.7 & 1.8 & 96.5 & 1.3 & 90.9 & 1.0 & 81.9 & 1.1 & 86.3 \\
    21 & 1.6 & 94.4 & 1.1 & 85.7 & 0.8 & 75.0 & 0.9 & 79.6 & 2.0 & 98.1 & 1.5 & 95.3 & 1.2 & 88.2 & 1.3 & 92.2 \\
    22 & 1.8 & 96.5 & 1.3 & 91.0 & 1.0 & 82.1 & 1.1 & 86.5 & 2.2 & 99.6 & 1.7 & 97.7 & 1.4 & 94.3 & 1.6 & 97.0 \\
    23 & 2.0 & 98.1 & 1.5 & 95.3 & 1.2 & 88.3 & 1.3 & 92.4 & 2.4 & 100.0 & 1.9 & 99.4 & 1.6 & 98.6 & 1.8 & 99.4 \\
    24 & 2.2 & 99.6 & 1.7 & 97.7 & 1.4 & 94.3 & 1.5 & 97.0 &  &  &  &  &  &  &  &  \\
    25 & 2.4 & 100.0 & 1.9 & 99.4 & 1.6 & 98.6 & 1.7 & 99.4 &  &  &  &  &  &  &  &  \\ \midrule
    M & 12.6 &  & 15.5 &  & 16.8 &  & 16.4 &  & 10.8 &  & 13.6 &  & 15.0 &  & 14.6 &  \\
    SD & 5.2 &  & 5.0 &  & 5.2 &  & 4.9 &  & 5.0 &  & 4.8 &  & 5.0 &  & 4.7 &  \\
    \bottomrule
    \end{tabular}
    
\end{table}

\FloatBarrier

\section{{Discussion and Conclusion}}

To understand the impact of the ever increasing Computer Science and Computational Thinking initiatives in K-12 formal education, assessments that (i) are useful to researchers, educators and practitioners, (ii) may be used in longitudinal studies, and (iii) provide the means of transitioning between assessments (e.g. through equivalency scales), are of particular importance. \\

In the present context we were interested in addressing this issue in the case of primary school with the competent CT test, a derivative of the Beginners' CT test by \citet{zapata-caceres_computational_2020} and the parent CT test by \citet{roman-gonzalez_which_2017}. This study therefore looked to expand on the validation of the cCTt to determine whether it could be employed in multi-year longitudinal studies between grades 3 and 6 (ages 7-11), provide student proficiency profiles, and determine at which point a transition to the CT test should be envisioned and how.  
While the parent CTt, which was validated for students in grades 5-10, may have been envisioned to continue to monitor students' progress, no equivalency scale exists yet between these two instruments, nor, to the best of our knowledge, between any other CT instruments. 
Therefore, using i) data from the administration of the cCTt between November 2021 and January 2022 to 1209 grade 5-6 students (585 in grade 5, 624 in grade 6) and ii) and data acquired from the administration of the cCTt in January 2021 \citep{el-hamamsy_datasetcCTt_2022} to 1457 grade 3-4 students (709 in grade 3, 748 in grade 4), the present study assessed the psychometric properties of the cCTt in grades 5-6 and compared them with grades 3-4 to establish the limits of validity of the cCTt for these age groups. The psychometric analysis considered conjointly the results of Classical Test Theory and Item Response Theory to evaluate the properties of the cCTt for students in grades 3-6. {The study further adds to the initial validation of the cCTt by including :}
\begin{itemize}
    \item {Relevant gender-based measurement invariance analyses to establish the gender-fairness of the cCTt and its applicability in gender studies.}
    \item {Student proficiency profiles and normalized z-scoring to as a first step towards establishing inter-test and inter-study comparisons.}
\end{itemize}

\subsection{Validity and reliability of the cCTt in grades 3-6 and the link with cognitive and developmental maturation.}

The results from the psychometric analysis confirm that the cCTt is valid and reliable for students in grades 3-6 and provides distinct proficiency profiles that describe the {``computational thinking tasks that students on a specific level are systematically able to master but which cannot be mastered by students on a lower level''} \citep[p.560]{guggemos_computational_2022}. Nonetheless, a ceiling effect starts to appear in grades 5-6 as the students perform well on the easier CT-concepts pertaining to sequences and loops. It would therefore be interesting to propose {additional} items pertaining to more advanced concepts {(e.g. complex loops, if-else statements, while statements, combinations, or in terms of other CT dimensions such as debugging which is present in the CTt)} to improve the reliability of the instrument for grades 5-6. 
Furthermore, the significant difference in scores across grades further stresses the importance of having targeted grade specific instruments to improve the validity and reliability of proposed assessments\footnote{{In this case, a grade 5-6 specific variant of the cCTt could reduce the number of items pertaining to sequences and loops, in favour of more complex items}}. As the BCTt validation \citep{zapata-caceres_computational_2020}, the BCTt - cCTt comparison \citep{el-hamamsy_comparing_2022}, and development of the TechCheck and its variants \citep{relkin_techcheck_2020, relkin_techcheck-k_2021, relkin_cross-grade_2022} showed, it is difficult to have a single assessment which is valid and reliable for a broad age range in primary school. This is unsurprising given the rapid cognitive development students undergo at this time of their lives. Indeed, as stated by \citet{el-hamamsy_comparing_2022}, CT is correlated with other cognitive abilities (e.g. numerical, verbal, non-verbal, \citealt{tsarava_cognitive_2022}) that are related to students' maturation, increase in working memory \citep{gathercole_structure_2004, cowan_working_2016}, and executive functions \citep{arfe_coding_2019, robertson_relationship_2020, robledo-castro_effects_2023}, thus improving their capacity to solve complex computational problems. As such, it is both unsurprising to see significant improvements over time, and difficult to have a single instrument that can be reliably employed in multi-year longitudinal studies.
Indeed, a corollary finding from the present analysis is that students, as they get older, have a good mastery of easier CT-concepts, but appear to still have a possible margin of progression for more advanced concepts such as conditional statements, while statements and in particular their combination. Indeed, it would appear that grade 5-6 students require targeted instruction to progress on these more advanced CT-concepts, thus providing insight for interventions in this age group. From a cognitive development perspective, it therefore appears that certain CT-concepts are easier than others, which helps derive a developmental progression from sequences, to loops, to conditionals and while statements. The findings from the study, and in particular the student profiles established, can therefore contribute to further tailoring CT assessments for the successive stages of their cognitive development.

\subsection{The importance of providing means of {comparing scores across grades and} transitioning between instruments across grades, including between the cCTt and CTt in grades 5-6.} 

Given that the findings established the relevance of developing more grade specific instruments across primary school, it is all the more critical for researchers and practitioners to have means of seamlessly transitioning between instruments of a given assessment family. Therefore, provided the validity of both instruments in grades 5-6, the findings confirm that grades 5-6 is an interesting point to transition between the cCTt and the CTt. Therefore, future work should consider comparing the cCTt and the CTt in grades 5-6, as was done for BCTt and cCTt for grades 3-4 in \citet{el-hamamsy_comparing_2022}. Such a comparison should be performed with a comparable group of students in order to establish equivalency scales which would help assess students' CT development in the long run. These equivalency scales can be achieved using Z-scoring (and percentiles) as was done here and in other studies \citep{roman-gonzalez_which_2017, relkin_cross-grade_2022}, as it provides normalised cCTt scores across grades which makes it possible to compare between grades. We argue that such percentiles should be grade-specific (without aggregating students in several grades), and established using comparable populations (i.e. from similar educational systems) both intra- and inter-assessments in order to provide a reliable means of comparing results across grades and passing from one instrument to another. {For example, using the normalized z-scoring approach for inter-grade comparison is particularly relevant in the case of the cCTt when transitioning from grades 3-4 to 5-6 given the existing DIF found between these sets of grades.}

\subsection{The importance of gender-fairness analyses in the CT literature.} 
Finally, the present study contributes to the literature by guaranteeing that the cCTt is a fair assessment with respect to gender through gender Differential Item Functioning. {Therefore, boys and girls with the same ability have the same probability of responding correctly to each cCTt item. The direct implication is that the cCTt can be used for analyses where the objective is to determine whether or not there are performance related gender-gaps, for instance to determine how these evolve following an intervention}. Given the present scarcity of such types of analyses in the CT assessment literature, we propose that gender Differential Item Functioning should become more standard practice in CT assessment validations to ensure that the proposed instruments are of relevance to the educator and research communities, particularly for those looking to address gender gaps in computing.\\

\subsection{{Limitations}}

A number of limitations can be raised concerning this study. 

Firstly, the validity of the cCTt for grades 5-6 was compared with data acquired a year prior for a different group of students in grades 3-4. While the measurements took place at the same point of the academic year, there might be certain contextual elements which may impact the students' results and thus the suitability of the comparison. In particular, the grade 5 students appear to be performing better than the grade 6 students, which is somewhat unexpected (although it may be related to the ceiling effect that we begin to observe in grades 5-6). Indeed other studies have found that students tend to progress on CT-abilities as they get older, without having received any CT-specific instruction \citep{roman-gonzalez_which_2017, relkin_techcheck_2020, relkin_techcheck-k_2021, piatti_ct-cube_2022}, in alignment with the consideration that Computational Thinking can be considered as a universal skill \citep{moreno-leon_computational_2018}. It would thus be interesting to collect data from another subset of students from grades 3-6 at the same point in time and replicate the study.

Secondly, as all the data was collected in a single region, the performance of the students in the sample may differ from that of students in other regions and countries, due to inherent differences in the curricula. It would thus be interesting to expand the validation to students in other countries to determine to what extent the results generalise or are influenced by local curricula. This would also provide the opportunity to conduct Differential Item Functioning across different countries to establish to what extent the cCTt is generalisable \citep{rachmatullah_toward_2022}.

Finally, the IRT analysis employed the same model for all grades (2-PL) in order to facilitate their comparison, although the 3-PL model may have been better suited for certain grades. Furthermore, there was a small mis-specification of the unidimensionality criteria. It thus remains likely that the discrimination parameters were slightly overestimated. 

More generally, paper-based assessments, such as the cCTt and those presented in the literature review, should be considered within a systems of assessments \citep{grover_systems_2015, roman-gonzalez_combining_2019, guggemos_computational_2022} to gain a more comprehensive picture of students' CT competence. This is because paper-based assessments tend to lack insight into :
\begin{itemize}
    \item CT-processes, which is generally acquired through educational data mining.
    \item CT-perspectives, which is generally acquired through self-assessment scales such as the Computational Thinking Scale by \citet{korkmaz_validity_2017} as was done by \citet{guggemos_computational_2022}.
    \item the link between CT and other abilities, such as numerical, verbal reasoning, and non verbal visuo-spatial abilities as was done by \citet{tsai_development_2022} or spatial, reasoning, and problem solving abilities as was done by \citet{roman-gonzalez_which_2017}.
\end{itemize}

\subsection{{Concluding remarks}}

To conclude, in addition to validating and providing recommendations for the use of the cCTt in grades 3-6 (ages 7-11), the findings of the present study contribute to improving the design of longitudinal research on the acquisition of CT-concepts. In combination with insight into cognitive process at play (e.g. CT-practices such as abstraction, algorithmic thinking, decomposition, evaluation and generalisation, \citealp{selby_computational_2013}) and in accordance with the theory of constructive alignment \citep{biggs_enhancing_1996}, it should be possible to provide guidelines for the design of developmentally appropriate learning objectives, assessments, and interventions for each level of primary school that account for both the type of cognitive processes and concepts that students are able to engage with at a given age. 

\ifdefined \Anonymous

\else
\section*{Data availability}

The data is available on Zenodo (doi: 10.5281/zenodo.7983525, \citealp{elhamamsy_cCTtdataset36_2023}).

\section*{Ethics}

The researchers were granted ethical approval to conduct the study by the head of the Department of Education and by 
the Human Research Ethics Committee of EPFL (project HREC 033-2019).

\section*{Conflicts of Interest}

The authors declare that there were no conflicts of interest involved with the realisation of the present study. 

\section*{Acknowledgements}

We would like to thank all the participants and the members of the different institutions (Department of Education - DEF, the University of Teacher Education – HEP Vaud, the teams from the two universities - EPFL and Unil) for supporting the EduNum project led by the minister of education of the Canton Vaud. 
This work was supported by i) the NCCR Robotics, a National Centre of Competence in Research, funded by the Swiss National Science Foundation (grant number 51NF40\_185543), ii) the Madrid Regional Government through the project e-Madrid-CM (P2018/TCS-4307) which is co-financed by the Structural Funds (FSE and FEDER). 

\fi

\bibliographystyle{apa_}

\bibliography{0-bib}

\begin{thebibliography}{}

\bibitem[\protect\astroncite{Alavi et~al.}{2020}]{alavi_chi-square_2020}
Alavi, M., Visentin, D.~C., Thapa, D.~K., Hunt, G.~E., Watson, R., and Cleary,
  M. (2020).
\newblock Chi-square for model fit in confirmatory factor analysis.
\newblock {\em Journal of Advanced Nursing}, 76(9):2209--2211.

\bibitem[\protect\astroncite{Araujo et~al.}{2017}]{araujo_exploring_2017}
Araujo, A. L. S.~O., Santos, J.~S., Andrade, W.~L., Guerrero, D. D.~S., and
  Dagienė, V. (2017).
\newblock Exploring computational thinking assessment in introductory
  programming courses.
\newblock In {\em 2017 {IEEE} {Frontiers} in {Education} {Conference} ({FIE})},
  pages 1--9.

\bibitem[\protect\astroncite{Arfé et~al.}{2019}]{arfe_coding_2019}
Arfé, B., Vardanega, T., Montuori, C., and Lavanga, M. (2019).
\newblock Coding in {Primary} {Grades} {Boosts} {Children}’s {Executive}
  {Functions}.
\newblock {\em Frontiers in Psychology}, 10.

\bibitem[\protect\astroncite{Awopeju and Afolabi}{2016}]{o_a_comparative_2016}
Awopeju, O. and Afolabi, E. (2016).
\newblock Comparative {Analysis} of {Classical} {Test} {Theory} and {Item}
  {Response} {Theory} {Based} {Item} {Parameter} {Estimates} of {Senior}
  {School} {Certificate} {Mathematics} {Examination}.
\newblock {\em ESJ}, 12(28):263.

\bibitem[\protect\astroncite{Baker}{2001}]{baker_basics_2001}
Baker, F.~B. (2001).
\newblock {\em The basics of item response theory}.
\newblock ERIC Clearinghouse on Assessment and Evaluation, College Park, Md.,
  2nd ed edition.

\bibitem[\protect\astroncite{Bean and Bowen}{2021}]{bean_item_2021}
Bean, G.~J. and Bowen, N.~K. (2021).
\newblock Item {Response} {Theory} and {Confirmatory} {Factor} {Analysis}:
  {Complementary} {Approaches} for {Scale} {Development}.
\newblock {\em Journal of Evidence-Based Social Work}, 18(6):597--618.

\bibitem[\protect\astroncite{Bellettini et~al.}{2015}]{bellettini_how_2015}
Bellettini, C., Lonati, V., Malchiodi, D., Monga, M., Morpurgo, A., and
  Torelli, M. (2015).
\newblock How {Challenging} are {Bebras} {Tasks}? {An} {IRT} {Analysis} {Based}
  on the {Performance} of {Italian} {Students}.
\newblock In {\em Proceedings of the 2015 {ACM} {Conference} on {Innovation}
  and {Technology} in {Computer} {Science} {Education}}, {ITiCSE} '15, pages
  27--32, New York, NY, USA. Association for Computing Machinery.

\bibitem[\protect\astroncite{Bers et~al.}{2022}]{bers_state_2022}
Bers, M.~U., Strawhacker, A., and Sullivan, A. (2022).
\newblock The state of the field of computational thinking in early childhood
  education.
\newblock {OECD} {Education} {Working} {Papers} 274.

\bibitem[\protect\astroncite{Biggs}{1996}]{biggs_enhancing_1996}
Biggs, J. (1996).
\newblock Enhancing teaching through constructive alignment.
\newblock {\em Higher education}, 32(3):347--364.

\bibitem[\protect\astroncite{Bland and Altman}{1997}]{bland1997statistics}
Bland, J.~M. and Altman, D.~G. (1997).
\newblock Statistics notes: Cronbach's alpha.
\newblock {\em Bmj}, 314(7080):572.

\bibitem[\protect\astroncite{Brennan and Resnick}{2012}]{brennan_new_2012}
Brennan, K. and Resnick, M. (2012).
\newblock New frameworks for studying and assessing the development of
  computational thinking.
\newblock page~25.

\bibitem[\protect\astroncite{Chae et~al.}{2019}]{chae_relationship_2019}
Chae, Y.-m., Park, S.~G., and Park, I. (2019).
\newblock The relationship between classical item characteristics and item
  response time on computer-based testing.
\newblock {\em Korean J Med Educ}, 31(1):1--9.

\bibitem[\protect\astroncite{Chen et~al.}{2017}]{chen_assessing_2017}
Chen, G., Shen, J., Barth-Cohen, L., Jiang, S., Huang, X., and Eltoukhy, M.
  (2017).
\newblock Assessing elementary students’ computational thinking in everyday
  reasoning and robotics programming.
\newblock {\em Comput Educ}, 109:162--175.

\bibitem[\protect\astroncite{Christensen
  et~al.}{2017}]{christensen_critical_2017}
Christensen, K.~B., Makransky, G., and Horton, M. (2017).
\newblock Critical {Values} for {Yen}’s \textit{{Q}} $_{\textrm{3}}$ :
  {Identification} of {Local} {Dependence} in the {Rasch} {Model} {Using}
  {Residual} {Correlations}.
\newblock {\em Applied Psychological Measurement}, 41(3):178--194.

\bibitem[\protect\astroncite{Commission et~al.}{2022}]{european_reviewing_2022}
Commission, E., Centre, J.~R., Bocconi, S., Chioccariello, A., Kampylis, P.,
  Dagien{\.e}, V., Wastiau, P., Engelhardt, K., Earp, K., Horvath, M.,
  Jasut{\.e}, M., Malagoli, C., Masiulionyt{\.e}-Dagien{\.e}, V.,
  Stupurien{\.e}, G., and Giannoutsou, N. (2022).
\newblock {\em Reviewing computational thinking in compulsory education : state
  of play and practices from computing education}.

\bibitem[\protect\astroncite{Cowan}{2016}]{cowan_working_2016}
Cowan, N. (2016).
\newblock Working {Memory} {Maturation}: {Can} {We} {Get} at the {Essence} of
  {Cognitive} {Growth}?
\newblock {\em Perspect Psychol Sci}, 11(2):239--264.
\newblock Publisher: SAGE Publications Inc.

\bibitem[\protect\astroncite{Cutumisu et~al.}{2019}]{cutumisu_scoping_2019}
Cutumisu, M., Adams, C., and Lu, C. (2019).
\newblock A {Scoping} {Review} of {Empirical} {Research} on {Recent}
  {Computational} {Thinking} {Assessments}.
\newblock {\em J Sci Educ Technol}, 28(6):651--676.

\bibitem[\protect\astroncite{Dai et~al.}{2020}]{dai_comparison_2020}
Dai, B., Zhang, W., Wang, Y., and Jian, X. (2020).
\newblock Comparison of {Trust} {Assessment} {Scales} {Based} on {Item}
  {Response} {Theory}.
\newblock {\em Frontiers in Psychology}, 11:10.

\bibitem[\protect\astroncite{Dai et~al.}{2022}]{subscore_Q3}
Dai, S., Wang, X., and Svetina, D. (2022).
\newblock {\em subscore: Computing Subscores in Classical Test Theory and Item
  Response Theory}.
\newblock R package version 3.3.

\bibitem[\protect\astroncite{De~Ayala and Little}{2022}]{de_ayala_theory_2022}
De~Ayala, R.~J. and Little, T.~D. (2022).
\newblock {\em The theory and practice of item response theory}.
\newblock Methodology in the social sciences. The Guilford Press, New York,
  second edition.

\bibitem[\protect\astroncite{De~Champlain}{2010}]{de_champlain_primer_2010}
De~Champlain, A.~F. (2010).
\newblock A primer on classical test theory and item response theory for
  assessments in medical education.
\newblock {\em Medical Education}, 44(1):109--117.
\newblock \_eprint:
  https://onlinelibrary.wiley.com/doi/pdf/10.1111/j.1365-2923.2009.03425.x.

\bibitem[\protect\astroncite{del Olmo-Muñoz
  et~al.}{2020}]{del_olmo-munoz_computational_2020}
del Olmo-Muñoz, J., Cózar-Gutiérrez, R., and González-Calero, J.~A. (2020).
\newblock Computational thinking through unplugged activities in early years of
  {Primary} {Education}.
\newblock {\em Comput Educ}, 150:103832.

\bibitem[\protect\astroncite{DeVellis}{2006}]{devellis_classical_2006}
DeVellis, R.~F. (2006).
\newblock Classical {Test} {Theory}.
\newblock {\em Medical Care}, 44(11):S50--S59.

\bibitem[\protect\astroncite{El-Hamamsy
  et~al.}{2023}]{elhamamsy_cCTtdataset36_2023}
El-Hamamsy, L., , Zufferey, B. B. J.~D., and Mondada, F. (2023).
\newblock {Extended dataset for the validation the competent Computational
  Thinking test in grades 3-6}.

\bibitem[\protect\astroncite{El-Hamamsy et~al.}{2022a}]{el-hamamsy_tacs_2022}
El-Hamamsy, L., Bruno, B., Avry, S., Chessel-Lazzarotto, F., Zufferey, J.~D.,
  and Mondada, F. (2022a).
\newblock The tacs model: Understanding primary school teachers’ adoption of
  computer science pedagogical content.
\newblock {\em ACM Trans. Comput. Educ.}

\bibitem[\protect\astroncite{El-Hamamsy
  et~al.}{2021}]{el-hamamsy_computer_2021}
El-Hamamsy, L., Chessel-Lazzarotto, F., Bruno, B., Roy, D., Cahlikova, T.,
  Chevalier, M., Parriaux, G., Pellet, J.-P., Lanarès, J., Zufferey, J.~D.,
  and Mondada, F. (2021).
\newblock A computer science and robotics integration model for primary school:
  evaluation of a large-scale in-service {K}-4 teacher-training program.
\newblock {\em Education and Information Technologies}, 26(3):2445--2475.

\bibitem[\protect\astroncite{El-Hamamsy
  et~al.}{2022b}]{elhamamsy_competent_2022}
El-Hamamsy, L., Zapata-C{\'a}ceres, M., Barroso, E.~M., Mondada, F., Zufferey,
  J.~D., and Bruno, B. (2022b).
\newblock The competent computational thinking test: Development and validation
  of an unplugged computational thinking test for upper primary school.
\newblock {\em Journal of Educational Computing Research}, page
  07356331221081753.

\bibitem[\protect\astroncite{El-Hamamsy
  et~al.}{2022c}]{el-hamamsy_datasetcCTt_2022}
El-Hamamsy, L., Zapata-Cáceres, M., Barroso, E.~M., Mondada, F., Zufferey,
  J.~D., and Bruno, B. (2022c).
\newblock {Dataset for the validation of a Computational Thinking test for
  upper primary school (grades 3-4)}.

\bibitem[\protect\astroncite{El-Hamamsy
  et~al.}{2022d}]{el-hamamsy_comparing_2022}
El-Hamamsy, L., Zapata-Cáceres, M., Marcelino, P., Dehler~Zufferey, J., Bruno,
  B., Martín-Barroso, E., and Román-González, M. (2022d).
\newblock Comparing the psychometric properties of two primary school
  computational thinking (ct) assessments for grades 3 and 4: the beginners' ct
  test (bctt) and the competent ct test (cctt).
\newblock {\em Frontiers in Psychology}.

\bibitem[\protect\astroncite{Field et~al.}{2012}]{field_discovering_2012}
Field, A.~P., Miles, J., and Field, Z. (2012).
\newblock {\em Discovering statistics using {R}}.
\newblock Sage, London ; Thousand Oaks, Calif.

\bibitem[\protect\astroncite{Gane et~al.}{2021}]{gane_design_2021}
Gane, B.~D., Israel, M., Elagha, N., Yan, W., Luo, F., and Pellegrino, J.~W.
  (2021).
\newblock Design and validation of learning trajectory-based assessments for
  computational thinking in upper elementary grades.
\newblock {\em Computer Science Education}, 31(2):141--168.

\bibitem[\protect\astroncite{Gathercole
  et~al.}{2004}]{gathercole_structure_2004}
Gathercole, S.~E., Pickering, S.~J., Ambridge, B., and Wearing, H. (2004).
\newblock The {Structure} of {Working} {Memory} {From} 4 to 15 {Years} of
  {Age}.
\newblock {\em Developmental Psychology}, 40:177--190.
\newblock Place: US Publisher: American Psychological Association.

\bibitem[\protect\astroncite{Gravetter et~al.}{2020}]{gravetter2020essentials}
Gravetter, F.~J., Wallnau, L.~B., Forzano, L.-A.~B., and Witnauer, J.~E.
  (2020).
\newblock {\em Essentials of statistics for the behavioral sciences}.
\newblock Cengage Learning.

\bibitem[\protect\astroncite{Grover et~al.}{2015}]{grover_systems_2015}
Grover, S., Pea, R., and Cooper, S. (2015).
\newblock Systems of assessments” for deeper learning of computational
  thinking in k-12.
\newblock In {\em Proceedings of the 2015 annual meeting of the American
  educational research association}, pages 15--20.

\bibitem[\protect\astroncite{Guggemos
  et~al.}{2022}]{guggemos_computational_2022}
Guggemos, J., Seufert, S., and Román-González, M. (2022).
\newblock Computational {Thinking} {Assessment} – {Towards} {More} {Vivid}
  {Interpretations}.
\newblock {\em Tech Know Learn}.

\bibitem[\protect\astroncite{Hambleton and
  Jones}{1993}]{hambleton_comparison_1993}
Hambleton, R.~K. and Jones, R.~W. (1993).
\newblock Comparison of classical test theory and item response theory and
  their applications to test development.
\newblock {\em Educational measurement: issues and practice}, 12(3):38--47.

\bibitem[\protect\astroncite{Hambleton
  et~al.}{1991}]{hambleton_fundamentals_1991}
Hambleton, R.~K., Swaminathan, H., and Rogers, H.~J. (1991).
\newblock {\em Fundamentals of {Item} {Response} {Theory}}.
\newblock SAGE.
\newblock Google-Books-ID: gW05DQAAQBAJ.

\bibitem[\protect\astroncite{Hinton et~al.}{2014}]{hinton_spss_2014}
Hinton, P., McMurray, I., and Brownlow, C. (2014).
\newblock {\em {SPSS} explained}.
\newblock Routledge.

\bibitem[\protect\astroncite{Hsu et~al.}{2018}]{hsu_how_2018}
Hsu, T.-C., Chang, S.-C., and Hung, Y.-T. (2018).
\newblock How to learn and how to teach computational thinking: {Suggestions}
  based on a review of the literature.
\newblock {\em Comput Educ}, 126:296--310.

\bibitem[\protect\astroncite{Hu and Bentler}{1999}]{hu_cutoff_1999}
Hu, L.-t. and Bentler, P.~M. (1999).
\newblock Cutoff criteria for fit indexes in covariance structure analysis:
  {Conventional} criteria versus new alternatives.
\newblock {\em Structural Equation Modeling: A Multidisciplinary Journal},
  6(1):1--55.

\bibitem[\protect\astroncite{Hubwieser and
  Mühling}{2014}]{hubwieser_playing_2014}
Hubwieser, P. and Mühling, A. (2014).
\newblock Playing {PISA} with bebras.
\newblock In {\em Proceedings of the 9th {Workshop} in {Primary} and
  {Secondary} {Computing} {Education}}, {WiPSCE} '14, pages 128--129, New York,
  NY, USA. Association for Computing Machinery.

\bibitem[\protect\astroncite{Irribarra and Freund}{2014}]{wrightMap}
Irribarra, D.~T. and Freund, R. (2014).
\newblock {\em Wright Map: IRT item-person map with ConQuest integration}.

\bibitem[\protect\astroncite{Jabrayilov
  et~al.}{2016}]{jabrayilov_comparison_2016}
Jabrayilov, R., Emons, W. H.~M., and Sijtsma, K. (2016).
\newblock Comparison of {Classical} {Test} {Theory} and {Item} {Response}
  {Theory} in {Individual} {Change} {Assessment}.
\newblock {\em Appl Psychol Meas}, 40(8):559--572.

\bibitem[\protect\astroncite{Kong and Lai}{2022}]{kong_validating_2022}
Kong, S.-C. and Lai, M. (2022).
\newblock Validating a computational thinking concepts test for primary
  education using item response theory: {An} analysis of students’ responses.
\newblock {\em Computers \& Education}, 187:104562.

\bibitem[\protect\astroncite{Korkmaz et~al.}{2017}]{korkmaz_validity_2017}
Korkmaz, O., Cakir, R., and Ozden, M.~Y. (2017).
\newblock A validity and reliability study of the computational thinking scales
  ({CTS}).
\newblock {\em Computers in Human Behavior}, 72:558--569.

\bibitem[\protect\astroncite{Kyriazos}{2018}]{kyriazos_applied_2018}
Kyriazos, T.~A. (2018).
\newblock Applied {Psychometrics}: {Writing}-{Up} a {Factor} {Analysis}
  {Construct} {Validation} {Study} with {Examples}.
\newblock {\em Psychology}, 9(11):2503--2530.

\bibitem[\protect\astroncite{Li et~al.}{2021}]{li_development_2021}
Li, Y., Xu, S., and Liu, J. (2021).
\newblock Development and {Validation} of {Computational} {Thinking}
  {Assessment} of {Chinese} {Elementary} {School} {Students}.
\newblock {\em Journal of Pacific Rim Psychology}, 15.

\bibitem[\protect\astroncite{Magis et~al.}{2020}]{magis2020package}
Magis, D., Beland, S., Raiche, G., and Magis, M.~D. (2020).
\newblock Package ‘difr’.

\bibitem[\protect\astroncite{Magis et~al.}{2010}]{difR_package}
Magis, D., Beland, S., Tuerlinckx, F., and {De Boeck}, P. (2010).
\newblock A general framework and an r package for the detection of dichotomous
  differential item functioning.
\newblock {\em Behavior Research Methods}, 42:847--862.

\bibitem[\protect\astroncite{Magis et~al.}{2011}]{magis_generalized_2011}
Magis, D., Raîche, G., Béland, S., and Gérard, P. (2011).
\newblock A {Generalized} {Logistic} {Regression} {Procedure} to {Detect}
  {Differential} {Item} {Functioning} {Among} {Multiple} {Groups}.
\newblock {\em International Journal of Testing}, 11(4):365--386.

\bibitem[\protect\astroncite{Marinus et~al.}{2018}]{marinus_unravelling_2018}
Marinus, E., Powell, Z., Thornton, R., McArthur, G., and Crain, S. (2018).
\newblock Unravelling the {Cognition} of {Coding} in 3-to-6-year {Olds}: {The}
  development of an assessment tool and the relation between coding ability and
  cognitive compiling of syntax in natural language.
\newblock In {\em Proceedings of the 2018 {ACM} {Conference} on {International}
  {Computing} {Education} {Research}}, pages 133--141, Espoo Finland. ACM.

\bibitem[\protect\astroncite{Master et~al.}{2021}]{master_gender_2021}
Master, A., Meltzoff, A.~N., and Cheryan, S. (2021).
\newblock Gender stereotypes about interests start early and cause gender
  disparities in computer science and engineering.
\newblock {\em PNAS}, 118(48).

\bibitem[\protect\astroncite{Moreno-León
  et~al.}{2018}]{moreno-leon_computational_2018}
Moreno-León, J., Román-González, M., and Robles, G. (2018).
\newblock On computational thinking as a universal skill: {A} review of the
  latest research on this ability.
\newblock In {\em 2018 {IEEE} {Global} {Engineering} {Education} {Conference}
  ({EDUCON})}, pages 1684--1689.

\bibitem[\protect\astroncite{Mouza et~al.}{2020}]{mouza_multiyear_2020}
Mouza, C., Pan, Y.-C., Yang, H., and Pollock, L. (2020).
\newblock A {Multiyear} {Investigation} of {Student} {Computational} {Thinking}
  {Concepts}, {Practices}, and {Perspectives} in an {After}-{School}
  {Computing} {Program}.
\newblock {\em Journal of Educational Computing Research}, 58(5):1029--1056.
\newblock Publisher: SAGE Publications Inc.

\bibitem[\protect\astroncite{OECD}{2014}]{oecd_pisa_2014}
OECD (2014).
\newblock {PISA} 2012 {Technical} {Report}.
\newblock Technical report, OECD Publishing.

\bibitem[\protect\astroncite{OECD}{2017}]{oecd_pisa_2017}
OECD (2017).
\newblock {PISA} 2015 {Technical} {Report}.
\newblock Technical report, OECD Publishing.

\bibitem[\protect\astroncite{Parker et~al.}{2021}]{parker_development_2021}
Parker, M.~C., Kao, Y.~S., Saito-Stehberger, D., Franklin, D., Krause, S.,
  Richardson, D., and Warschauer, M. (2021).
\newblock Development and preliminary validation of the assessment of computing
  for elementary students (aces).
\newblock In {\em Proc. 52nd ACM Tech. Symp. Comput. Sci. Educ.}, SIGCSE '21,
  page 10–16, New York, NY, USA. Association for Computing Machinery.

\bibitem[\protect\astroncite{Piatti et~al.}{2022}]{piatti_ct-cube_2022}
Piatti, A., Adorni, G., El-Hamamsy, L., Negrini, L., Assaf, D., Gambardella,
  L., and Mondada, F. (2022).
\newblock The {CT}-cube: {A} framework for the design and the assessment of
  computational thinking activities.
\newblock {\em Computers in Human Behavior Reports}, page 100166.

\bibitem[\protect\astroncite{Prudon}{2015}]{prudon_confirmatory_2015}
Prudon, P. (2015).
\newblock Confirmatory {Factor} {Analysis} as a {Tool} in {Research} {Using}
  {Questionnaires}: {A} {Critique},.
\newblock {\em Comprehensive Psychology}, 4:03.CP.4.10.

\bibitem[\protect\astroncite{Putnick and
  Bornstein}{2016}]{putnick2016measurement}
Putnick, D.~L. and Bornstein, M.~H. (2016).
\newblock Measurement invariance conventions and reporting: The state of the
  art and future directions for psychological research.
\newblock {\em Developmental review}, 41:71--90.

\bibitem[\protect\astroncite{{R Core Team}}{2019}]{r_core_team_r_2019}
{R Core Team} (2019).
\newblock {\em R: {A} {Language} and {Environment} for {Statistical}
  {Computing}}.
\newblock R Foundation for Statistical Computing, Vienna, Austria.

\bibitem[\protect\astroncite{Rachmatullah
  et~al.}{2022}]{rachmatullah_toward_2022}
Rachmatullah, A., Vandenberg, J., and Wiebe, E. (2022).
\newblock Toward {More} {Generalizable} {CS} and {CT} {Instruments}:
  {Examining} the {Interaction} of {Country} and {Gender} at the {Middle}
  {Grades} {Level}.
\newblock In {\em Proceedings of the 27th {ACM} {Conference} on on {Innovation}
  and {Technology} in {Computer} {Science} {Education} {Vol}. 1}, {ITiCSE} '22,
  pages 179--185, New York, NY, USA. Association for Computing Machinery.

\bibitem[\protect\astroncite{Relkin}{2022}]{relkin_cross-grade_2022}
Relkin, E. (2022).
\newblock Cross-grade {Comparison} of {Computational} {Thinking} in {Young}
  {Children} {Using} {Normalized} {Unplugged} {Assessment} {Scores}.
\newblock In {\em Proceedings of the 53rd {ACM} {Technical} {Symposium} on
  {Computer} {Science} {Education} {V}. 2}, {SIGCSE} 2022, page 1175, New York,
  NY, USA. Association for Computing Machinery.

\bibitem[\protect\astroncite{Relkin and Bers}{2021}]{relkin_techcheck-k_2021}
Relkin, E. and Bers, M. (2021).
\newblock Techcheck-k: A measure of computational thinking for kindergarten
  children.
\newblock In {\em 2021 IEEE Global Engineering Education Conference (EDUCON)},
  pages 1696--1702.

\bibitem[\protect\astroncite{Relkin et~al.}{2020}]{relkin_techcheck_2020}
Relkin, E., de~Ruiter, L., and Bers, M.~U. (2020).
\newblock {TechCheck}: {Development} and {Validation} of an {Unplugged}
  {Assessment} of {Computational} {Thinking} in {Early} {Childhood}
  {Education}.
\newblock {\em J Sci Educ Technol}, 29(4):482--498.

\bibitem[\protect\astroncite{Relkin et~al.}{2023}]{relkin_normative_2023}
Relkin, E., Johnson, S.~K., and Bers, M.~U. (2023).
\newblock A {Normative} {Analysis} of the {TechCheck} {Computational}
  {Thinking} {Assessment}.
\newblock {\em Educational Technology \& Society}, 26(2):118--130.
\newblock Publisher: International Forum of Educational Technology \& Society,
  National Taiwan Normal University, Taiwan.

\bibitem[\protect\astroncite{Revelle}{2021}]{revelle_psych_2021}
Revelle, W. (2021).
\newblock {\em psych: {Procedures} for {Psychological}, {Psychometric}, and
  {Personality} {Research}}.
\newblock Northwestern University, Evanston, Illinois.

\bibitem[\protect\astroncite{Rizopoulos}{2006}]{rizopoulos_ltm_2006}
Rizopoulos, D. (2006).
\newblock ltm: {An} {R} package for {Latent} {Variable} {Modelling} and {Item}
  {Response} {Theory} {Analyses}.
\newblock {\em Journal of Statistical Software}, 17(5):1--25.

\bibitem[\protect\astroncite{Robertson
  et~al.}{2020}]{robertson_relationship_2020}
Robertson, J., Gray, S., Toye, M., and Booth, J. (2020).
\newblock The relationship between {Executive} {Functions} and {Computational}
  {Thinking}.
\newblock {\em International Journal of Computer Science Education in Schools},
  3(4):35--49.
\newblock Number: 4.

\bibitem[\protect\astroncite{Robitzsch et~al.}{2022}]{TAM_4.1-4}
Robitzsch, A., Kiefer, T., and Wu, M. (2022).
\newblock {\em TAM: Test Analysis Modules}.
\newblock R package version 4.1-4.

\bibitem[\protect\astroncite{Robledo-Castro
  et~al.}{2023}]{robledo-castro_effects_2023}
Robledo-Castro, C., Castillo-Ossa, L.~F., and Hederich-Martínez, C. (2023).
\newblock Effects of a computational thinking intervention program on executive
  functions in children aged 10 to 11.
\newblock {\em International Journal of Child-Computer Interaction}, 35:100563.

\bibitem[\protect\astroncite{Rojas-López and
  García-Peñalvo}{2018}]{rojas-lopez_learning_2018}
Rojas-López, A. and García-Peñalvo, F.~J. (2018).
\newblock Learning {Scenarios} for the {Subject} {Methodology} of {Programming}
  {From} {Evaluating} the {Computational} {Thinking} of {New} {Students}.
\newblock {\em IEEE Rev. Iberoam. de Tecnol. del Aprendiz.}, 13(1):30--36.

\bibitem[\protect\astroncite{Rom{\'a}n-Gonz{\'a}lez
  et~al.}{2019}]{roman-gonzalez_combining_2019}
Rom{\'a}n-Gonz{\'a}lez, M., Moreno-Le{\'o}n, J., and Robles, G. (2019).
\newblock Combining {Assessment} {Tools} for a {Comprehensive} {Evaluation} of
  {Computational} {Thinking} {Interventions}.
\newblock In Kong, S.-C. and Abelson, H., editors, {\em Computational
  {Thinking} {Education}}, pages 79--98. Springer, Singapore.

\bibitem[\protect\astroncite{Rom{\'a}n-Gonz{\'a}lez
  et~al.}{2017}]{roman-gonzalez_which_2017}
Rom{\'a}n-Gonz{\'a}lez, M., P{\'e}rez-Gonz{\'a}lez, J.-C., and
  Jim{\'e}nez-Fern{\'a}ndez, C. (2017).
\newblock Which cognitive abilities underlie computational thinking?
  {Criterion} validity of the {Computational} {Thinking} {Test}.
\newblock {\em Comput. Hum. Behav}, 72:678--691.

\bibitem[\protect\astroncite{Román~González}{2016}]{roman_gonzalez_code_2016}
Román~González, M. (2016).
\newblock {\em Code literacy and computational thinking in {Primary} and
  {Secondary} {Education}: validation of an instrument and evaluation of
  programs}.
\newblock PhD thesis, National University of Distance Education, Spain.
\newblock Publisher: Universidad Nacional de Educación a Distancia (España).
  Escuela Internacional de Doctorado. Programa de Doctorado en Educación.

\bibitem[\protect\astroncite{Rosseel}{2012}]{rosseel_lavaan_2012}
Rosseel, Y. (2012).
\newblock lavaan: {An} {R} {Package} for {Structural} {Equation} {Modeling}.
\newblock {\em J. Stat. Softw.}, 48(2):1--36.

\bibitem[\protect\astroncite{Satorra and Bentler}{2010}]{satorra2010ensuring}
Satorra, A. and Bentler, P.~M. (2010).
\newblock Ensuring positiveness of the scaled difference chi-square test
  statistic.
\newblock {\em Psychometrika}, 75(2):243--248.

\bibitem[\protect\astroncite{Selby and
  Woollard}{2013}]{selby_computational_2013}
Selby, C. and Woollard, J. (2013).
\newblock Computational thinking: the developing definition.

\bibitem[\protect\astroncite{Souza et~al.}{2017}]{souza_psychometric_2017}
Souza, A. C.~d., Alexandre, N. M.~C., and Guirardello, E. d.~B. (2017).
\newblock Psychometric properties in instruments evaluation of reliability and
  validity.
\newblock {\em Epidemiologia e Serviços de Saúde}, 26(3):649--659.

\bibitem[\protect\astroncite{Sovey et~al.}{2022}]{sovey_gender_2022}
Sovey, S., Osman, K., and Matore, M. E. E.~M. (2022).
\newblock Gender differential item functioning analysis in measuring
  computational thinking disposition among secondary school students.
\newblock {\em Frontiers in Psychiatry}, 13.

\bibitem[\protect\astroncite{Sullivan and Bers}{2016}]{sullivan_girls_2016}
Sullivan, A. and Bers, M.~U. (2016).
\newblock Girls, {Boys}, and {Bots}: {Gender} {Differences} in {Young}
  {Children}’s {Performance} on {Robotics} and {Programming} {Tasks}.
\newblock {\em JITE:IIP}, 15:145--165.

\bibitem[\protect\astroncite{Taherdoost}{2016}]{taherdoost_validity_2016}
Taherdoost, H. (2016).
\newblock Validity and {Reliability} of the {Research} {Instrument}; {How} to
  {Test} the {Validation} of a {Questionnaire}/{Survey} in a {Research}.
\newblock {\em SSRN Journal}.

\bibitem[\protect\astroncite{Tang et~al.}{2020}]{tang_assessing_2020}
Tang, X., Yin, Y., Lin, Q., Hadad, R., and Zhai, X. (2020).
\newblock Assessing computational thinking: {A} systematic review of empirical
  studies.
\newblock {\em Comput Educ}, 148.

\bibitem[\protect\astroncite{Teresi}{2006}]{teresi2006overview}
Teresi, J.~A. (2006).
\newblock Overview of quantitative measurement methods: Equivalence,
  invariance, and differential item functioning in health applications.
\newblock {\em Medical care}, 44(11):S39--S49.

\bibitem[\protect\astroncite{Tikva and Tambouris}{2021}]{tikva_mapping_2021}
Tikva, C. and Tambouris, E. (2021).
\newblock Mapping computational thinking through programming in {K}-12
  education: {A} conceptual model based on a systematic literature {Review}.
\newblock {\em Comput Educ}, 162.

\bibitem[\protect\astroncite{Tsai et~al.}{2022}]{tsai_development_2022}
Tsai, M.-J., Chien, F., Lee, S. W.-Y., Hsu, C.-Y., and Liang, J.-C. (2022).
\newblock Development and {Validation} of the {Computational} {Thinking} {Test}
  for {Elementary} {School} {Students} ({CTT}-{ES}): {Correlate} {CT}
  {Competency} {With} {CT} {Disposition}.
\newblock {\em J. Educ. Comput. Res.}

\bibitem[\protect\astroncite{Tsarava et~al.}{2022}]{tsarava_cognitive_2022}
Tsarava, K., Moeller, K., Román-González, M., Golle, J., Leifheit, L., Butz,
  M.~V., and Ninaus, M. (2022).
\newblock A cognitive definition of computational thinking in primary
  education.
\newblock {\em Comput Educ}, 179:104425.

\bibitem[\protect\astroncite{Varma}{2006}]{varma2006preliminary}
Varma, S. (2006).
\newblock Preliminary item statistics using point-biserial correlation and
  p-values.
\newblock {\em Educational Data Systems Inc.: Morgan Hill CA. Retrieved},
  16(07).

\bibitem[\protect\astroncite{Vincent and Shanmugam}{2020}]{vincent_role_2020}
Vincent, W. and Shanmugam, S. K.~S. (2020).
\newblock The {Role} of {Classical} {Test} {Theory} to {Determine} the
  {Quality} of {Classroom} {Teaching} {Test} {Items}:.
\newblock {\em Pedagogia : Jurnal Pendidikan}, 9(1):5--34.

\bibitem[\protect\astroncite{Weintrop et~al.}{2021}]{weintrop_assessing_2021}
Weintrop, D., Wise~Rutstein, D., Bienkowski, M., and McGee, S. (2021).
\newblock Assessing computational thinking: an overview of the field.
\newblock {\em Comput. Sci. Educ.}, 31(2):113--116.

\bibitem[\protect\astroncite{Willse}{2018}]{CTT_R}
Willse, J.~T. (2018).
\newblock {\em CTT: Classical Test Theory Functions}.
\newblock R package version 2.3.3.

\bibitem[\protect\astroncite{Wing}{2006}]{wing_computational_2006}
Wing, J.~M. (2006).
\newblock Computational {Thinking}.
\newblock {\em Commun. ACM}, 49(3):33--35.

\bibitem[\protect\astroncite{Xia and Yang}{2019}]{xia_rmsea_2019}
Xia, Y. and Yang, Y. (2019).
\newblock {RMSEA}, {CFI}, and {TLI} in structural equation modeling with
  ordered categorical data: {The} story they tell depends on the estimation
  methods.
\newblock {\em Behav Res}, 51(1):409--428.

\bibitem[\protect\astroncite{Xie et~al.}{2019}]{xie_item_2019}
Xie, B., Davidson, M.~J., Li, M., and Ko, A.~J. (2019).
\newblock An {Item} {Response} {Theory} {Evaluation} of a
  {Language}-{Independent} {CS1} {Knowledge} {Assessment}.
\newblock In {\em Proceedings of the 50th {ACM} {Technical} {Symposium} on
  {Computer} {Science} {Education}}, pages 699--705, Minneapolis MN USA. ACM.

\bibitem[\protect\astroncite{Xu et~al.}{2021}]{xu_neural_2021}
Xu, S., Li, Y., and Liu, J. (2021).
\newblock The {Neural} {Correlates} of {Computational} {Thinking}:
  {Collaboration} of {Distinct} {Cognitive} {Components} {Revealed} by {fMRI}.
\newblock {\em Cerebral Cortex}, (bhab182).

\bibitem[\protect\astroncite{Xu et~al.}{2022}]{xu_relations_2022}
Xu, W., Geng, F., Wang, L., and Li, Y. (2022).
\newblock Relations of computational thinking to reasoning thinking and
  creative thinking in young children: mediating role of arithmetic fluency.
\newblock {\em Thinking Skills and Creativity}, page 101041.

\bibitem[\protect\astroncite{Yadav et~al.}{2022}]{yadav_computational_2022}
Yadav, A., Ocak, C., and Oliver, A. (2022).
\newblock Computational {Thinking} and {Metacognition}.
\newblock {\em TechTrends}.

\bibitem[\protect\astroncite{Yen}{1984}]{yen_effects_1984}
Yen, W.~M. (1984).
\newblock Effects of {Local} {Item} {Dependence} on the {Fit} and {Equating}
  {Performance} of the {Three}-{Parameter} {Logistic} {Model}.
\newblock {\em Applied Psychological Measurement}, 8(2):125--145.

\bibitem[\protect\astroncite{Zapata-C{\'a}ceres
  et~al.}{2020}]{zapata-caceres_computational_2020}
Zapata-C{\'a}ceres, M., Mart{\'\i}n-Barroso, E., and Rom{\'a}n-Gonz{\'a}lez, M.
  (2020).
\newblock Computational {Thinking} {Test} for {Beginners}: {Design} and
  {Content} {Validation}.
\newblock In {\em IEEE Glob. Eng. Educ. Conf. EDUCON}, pages 1905--1914.

\end{thebibliography}

\appendix

\pagebreak

\section{{Detailed Classical Test Theory analysis}}
\label{app:CTT}

\begin{landscape}
\begin{table}[!htb]
\centering
\caption{Classical Test Theory analysis according to the students' scores. Values in red either exceed the maximum threshold or are below the minimum threshold. Corresponding items could be considered for revision according to Classical Test Theory.}
\label{tab:CTT_full_cCTt}
\footnotesize
\begin{tabular}{l|r|cccc|cccc|cccc|cccc}
\toprule
 & \textbf{Metric} & \multicolumn{4}{c|}{\textbf{\begin{tabular}[c]{@{}c@{}}Difficulty \\ (or   p-values, mean)\end{tabular}}} & \multicolumn{4}{c|}{\textbf{Standard Deviation}} & \multicolumn{4}{c|}{\textbf{\begin{tabular}[c]{@{}c@{}}Point-biserial correlation \\ (or item discrimination)\end{tabular}}} & \multicolumn{4}{c}{\textbf{Drop alpha}} \\
 & \textbf{Grade} & 3 & 4 & 5 & 6 & 3 & 4 & 5 & 6 & 3 & 4 & 5 & 6 & 3 & 4 & 5 & 6 \\ 
 & &  &  &  &  &  &  &  &  &  &  &  &  & $\alpha=.84$ & $\alpha=.84$ & $\alpha=.83$ & $\alpha=.82$ \\ \midrule
 \textbf{Item} & &  &  &  &  &  &  &  &  &  &  &  &  &  &  \\
1 & & {\color[HTML]{CB0000} .91} & {\color[HTML]{CB0000} .97} & {\color[HTML]{CB0000} .99} & {\color[HTML]{CB0000} .97} & .28 & .17 & .16 & .18 & .25 & .22 & {\color[HTML]{CB0000} .08} & {\color[HTML]{CB0000} .19} & .84 & .84 & .83 & .81 \\
2  &  & {\color[HTML]{CB0000} .88} & {\color[HTML]{CB0000} .94} & {\color[HTML]{CB0000} 1} & {\color[HTML]{CB0000} .96} & .33 & .24 & .11 & .19 & .37 & .31 & {\color[HTML]{CB0000} .01} & {\color[HTML]{CB0000} .14} & .84 & .84 & .83 & .81 \\
3  &  & .77 & .83 & {\color[HTML]{CB0000} .89} & .79 & .42 & .38 & .33 & .42 & .33 & .25 & .36 & .28 & .84 & .84 & .82 & .81 \\
4  &  & .71 & .79 & {\color[HTML]{CB0000} .88} & {\color[HTML]{CB0000} .85} & .46 & .41 & .34 & .36 & .36 & .34 & {\color[HTML]{CB0000} .19} & {\color[HTML]{CB0000} .19} & .84 & .84 & .83 & .81 \\
5  &  & .65 & .8 & .81 & .74 & .48 & .4 & .4 & .44 & .43 & .4 & .31 & .33 & .84 & .84 & .83 & .81 \\
6  &  & .82 & {\color[HTML]{CB0000} .9} & {\color[HTML]{CB0000} .98} & {\color[HTML]{CB0000} .93} & .38 & .3 & .17 & .26 & .42 & .33 & .28 & .24 & .84 & .84 & .83 & .81 \\
7  &  & .59 & .73 & .74 & .81 & .49 & .44 & .46 & .41 & .38 & .38 & .3 & .31 & .84 & .84 & .83 & .81 \\
8  &  & .59 & .78 & {\color[HTML]{CB0000} .88} & .84 & .49 & .41 & .35 & .35 & .46 & .44 & .31 & .28 & .83 & .84 & .83 & .81 \\
9  &  & .7 & .83 & {\color[HTML]{CB0000} .91} & {\color[HTML]{CB0000} .86} & .46 & .38 & .3 & .34 & .37 & .34 & .31 & .31 & .84 & .84 & .83 & .81 \\
10  &  & .35 & .54 & .72 & .67 & .48 & .5 & .46 & .48 & .45 & .48 & .46 & .35 & .83 & .83 & .82 & .81 \\
11  &  & .41 & .58 & .75 & .75 & .49 & .49 & .45 & .44 & .5 & .5 & .53 & .56 & .83 & .83 & .82 & .8 \\
12  &  & .55 & .7 & {\color[HTML]{CB0000} .88} & {\color[HTML]{CB0000} .85} & .5 & .46 & .34 & .37 & .45 & .49 & .49 & .47 & .83 & .83 & .82 & .8 \\
13  &  & .41 & .62 & .78 & .82 & .49 & .49 & .43 & .4 & .51 & .57 & .52 & .51 & .83 & .83 & .82 & .8 \\
14  &  & .36 & .56 & .66 & .69 & .48 & .5 & .48 & .46 & .34 & .42 & .36 & .35 & .84 & .84 & .82 & .81 \\
15  &  & .33 & .57 & .75 & .74 & .47 & .5 & .45 & .45 & .44 & .53 & .44 & .51 & .83 & .83 & .82 & .8 \\
16   & & .55 & .67 & .79 & .81 & .5 & .47 & .43 & .41 & .44 & .4 & .36 & .35 & .83 & .84 & .82 & .81 \\
17  &  & {\color[HTML]{CB0000} .14} & {\color[HTML]{CB0000} .14} & .37 & .34 & .34 & .35 & .48 & .47 & .31 & .23 & .42 & .39 & .84 & .84 & .82 & .81 \\
18  & & .58 & .7 & .81 & .84 & .49 & .46 & .41 & .37 & .4 & .39 & .33 & .32 & .84 & .84 & .82 & .81 \\
19 &  & .44 & .6 & .76 & .71 & .5 & .49 & .45 & .46 & .37 & .37 & .37 & .36 & .84 & .84 & .82 & .81 \\
20  & & .32 & .39 & .52 & .47 & .47 & .49 & .5 & .5 & .32 & .4 & .35 & .33 & .84 & .84 & .82 & .81 \\
21  & & .46 & .56 & .81 & .79 & .5 & .5 & .4 & .41 & .41 & .48 & .45 & .53 & .84 & .83 & .82 & .8 \\
22  & & .38 & .41 & .69 & .55 & .49 & .49 & .47 & .5 & .34 & .36 & .44 & .23 & .84 & .84 & .82 & .81 \\
23  & & .38 & .48 & .71 & .76 & .49 & .5 & .45 & .43 & .33 & .28 & .25 & .39 & .84 & .84 & .83 & .81 \\
24  & & {\color[HTML]{CB0000} .12} & {\color[HTML]{CB0000} .12} & .35 & .32 & .33 & .33 & .47 & .46 & .23 & .24 & .5 & .39 & .84 & .84 & .82 & .81 \\
25  & & {\color[HTML]{CB0000} .22} & .28 & .59 & .51 & .41 & .45 & .5 & .5 & .35 & .41 & .52 & .35 & .84 & .84 & .82 & .81\\ \bottomrule
\end{tabular}
\end{table}
\end{landscape}

\FloatBarrier

\section{{Multi dimensional Confirmatory Factor Analysis for construct validity}}
\label{app:CFA_multidimensional}

{To verify the construct validity of the cCTt, we employed CFA with a robust estimator which is adapted to binary data (WLSMV). In particular, we consider that each block of questions relates to a latent factor. This analysis was conducted per grade to ensure (i) that the test's questions aligned with a coherent underlying concept, in line with those used to design the test, and (ii) to determine whether any changes exist between students according to their grades.}

{The data was considered suitable for factor analysis according to the Kaiser-Meyer-Olkin (KMO) measure of sampling adequacy which was above .8 ($>.5$), and Bartlett's test of sphericity which is significant with p-values below .05 \citep{field_discovering_2012}. Multiple fit indices are then provided for the cCTt-25 model for the different data subsets (see Table~\ref{tab:CFA_init}) and are within the acceptable limits where the $\chi^2$ test, CFI, TLI and RMSEA are concerned for grades 3, 4 and 6, although not for grade 5. Considering the proposed modification indices, and removing Q2 and Q1 (which according to the point-biserial correlation also needed to be revised), acceptable limits are once more attained. This reinforces the importance of considering these questions as ``warm up'' questions and discarding them from the total score when doing analyses for students in grade 5. We do however note that the SRMR values are acceptable (.1) but that these may be improved for grades 5 and 6. Indeed high values of SRMR imply that there is a difference between the observed correlations and predicted correlations of the residuals and that there are some correlations that we do not capture with the measurement model}\footnote{See \href{https://www.researchgate.net/post/The-SRMR-is-0141-in-CFA-but-other-indices-are-ok-What-is-the-implication}{discussion on research gate here}}.

{The factor loadings are significant ($p<.001$) for all items for all grades for all values except for Q2 in grade 6 where $p<.01$. The factor loadings range from $.53-.84$ ($\mu=.67\pm.09$) for students in grade 3, $.49-.90$ ($\mu=.71\pm.11$) in grade 4, $.41-.88$ ($\mu=.66\pm.13$) in grade 5 (without Q1 and Q2) and $.36-.94$ ($\mu=.63\pm.15$) for grade 6. The correlations between factors are in Table~\ref{tab:factor_correlations}. 
}

\begin{table}[h]
\centering
\caption{Confirmatory Factor Analysis for the cCTt-25, with latent variables and observed variables corresponding to CT-concepts shown in Table~\ref{tab:cCTt_description}. \\
Abbreviations: Kaiser, Meyer, Olkin sampling adequacy (KMO); Comparative Fit Index (CFI); Tucker-Lewis Index (TLI); Root Mean Square Error of Approximation (RMSEA); Standardized Root Mean Square Residual (SRMR) 
}
\label{tab:CFA_init}
\footnotesize
\begin{tabular}{lllllllll}
\toprule
 & \multicolumn{2}{l}{\begin{tabular}[c]{@{}l@{}}Initial conditions\end{tabular}} & \multicolumn{6}{l}{Robust model fit indices} \\ \midrule
 & KMO & Bartlett's test of sphericity & $\chi^2$ & $\chi^2/df$ & CFI & TLI & RMSEA & SRMR \\ \midrule

Grade 3 & $.88$ & {$\chi^2(300) = 3101$}, & {$\chi^2(260) =  378$}, & $1.45$ &  $.98$ & $.97$ & $.025$ & $.061$ \\
(Q1-25) & & $p<.001$ & $p=.000$ &  \\ \\

Grade 4 & $.88$ & {$\chi^2(300) = 3614$}, & {$\chi^2(260) = 403$}, & $1.55$ & $.98$ & $.97$ & $.027$ & $.070$ \\
(Q1-25)& &$p < .001$ & $p=.000$ & \\ \\

Grade 5 & $.86$ & {$\chi^2(300) = 2552$},  & {$\chi^2(260) = 512$}, & $1.97$ & $.88$ & $.87$ & $.053$ & $.119$ \\
(Q1-25) &  & $p < .001$ & $p=.000$ \\ \\

Grade 5  & $.86$ & {$\chi^2(276) = 2510$}, & {$\chi^2(237) = 316$}, & $1.33$ & $.96$ & $.95$ & $.031$ & $.106$ \\
(Q1+Q3-25) & & $p < .001$ & $p=.000$ &  \\ \\

Grade 5  & $.87$ & {$\chi^2(253) = 2438$}, & {$\chi^2(215) = 269$},  & $1.25$ & $.97$ & $.97$ & $.027$ & $.094$ \\
(Q3-25) & & $p < .001$ & $p=.000$ & \\ \\

Grade 6 & $.85$ & {$\chi^2(300) = 2505$},  & {$\chi^2(260) = 300$}, & $1.15$ & $.98$ & $.98$ & $.021$ & $.092$ \\ 
(Q1-25) & & $p < .001$ & $p=.043$ &  \\ 
 \bottomrule
\end{tabular}
\end{table}

\begin{table}[htpb]
\centering
\caption{Factor Loadings (beta) with significance}
\footnotesize
\label{tab:CFA_factor_loadings}
\begin{tabular}{cccccccclclccc}
\toprule
\textbf{Latent} & \multirow{2}{*}{\textbf{Item}} & \multicolumn{2}{c}{\textbf{Grade 3}} & \multicolumn{2}{c}{\textbf{Grade 4}} & \multicolumn{2}{c}{\textbf{Grade 5}} & \multicolumn{2}{c}{\textbf{Grade 5 (no Q2)}} & \multicolumn{2}{c}{\textbf{Grade 5 (no Q1, Q2)}} & \multicolumn{2}{c}{\textbf{Grade 6}} \\
\textbf{Factor} &  & \textbf{Beta} & \textbf{sig} & \textbf{Beta} & \textbf{sig} & \textbf{Beta} & \textbf{sig} & \multicolumn{1}{c}{\textbf{Beta}} & \textbf{sig} & \multicolumn{1}{c}{\textbf{Beta}} & \textbf{sig} & \textbf{Beta} & \textbf{sig} \\ \midrule
f1 & Q1 & .63 & *** & .79 & *** & .57 & *** & .33 & *** &  &  & .57 & *** \\
f1 & Q2 & .84 & *** & .86& *** & .61 & *** &  &  &  &  & .41 & ** \\
f1 & Q3 & .63 & *** & .58 & *** & .83 & *** & .74 & *** & .77 & *** & .53 & *** \\
f1 & Q4 & .66 & *** & .73 & *** & .55 & *** & .42 & *** & .41 & *** & .36 & *** \\  \midrule
f2 & Q5 & .71 & *** & .75 & *** & .55 & *** & .55 & *** & .56 & *** & .64 & *** \\
f2 & Q6 & .80 & *** & .74 & *** & .91 & *** & .89 & *** & .88 & *** & .59 & *** \\
f2 & Q7 & .61 & *** & .66 & *** & .45 & *** & .47 & *** & .47 & *** & .60 & *** \\
f2 & Q8 & .74 & *** & .80 & *** & .62 & *** & .59 & *** & .59 & *** & .56 & *** \\  \midrule
f3 & Q9 & .59 & *** & .58 & *** & .63 & *** & .58 & *** & .59 & *** & .59 & *** \\
f3 & Q10 & .71 & *** & .70 & *** & .65 & *** & .66 & *** & .66 & *** & .56 & *** \\
f3 & Q11 & .76 & *** & .78 & *** & .80 & *** & .81 & *** & .80 & *** & .89 & *** \\
f3 & Q12 & .72 & *** & .79 & *** & .86 & *** & .84 & *** & .83 & *** & .86 & *** \\
f3 & Q13 & .80 & *** & .86 & *** & .78 & *** & .79 & *** & .80 & *** & .84 & *** \\
f3 & Q14 & .53 & *** & .63 & *** & .53 & *** & .55 & *** & .55 & *** & .52 & *** \\
f3 & Q15 & .70 & *** & .79 & *** & .67 & *** & .67 & *** & .67 & *** & .80 & *** \\  \midrule
f4 & Q16 & .73 & *** & .68 & *** & .63 & *** & .61 & *** & .61 & *** & .67 & *** \\
f4 & Q17 & .66 & *** & .50 & *** & .66 & *** & .72 & *** & .71 & *** & .69 & *** \\
f4 & Q18 & .69 & *** & .67 & *** & .61 & *** & .58 & *** & .59 & *** & .62 & *** \\
f4 & Q19 & .64 & *** & .65 & *** & .63 & *** & .61 & *** & .62 & *** & .61 & *** \\  \midrule
f5 & Q20 & .54 & *** & .68 & *** & .53 & *** & .58 & *** & .57 & *** & .57 & *** \\
f5 & Q21 & .68 & *** & .84 & *** & .79 & *** & .77 & *** & .78 & *** & .94 & *** \\
f5 & Q22 & .55 & *** & .63 & *** & .71 & *** & .71 & *** & .72 & *** & .36 & *** \\
f5 & Q23 & .56 & *** & .49 & *** & .44 & *** & .42 & *** & .42 & *** & .70 & *** \\  \midrule
f6 & Q24 & .54 & *** & .62 & *** & .82 & *** & .82 & *** & .82 & *** & .68 & *** \\
f6 & Q25 & .73 & *** & .90 & *** & .78 & *** & .77 & *** & .78 & *** & .55 & *** \\ \bottomrule
\end{tabular}
\end{table}

\begin{table}[h]
\centering
\caption{Factor correlations}
\label{tab:factor_correlations}
\footnotesize
\begin{tabular}{ccccccccccccccccc}
\toprule
\multicolumn{2}{c}{Factor 1} & f1 & f1 & f1 & f1 & f1 & f2 & f2 & f2 & f2 & f3 & f3 & f3 & f4 & f4 & f5 \\
\multicolumn{2}{c}{Factor 2} & f2 & f3 & f4 & f5 & f6 & f3 & f4 & f5 & f6 & f4 & f5 & f6 & f5 & f6 & f6 \\ \midrule
\multirow{2}{*}{Grade 3} & & .73 & .66 & .54 & .54 & .40 & .73 & .55 & .59 & .46 & .58 & .60 & .60 & .85 & .70 & .89 \\
(Q1-25) & & *** & *** & *** & *** & *** & *** & *** & *** & *** & *** & *** & *** & *** & *** & *** \\ \midrule
\multirow{2}{*}{Grade 4} & & .61 & .62 & .55 & .43 & .34 & .75 & .46 & .53 & .41 & .60 & .56 & .47 & .78 & .70 & .77 \\
(Q1-25) & & *** & *** & *** & *** & *** & *** & *** & *** & *** & *** & *** & *** & *** & *** & *** \\ \midrule
\multirow{2}{*}{Grade 5} & & .88 & .57 & .33 & .45 & .14 & .88 & .78 & .67 & .68 & .76 & .72 & .79 & .80 & .88 & .95 \\
(Q1-25) & & *** & *** & *** & *** & * & *** & *** & *** & *** & *** & *** & *** & *** & *** & *** \\ \midrule
\multirow{2}{*}{Grade 5} & & .91 & .76 & .46 & .71 & .83 & .87 & .79 & .67 & .68 & .75 & .72 & .79 & .79 & .87 & .94 \\
(Q1+Q3-25) & & *** & *** & ** & *** & *** & *** & *** & *** & *** & *** & *** & *** & *** & *** & *** \\ \midrule
\multirow{2}{*}{Grade 5} & & .83 & .78 & .55 & .57 & .74 & .87 & .79 & .67 & .67 & .75 & .72 & .79 & .79 & .86 & .93 \\
(Q3-25) & & *** & *** & *** & *** & *** & *** & *** & *** & *** & *** & *** & *** & *** & *** & *** \\ \midrule
\multirow{2}{*}{Grade 6} & & .87 & .76 & .63 & .57 & .64 & .70 & .49 & .63 & .39 & .62 & .59 & .77 & .79 & .97 & .87 \\
(Q1-25) & & *** & *** & *** & *** & *** & *** & *** & *** & ** & *** & *** & *** & *** & *** & *** \\ \bottomrule
\end{tabular}
\end{table}

\pagebreak

\section{{Grade-specific IRT Item Difficulty and Discrimination Indices}}
\label{app:IRT_parameters}

\begin{table}[h]
    \centering
    \footnotesize
    \caption{2-PL IRT Difficulty (Dffclt) and Discrimination (Dscrmn) Parameters per grade. Please note that we also provide the equivalent difficulty for a 62\% of correct response probability for the student proficiency levels construction. }
    \label{tab:IRT_parameters}
    \begin{tabular}{c|ccP{1cm}|ccP{1cm}|ccP{1cm}|ccP{1cm}}
    \toprule
    & \multicolumn{3}{c}{Grade 3} & \multicolumn{3}{c}{Grade 4} & \multicolumn{3}{c}{Grade 5} & \multicolumn{3}{c}{Grade 6} \\ 
     & Dffclt & Dscrmn & Dffclt 62\% & Dffclt & Dscrmn & Dffclt 62\% & Dffclt & Dscrmn & Dffclt 62\% & Dffclt & Dscrmn & Dffclt 62\% \\ \midrule
    Q1 & -2.56 & 1.09 & -2.10 & -2.95 & 1.54 & -2.63 & -3.57 & 1.16 & -3.15 & -3.57 & 1.16 & -3.24 \\
    Q3 & -1.44 & 1.01 & -0.95 & -2.18 & 0.81 & -1.57 & -2.37 & 0.94 & -1.85 & -2.37 & 0.94 & -1.24 \\
    Q4 & -1.06 & 1.00 & -0.56 & -1.61 & 0.97 & -1.10 & -3.05 & 0.65 & -2.30 & -3.05 & 0.65 & -2.41 \\
    Q5 & -0.62 & 1.30 & -0.24 & -1.33 & 1.41 & -0.98 & -1.46 & 1.14 & -1.03 & -1.46 & 1.14 & -0.77 \\
    Q6 & -1.28 & 1.75 & -1.00 & -2.03 & 1.43 & -1.69 & -2.75 & 1.72 & -2.47 & -2.75 & 1.72 & -2.50 \\
    Q7 & -0.43 & 1.03 & 0.04 & -1.07 & 1.16 & -0.65 & -1.19 & 0.83 & -0.60 & -1.19 & 0.83 & -1.27 \\
    Q8 & -0.37 & 1.41 & -0.02 & -1.15 & 1.60 & -0.84 & -1.79 & 1.26 & -1.40 & -1.79 & 1.26 & -1.64 \\
    Q9 & -0.92 & 1.12 & -0.49 & -1.65 & 1.21 & -1.24 & -2.50 & 1.05 & -2.03 & -2.50 & 1.05 & -1.65 \\
    Q10 & 0.57 & 1.42 & 0.92 & -0.17 & 1.51 & 0.15 & -0.78 & 1.35 & -0.42 & -0.78 & 1.35 & -0.22 \\
    Q11 & 0.30 & 1.69 & 0.59 & -0.29 & 1.83 & -0.02 & -0.77 & 1.94 & -0.52 & -0.77 & 1.94 & -0.58 \\
    Q12 & -0.19 & 1.48 & 0.15 & -0.72 & 1.93 & -0.47 & -1.32 & 2.65 & -1.13 & -1.32 & 2.65 & -1.03 \\
    Q13 & 0.32 & 1.78 & 0.59 & -0.38 & 2.4 & -0.17 & -0.88 & 2.06 & -0.65 & -0.88 & 2.06 & -0.83 \\
    Q14 & 0.70 & 0.94 & 1.22 & -0.2 & 1.25 & 0.15 & -0.63 & 1.05 & -0.16 & -0.63 & 1.05 & -0.47 \\
    Q15 & 0.67 & 1.4 & 1.01 & -0.2 & 1.97 & 0.01 & -0.78 & 1.67 & -0.49 & -0.78 & 1.67 & -0.49 \\
    Q16 & -0.2 & 1.19 & 0.18 & -0.80 & 1.07 & -0.34 & -1.15 & 1.2 & -0.75 & -1.15 & 1.2 & -1.05 \\
    Q17 & 2.08 & 1.08 & 2.53 & 2.37 & 0.85 & 2.94 & 0.63 & 1.39 & 0.99 & 0.63 & 1.39 & 1.19 \\
    Q18 & -0.36 & 1.03 & 0.11 & -1.00 & 1.04 & -0.53 & -1.51 & 1.02 & -1.03 & -1.51 & 1.02 & -1.29 \\
    Q19 & 0.29 & 0.95 & 0.80 & -0.51 & 0.91 & 0.03 & -1.09 & 1.06 & -0.63 & -1.09 & 1.07 & -0.52 \\
    Q20 & 1.03 & 0.82 & 1.63 & 0.49 & 1.08 & 0.95 & 0.03 & 1.17 & 0.5 & 0.03 & 1.17 & 0.79 \\
    Q21 & 0.16 & 1.05 & 0.63 & -0.2 & 1.31 & 0.14 & -1.33 & 1.42 & -0.98 & -1.33 & 1.42 & -0.85 \\
    Q22 & 0.66 & 0.84 & 1.25 & 0.47 & 0.94 & 0.99 & -0.70 & 1.28 & -0.31 & -0.70 & 1.28 & 0.73 \\
    Q23 & 0.65 & 0.83 & 1.2 & 0.13 & 0.68 & 0.85 & -1.71 & 0.57 & -0.84 & -1.71 & 0.57 & -0.76 \\
    Q24 & 2.80 & 0.79 & 3.42 & 2.46 & 0.93 & 2.99 & 0.66 & 1.78 & 0.93 & 0.66 & 1.78 & 1.14 \\
    Q25 & 1.47 & 1.06 & 1.93 & 0.96 & 1.25 & 1.35 & -0.18 & 1.52 & 0.14 & -0.18 & 1.52 & 0.514 \\ \midrule
    M & 0.09 & 1.17 & 0.54 & -0.49 & 1.30 & -0.07 & -1.26 & 1.33 & -0.84 & -1.26 & 1.33 & -0.77 \\
    SD & 1.12 & 0.30 & 1.19 & 1.27 & 0.43 & 1.31 & 1.06 & 0.47 & 1.01 & 1.06 & 0.47 & 1.12 \\
    Min & -2.55 & 0.79 & -2.10 & -2.95 & 0.68 & -2.63 & -3.57 & 0.57 & -3.15 & -3.57 & 0.57 & -3.2 \\
    25\% & -0.48 & 0.99 & -0.08 & -1.19 & 0.96 & -0.88 & -1.73 & 1.05 & -1.20 & -1.73 & 1.05 & -1.27 \\
    50\% & 0.23 & 1.07 & 0.59 & -0.44 & 1.23 & -0.10 & -1.17 & 1.2 & -0.70 & -1.17 & 1.2 & -0.80 \\
    75\% & 0.67 & 1.42 & 1.22 & -0.10 & 1.52 & 0.33 & -0.76 & 1.56 & -0.39 & -0.76 & 1.56 & -0.41 \\
    MAX & 2.80 & 1.78 & 3.42 & 2.46 & 2.4 & 2.99 & 0.66 & 2.65 & 0.99 & 0.66 & 2.65 & 1.19 \\ \bottomrule
    \end{tabular}
\end{table}

\pagebreak

\section{{Grade-specific Gender Differential Item Functioning}}
\label{app:GenderDIF_FULL}

\begin{landscape}
\begin{table}[h]
\centering
\caption{Differential Item Functioning between genders (girls = focal). Abbreviations: Stat. = statistic, Adj. P. = adjusted p-value using Benjamini-Hochberg p-value correction).}
\label{tab:GenderDIF_FULL}
\footnotesize
\begin{tabular}{l|llP{1.15cm}|llP{1.15cm}P{1.15cm}|ll|llll}
\toprule
  & \multicolumn{3}{c|}{\textbf{Mantel-Haenszel (M.-H.) $\chi^2$}} & \multicolumn{4}{c|}{\textbf{Logistic Regression Likelihood Ratio Test (LRT)}} & \multicolumn{2}{c|}{\textbf{Generalized Lord's $\chi^2$}} & \multicolumn{4}{c}{\textbf{Synthesis}} \\
 & Stat. & Adj. P & Effect size (ETS Delta scale) & Stat. & Adj. P & $R^2$  (Nagelkerke's $R^2$) & Effect size (Jodoin \& Gierl) & Stat & Adj. P & M.-H. & LRT & Lord & \#DIF \\ \midrule
 Q1 & 1.02 & .3131 & Negligeable & 2.32 & .79 & .0027 & Negligeable & 1.44 & 0.757 & NoDIF & NoDIF & NoDIF & 0/3 \\
Q2 & 0.13 & .722 & Negligeable & 0.92 & .9237 & .0008 & Negligeable & 1.51 & 0.757 & NoDIF & NoDIF & NoDIF & 0/3 \\
Q3 & 0.74 & .389 & Negligeable & 1.77 & .7873 & .0006 & Negligeable & 2.05 & 0.757 & NoDIF & NoDIF & NoDIF & 0/3 \\
Q4 & 0.00 & .9714 & Negligeable & 0.30 & .9893 &.0001 & Negligeable & 0.81 & 0.7944 & NoDIF & NoDIF & NoDIF & 0/3 \\
Q5 & 0.42 & .5166 & Negligeable & 1.22 & .9069 & .0004 & Negligeable & 1.46 & 0.757 & NoDIF & NoDIF & NoDIF & 0/3 \\
Q6 & 0.18 & .6721 & Negligeable & 2.21 & .7873 & .0007 & Negligeable & 3.11 & 0.757 & NoDIF & NoDIF & NoDIF & 0/3 \\
Q7 & 0.02 & .8977 & Negligeable & 0.34 & .9893 & .0001 & Negligeable & 1.68 & 0.757 & NoDIF & NoDIF & NoDIF & 0/3 \\
Q8 & 5.18 & .0228* & Negligeable & 6.24 & .4432 & .0018 & Negligeable & 4.65 & 0.6103 & DIF & NoDIF & NoDIF & 1/3 \\
Q9 & 0.00 & .9773 & Negligeable & 0.16 & .9893 & .0001 & Negligeable & 1.08 & 0.7666 & NoDIF & NoDIF & NoDIF & 0/3 \\
Q10 & 0.38 & .5365 & Negligeable & 0.09 & .9893 & $<0.0001$ & Negligeable & 1.83 & 0.757 & NoDIF & NoDIF & NoDIF & 0/3 \\
Q11 & 2.49 & .1147 & Negligeable & 1.65 & .7873 & .0004 & Negligeable & 1.93 & 0.757 & NoDIF & NoDIF & NoDIF & 0/3 \\
Q12 & 1.14 & .2852 & Negligeable & 3.55 & .7154 & .0009 & Negligeable & 0.33 & 0.8661 & NoDIF & NoDIF & NoDIF & 0/3 \\
Q13 & 0.60 & .4394 & Negligeable & 2.17 & .7873 & .0005 & Negligeable & 6.21 & 0.3735 & NoDIF & NoDIF & NoDIF & 0/3 \\
Q14 & 6.14 & .0132* & Negligeable & 5.81 & .4432 & .0017 & Negligeable & 2.12 & 0.757 & DIF & NoDIF & NoDIF & 1/3 \\
Q15 & 2.27 & .1322 & Negligeable & 1.64 & .7873 & .0004 & Negligeable & 0.30 & 0.8661 & NoDIF & NoDIF & NoDIF & 0/3 \\
Q16 & 0.00 & .9787 & Negligeable & 0.14 & .9893 & $<0.0001$ & Negligeable & 1.26 & 0.757 & NoDIF & NoDIF & NoDIF & 0/3 \\
Q17 & 1.86 & .1721 & Negligeable & 2.14 & .7873 & .0007 & Negligeable & 1.46 & 0.757 & NoDIF & NoDIF & NoDIF & 0/3 \\
Q18 & 0.23 & .6282 & Negligeable & 0.82 & .9237 & .0002 & Negligeable & 3.04 & 0.757 & NoDIF & NoDIF & NoDIF & 0/3 \\
Q19 & 6.21 & .0127* & Negligeable & 6.51 & .4432 & .0019 & Negligeable & 8.95 & 0.1426 & DIF & NoDIF & NoDIF & 1/3 \\
Q20 & 2.65 & .1037 & Negligeable & 2.07 & .7873 & .0006 & Negligeable & 0.29 & 0.8661 & NoDIF & NoDIF & NoDIF & 0/3 \\
Q21 & 3.10 & .0781 & Negligeable & 3.52 & .7154 & .001 & Negligeable & 3.60 & 0.757 & NoDIF & NoDIF & NoDIF & 0/3 \\
Q22 & 1.04 & .3082 & Negligeable & 1.06 & .9207 & .0003 & Negligeable & 1.21 & 0.757 & NoDIF & NoDIF & NoDIF & 0/3 \\
Q23 & 0.04 & .838 & Negligeable & 0.02 & .9893 & $<0.0001$ & Negligeable & 0.97 & 0.7707 & NoDIF & NoDIF & NoDIF & 0/3 \\
Q24 & 1.82 & .1769 & Negligeable & 0.48 & .9893 & .0001 & Negligeable & 0.64 & 0.8232 & NoDIF & NoDIF & NoDIF & 0/3 \\
Q25 & 2.06 & .151 & Negligeable & 5.29 & .4432 & .0015 & Negligeable & 9.76 & 0.1426 & NoDIF & NoDIF & NoDIF & 0/3\\ \bottomrule
\end{tabular}

Please note that the DifR package does not provide the effect size for Lord's $\chi^2$ statistic.
\end{table}
\end{landscape}

\pagebreak

\section{{Grade-related Differential Item Functioning}}
\label{app:GradeDIF_FULL}

\begin{landscape}
\begin{table}[h]
\centering
\caption{Differential Item Functioning between grades 3-4 (reference) and grades 5-6 (focal). Abbreviations: Stat. = statistic, Adj. P. = adjusted p-value using Benjamini-Hochberg p-value correction).}
\label{tab:GradeDIF_FULL}
\footnotesize
\begin{tabular}{l|llP{1.15cm}|llP{1.15cm}P{1.15cm}|ll|llll}
\toprule
  & \multicolumn{3}{c|}{\textbf{Mantel-Haenszel (M.-H.) $\chi^2$}} & \multicolumn{4}{c|}{\textbf{Logistic Regression Likelihood Ratio Test (LRT)}} & \multicolumn{2}{c|}{\textbf{Generalized Lord's $\chi^2$}} & \multicolumn{4}{c|}{\textbf{Synthesis}} \\
 & Stat. & Adj. P & Effect size (ETS Delta scale) & Stat. & Adj. P & $R^2$  (Nagelkerke's $R^2$) & Effect size (Jodoin \& Gierl) & Stat & Adj. P & M.-H. & LRT & Lord & \#DIF \\ \midrule
Q1 & 61.4 & 0*** & Large & 9.1 & .0113* & 0.0096 & Negligible & 8.3 & .0217* & DIF & DIF & DIF & 3/3 \\
Q2 & 101.7 & 0*** & Large & 0.1 & .9651 & 0.0001 & Negligible & 25.8 & 0*** & DIF & No DIF & DIF & 2/3 \\
Q3 & 95.1 & 0*** & Large & 49.3 & 0*** & 0.0118 & Negligible & 30.4 & 0*** & DIF & DIF & DIF & 3/3 \\
Q4 & 134 & 0*** & Large & 0.2 & .9506 & 0 & Negligible & 29.0 & 0*** & DIF & No DIF & DIF & 2/3 \\
Q5 & 118.1 & 0*** & Large & 37.2 & 0*** & 0.008 & Negligible & 73.1 & 0*** & DIF & DIF & DIF & 3/3 \\
Q6 & 130.3 & 0*** & Large & 9.7 & .009** & 0.0027 & Negligible & 15.7 & .0008*** & DIF & DIF & DIF & 3/3 \\
Q7 & 137.9 & 0*** & Large & 21.7 & 0*** & 0 & Negligible & 38.0 & 0*** & DIF & DIF & DIF & 3/3 \\
Q8 & 186.9 & 0*** & Large & 15.1 & .0006*** & 0.0033 & Negligible & 28.9 & 0*** & DIF & DIF & DIF & 3/3 \\
Q9 & 113 & 0*** & Large & 81.5 & 0*** & 0.0214 & Negligible & 10.3 & .009** & DIF & DIF & DIF & 3/3 \\
Q10 & 117.6 & 0*** & Large & 181.6 & 0*** & 0 & Negligible & 31.3 & 0*** & DIF & DIF & DIF & 3/3 \\
Q11 & 133 & 0*** & Large & 148.6 & 0*** & 0 & Negligible & 19.0 & .0002*** & DIF & DIF & DIF & 3/3 \\
Q12 & 157.8 & 0*** & Large & 175.7 & 0*** & 0.0398 & Moderate & 1.7 & .4754 & DIF & DIF & No DIF & 2/3 \\
Q13 & 149.9 & 0*** & Large & 244.8 & 0*** & 0 & Negligible & 13.7 & .0019** & DIF & DIF & DIF & 3/3 \\
Q14 & 115.8 & 0*** & Large & 117.6 & 0*** & 0 & Negligible & 18.5 & .0002*** & DIF & DIF & DIF & 3/3 \\
Q15 & 162.4 & 0*** & Large & 174.7 & 0*** & 0 & Negligible & 16.7 & .0005*** & DIF & DIF & DIF & 3/3 \\
Q16 & 237.2 & 0*** & Large & 77.0 & 0*** & 0.0169 & Negligible & 11.6 & .0052** & DIF & DIF & DIF & 3/3 \\
Q17 & 251.1 & 0*** & Large & 55.0 & 0*** & 0 & Negligible & 0.7 & .7053 & DIF & DIF & No DIF & 2/3 \\
Q18 & 245.9 & 0*** & Large & 104.1 & 0*** & 0.0238 & Negligible & 5.0 & .1082 & DIF & DIF & No DIF & 2/3 \\
Q19 & 271.2 & 0*** & Large & 86.3 & 0*** & 0.577 & Large & 4.9 & .1082 & DIF & DIF & No DIF & 2/3 \\
Q20 & 162.9 & 0*** & Large & 44.1 & 0*** & 0 & Negligible & 34.7 & 0*** & DIF & DIF & DIF & 3/3 \\
Q21 & 253 & 0*** & Large & 247.7 & 0*** & 0.5883 & Large & 3.9 & .1663 & DIF & DIF & No DIF & 2/3 \\
Q22 & 193.9 & 0*** & Large & 88.8 & 0*** & 0 & Negligible & 3.1 & .2435 & DIF & DIF & No DIF & 2/3 \\
Q23 & 271.3 & 0*** & Large & 226.4 & 0*** & 0 & Negligible & 14.8 & .0012** & DIF & DIF & DIF & 3/3 \\
Q24 & 225.4 & 0*** & Large & 45.9 & 0*** & 0.5763 & Large & 9.6 & .0121* & DIF & DIF & DIF & 3/3 \\
Q25 & 277 & 0*** & Large & 104.9 & 0*** & 0 & Negligible & 1.4 & .5176 & DIF & DIF & No DIF & 2/3 \\ \bottomrule
\end{tabular}

Please note that the DifR package does not provide the effect size for Lord's $\chi^2$ statistic.

\end{table}
\end{landscape}

\begin{figure}[h]
    \centering
    \includegraphics[width=0.47\textwidth]{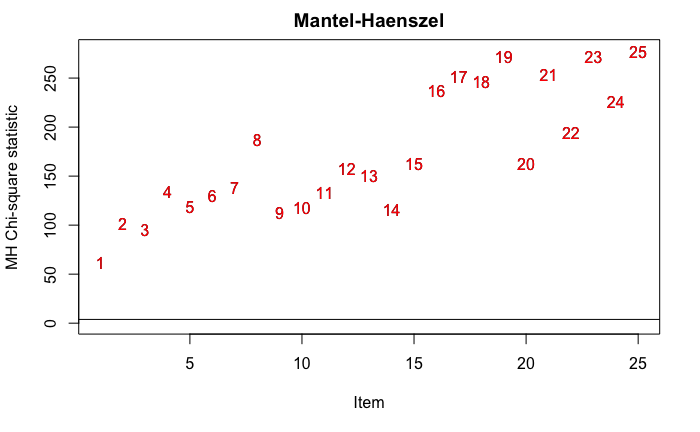}
    \includegraphics[width=0.47\textwidth]{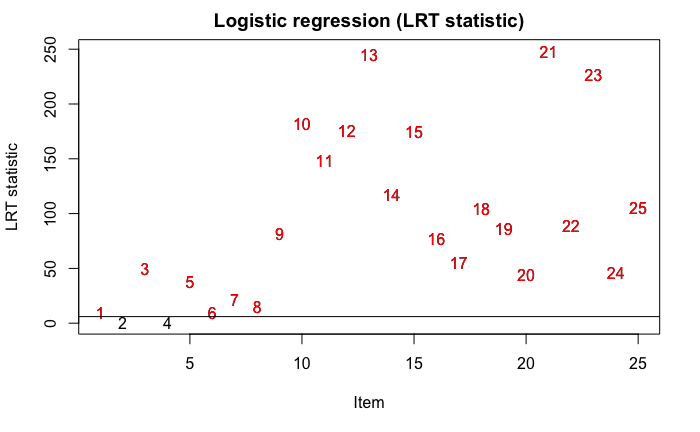}
    \includegraphics[width=0.47\textwidth]{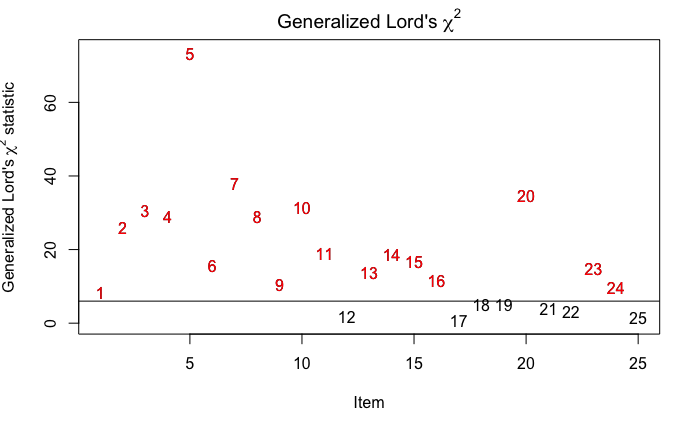}
    \caption{Results of the DIF according to the detection method employed. The higher the statistic for a given item, the more the item is DIF between grades 3-4 and 5-6.}
    \label{fig:GradeDIF}
\end{figure}

\FloatBarrier

\section{Grade-agnostic IRT Model and Wright map}
\label{app:IRT_grade_agnostic}

\begin{table}[h]
    \caption{Grade-agnostic cCTt (excluding item Q2) IRT Model comparison using ANOVA}
    \label{tab:IRT_model_fit_grade_agnostic}
    \footnotesize
    
    \centering
    \begin{tabular}{cccccccc}
    \toprule
    Grade & Model & AIC & BIC & Log Likelihood & LRT & df & p-value \\ \midrule
    All & 1PL & 62377.34 &	62524.58 &	-31163.67	 &  &  &   \\
     & 2PL & 61891.00 &	62173.69	& -30897.50	& 532.34	& 23 &	$<.001$ \\ \bottomrule
    \end{tabular}
\end{table}

\begin{table}[h]
    \centering
    \caption{Grade-agnostic cCTt (excluding item Q2) IRT 2PL Model Parameters}
    \label{tab:IRT_parameters_grade_agnostic}
    \begin{tabular}{cccc}
    \toprule
    & Dffclt & Dscrmn & Dffclt 62\% \\ \midrule 
    Q1 & -2.91 & 1.28 & -2.53 \\
    Q3 & -2.04 & 0.81 & -1.43 \\
    Q4 & -1.71 & 0.93 & -1.18 \\
    Q5 & -1.18 & 1.15 & -0.75 \\
    Q6 & -1.90 & 1.61 & -1.59 \\
    Q7 & -1.03 & 0.99 & -0.54 \\
    Q8 & -1.08 & 1.52 & -0.76 \\
    Q9 & -1.51 & 1.28 & -1.13 \\
    Q10 & -0.18 & 1.54 & 0.14 \\
    Q11 & -0.34 & 2.08 & -0.10 \\
    Q12 & -0.78 & 2.14 & -0.55 \\
    Q13 & -0.42 & 2.40 & -0.22 \\
    Q14 & -0.22 & 1.22 & 0.18 \\
    Q15 & -0.23 & 1.96 & 0.02 \\
    Q16 & -0.80 & 1.24 & -0.41 \\
    Q17 & 1.26 & 1.27 & 1.65 \\
    Q18 & -0.98 & 1.20 & -0.57 \\
    Q19 & -0.48 & 1.13 & -0.04 \\
    Q20 & 0.42 & 1.01 & 0.91 \\
    Q21 & -0.52 & 1.54 & -0.21 \\
    Q22 & 0.10 & 0.98 & 0.60 \\
    Q23 & -0.30 & 1.00 & 0.19 \\
    Q24 & 1.36 & 1.32 & 1.73 \\
    Q25 & 0.52 & 1.43 & 0.86 \\ \midrule
    M & -0.62 & 1.38 & -0.24 \\
    SD & 1.00 & 0.41 & 0.99 \\
    Min & -2.91 & 0.81 & -2.53 \\
    25\% & -1.10 & 1.10 & -0.75 \\
    50\% & -0.50 & 1.27 & -0.21 \\
    75\% & -0.21 & 1.53 & 0.18 \\
    MAX & 1.357 & 2.40 & 1.73 \\ \bottomrule
    \end{tabular}
\end{table}

\begin{figure}[h]
    \centering
    \includegraphics[width=0.8\textwidth]{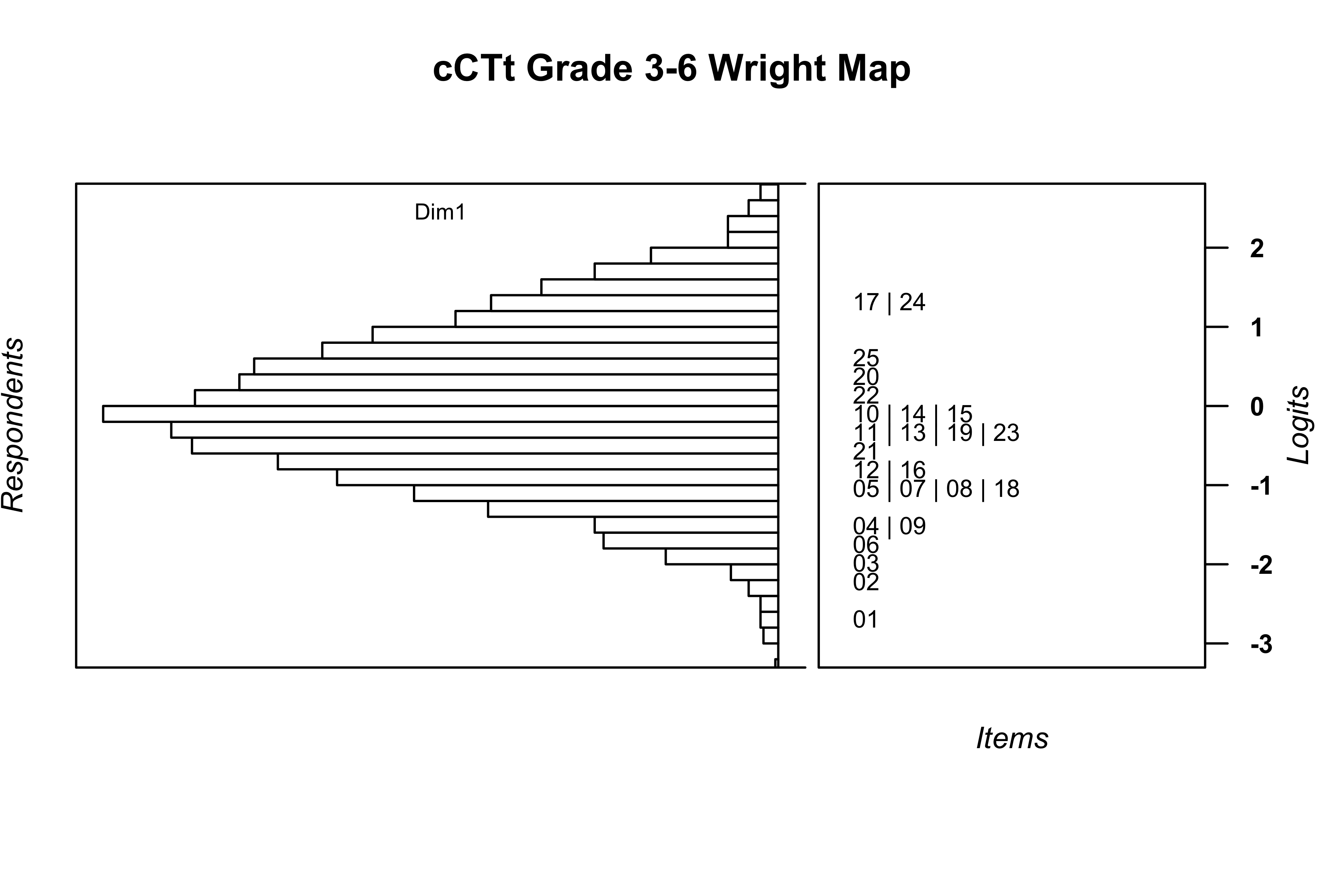}
    \caption{Grade-agnostic cCTt (excluding item Q2) IRT 2PL model Wright Map for all grade 3-6 students}
    \label{fig:WrightMapGradeAgnostic}
\end{figure}

\end{document}